\documentclass[reprint,amsmath,amssymb,aip,jcp]{revtex4-2}
\usepackage{dcolumn}
\usepackage{bm}
\usepackage[utf8]{inputenc}
\usepackage{url}
\usepackage{graphicx}
\usepackage{xcolor}
\usepackage{cancel}
\usepackage{ulem}
\usepackage{float}
\definecolor{link}{rgb}{0.0, 0.0, 0.0 }
\usepackage[bookmarks = true,
			pdfstartview = FitH,
			colorlinks = true,
			urlcolor=link,
			citecolor=link,
			linkcolor=link,
			hyperfootnotes=false]{hyperref}
\usepackage[capitalise]{cleveref}
\crefname{section}{Sec.}{Secs.}
\Crefname{section}{Section}{Sections}
\crefname{figure}{Fig.}{Figs.}
\crefname{equation}{Eq.}{Eqs.}
\AddToHook{cmd/appendix/before}{%
  \crefalias{section}{appendix}%
  \crefalias{subsection}{appendix}%
}
\usepackage{subfiles}
\begin{document}
\title{First-Principles Nonadiabatic Dynamics via the Multi-Orbital Anderson-Newns Model}
\author{Liwen Ko}
\author{Joonho Lee}
\email{joonholee@g.harvard.edu}
\affiliation{Department of Chemistry and Chemical Biology, Harvard University}
\date{\today}

\begin{abstract}
We develop a first-principles theory for nonadiabatic surface dynamics, providing a fit-free connection between density functional theory (DFT) calculations and the effective multi-orbital Anderson-Newns (AN) theory. Our theory contains two main advances.
First, we outline the multi-orbital AN theory with orbital overlap and derive closed-form expressions for the hybridization energy, electronic dynamics, and electronic friction.
Second, we describe a procedure to map the DFT Hamiltonian into the effective AN Hamiltonian with nuclear-position dependence. We obtain the electronic part of the AN Hamiltonian solely from the adsorbate-projected density-of-states matrix. We then define the bare nuclear potential as the difference between the total energy and the hybridization energy.
The theory is applied to H and CO on the Cu surface, where widely used assumptions about the hybridization function, including the wide-band limit, semi-elliptical forms, and separability in energy and nuclear coordinate, are found to fail, and the single-orbital description breaks down qualitatively for CO.
We expect this work to be broadly useful for first-principles modeling of coupled nuclear-electronic dynamics and chemical reactions at metallic surfaces.
\end{abstract}

\maketitle

\section{Introduction}
Adsorbate-surface systems, whether in the gas phase or in solution, are ubiquitous in scientific research and industrial processes. Some examples include the Haber-Bosch process,\cite{ertl1980surface} steam reforming,\cite{sanna2025steam} electrocatalysis,\cite{schmickler2010interfacial, eliaz2019physical} and batteries.\cite{schmickler2010interfacial, eliaz2019physical} Since the macroscopic surface contains a continuum of electronic states, adsorbate-surface systems are often highly nonadiabatic, especially when the surface is metallic. 
Nuclear motions of adsorbates are coupled to low-energy electron-hole excitations on the metallic surface, and the Born-Oppenheimer (BO) approximation breaks down.\cite{persson1980vibrational,dou2017born} 
\par
The Anderson-Newns (AN) Hamiltonian\cite{newns1969, anderson1961localized} has been widely used as a framework for describing nonadiabatic adsorbate-surface dynamics\cite{gardner2023efficient, gardner2023assessing,preston2025nonadiabatic, dan2026nonadiabatic, shenvi2009nonadiabatic,dou2017born, dou2015frictional,brandbyge1995electronically,mizielinski2005electronic} while modeling surface electrons as non-interacting fermions.
In addition, the AN Hamiltonian has also been applied to study electron transfer dynamics,\cite{nitzan2001electron,ghan2023interpreting, hertl2026first} model free energy surfaces in electrochemical interfaces,\cite{santos2006model, santos2009model} study Kondo physics in strongly correlated systems, \cite{shick2022spin,valli2020kondo,jacob2010dynamical} and interpret adsorption energy trends.\cite{greiner2018free, tiwari2021reactivity,vijay2022limits,vojvodic2014electronic, norskov2014fundamental}
The total Hamiltonian including nuclear degrees of freedom is
\begin{equation}
    \hat{H}_\text{tot} = \hat{T}_N + \hat{V}_N(R) + \underbrace{\hat{H}_\text{ad}(R)+\hat{H}_{\text{bath}}+\hat{H}_{\text{coup}}(R)}_{\hat{H}_{el}(R)},
\label{Eq:H_tot}
\end{equation}
where $R$ is the adsorbate nuclear coordinate. $\hat{T}_N$ and $\hat{V}_N(R)$ are the nuclear kinetic and potential energy operators, respectively. The remaining terms are given by 
\begin{equation}
    \hat{H}_{\text{ad}}(R) = \sum_{\mu,\nu\in\text{ad.}} \overline{H}_{\mu\nu}(R) a^\dagger_\mu a_\nu,
\label{Eq:H_ad_second_quantized}
\end{equation}
\begin{equation}
    \hat{H}_{\text{bath}} = \sum_{\sigma\in\text{surf.}} E_\sigma a_\sigma^\dagger a_\sigma,
\label{Eq:H_bath}
\end{equation}
and
\begin{equation}
    \hat{H}_{\text{coup}}(R) = \sum_{\substack{\mu\in\text{ad.}\\ \sigma\in\text{surf.}}} \big( \overline{V}_{\mu \sigma}(R)a_\mu^\dagger a_\sigma+\overline{V}_{\sigma\nu}(R)a_\sigma^\dagger a_\nu\big).
\label{Eq:H_coup}
\end{equation}
$a_\lambda$ is the electron annihilation operator associated with the orbital $|\overline{\phi}_\lambda\rangle$; the overbar indicates that the underlying orbital basis is orthonormal (see \cref{Sec:adsorbate_hamiltonian} for a summary of our notation). The orbital either belongs to the adsorbate (ad.) or the surface (surf.).
$\hat{H}_\text{ad}(R)$ is the adsorbate electronic Hamiltonian. 
In $\hat{H}_{\text{bath}}$, $E_\sigma$ is the band energy of the surface orbital, $\sigma$. We consider $E_\sigma$ values to lie on a fixed grid that is dense enough such that $E_\sigma$ is independent of $R$ and can represent the continuum of surface states.
In $\hat{H}_{\text{coup}}$, $\overline{V}_{\mu \sigma}(R)$ is the adsorbate-surface coupling matrix element. 
The adsorbate-surface interaction is characterized by the hybridization function,
\begin{equation}
    \Delta_{\mu\nu}(E,R) = \pi \sum_{\sigma\in\text{surf.}} \overline{V}_{\mu \sigma}(R) \overline{V}_{\sigma\nu}(R) \delta(E-E_\sigma),
\label{Eq:Delta_def}
\end{equation}
where $\mu$ and $\nu$ are adsorbate orbitals.
In the continuum limit of metals, the sum over surface states becomes an integral, and $\mathbf{\Delta}(E,R)$ becomes a smooth function. 

\par
The AN Hamiltonian used in previous studies falls into two broad categories: first-principles and phenomenological models.
Electronic structure studies of metal adatoms exhibiting strong-correlation effects \cite{shick2022spin,valli2020kondo,jacob2010dynamical} typically use first-principles AN Hamiltonians, which contain multiple adsorbate orbitals and include electron-electron (e-e) interaction explicitly.
The multi-orbital hybridization function is directly obtained from density functional theory (DFT)\cite{shick2022spin,valli2020kondo} or from dynamical mean-field theory (DMFT) self-consistency loops.\cite{jacob2010dynamical}
However, because the e-e interaction is treated explicitly, physical observables must be computed with many-body impurity solvers, such as quantum Monte Carlo\cite{valli2020kondo,jacob2010dynamical} or exact diagonalization.\cite{shick2022spin}
These studies therefore address electronic properties including correlation effects at fixed nuclear geometries.

\par
In nonadiabatic dynamics studies, where one is interested in how nuclear degrees of freedom couple to electronic degrees of freedom, a phenomenological, single-orbital AN Hamiltonian is used. \cite{gardner2023efficient, gardner2023assessing,preston2025nonadiabatic, dan2026nonadiabatic, shenvi2009nonadiabatic,dou2017born, dou2015frictional,brandbyge1995electronically,mizielinski2005electronic} Typically, only a single adsorbate orbital is considered (thus $\Delta(E,R)$  becomes a scalar quantity), and $\Delta(E,R)$ is commonly taken to be independent of $E$ (known as the wide-band limit) or to have simple functional forms. Explicit e-e interaction is rarely included, \cite{dou2017born} and the nuclear potential energy $V(R)$ is chosen to have simple functional forms such as the Morse potential or the harmonic potential. 
We note that this type of phenomenological single-orbital AN Hamiltonian has also been used for studying binding energy trends for catalysis, where $\Delta(E)$ is taken to be semi-elliptical, and the dependence on $R$ is usually not considered. \cite{greiner2018free, tiwari2021reactivity,vijay2022limits,vojvodic2014electronic, norskov2014fundamental}
The simplicity of the phenomenological AN Hamiltonian allows simulation of coupled nuclear-electron dynamics and understanding of binding-energy trends.
Furthermore, physical quantities such as the hybridization energy can be expressed in closed form under the single-orbital AN Hamiltonian. \cite{vojvodic2014electronic, norskov2014fundamental} 

\par
\begin{figure}
    \centering
    \includegraphics[width=0.75\linewidth]{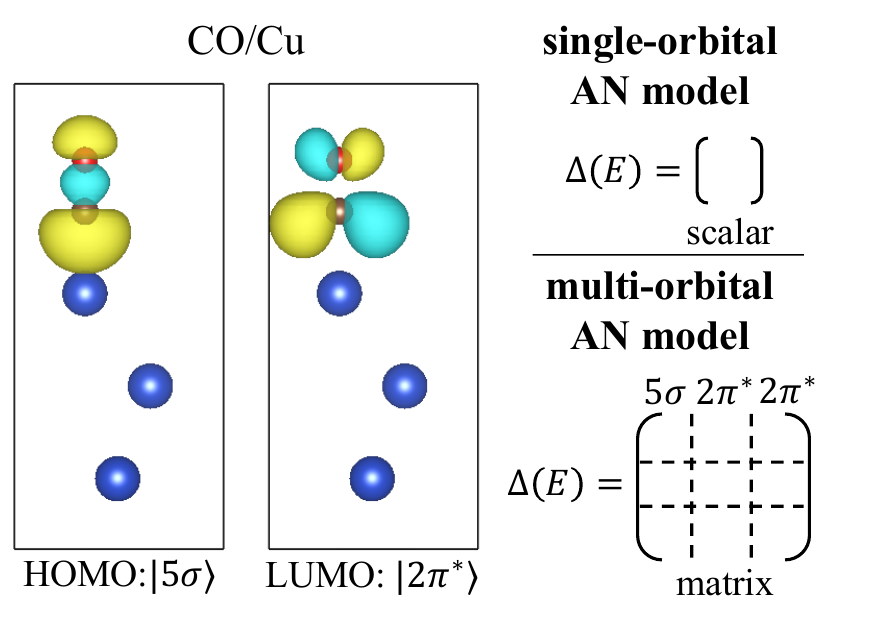}
    \caption{\textbf{Describing CO/Cu using single vs. multi-orbital AN model.} In the single-orbital AN model, only one adsorbate orbital interacts with the surface, and the interaction is characterized by a scalar hybridization function $\Delta(E)$. In the multi-orbital AN model, we consider all adsorbate frontier orbitals ($5\sigma$ HOMO and the doubly degenerate $2\pi^*$ LUMOs), so that $\mathbf{\Delta}(E)$ becomes matrix-valued, resulting in a more complete picture of the adsorbate-surface interaction. 
    }
\label{fig:single_vs_multi_orbital_schematics}
\end{figure}

However, previous studies using the phenomenological AN Hamiltonian suffer from two major limitations. First, representing the adsorbate with a single orbital can be questionable for more complex adsorbates. For example, CO adsorption on metal surfaces involves both the electron donation from the $5\sigma$ HOMO (highest occupied molecular orbital) and the back-donation into the $2\pi^*$ LUMOs (lowest unoccupied molecular orbital) \cite{blyholder1964molecular,aizawa1998first} (see \cref{fig:single_vs_multi_orbital_schematics}). Second, the form of $\Delta(E,R)$ is usually assumed and fitted, rather than derived from {\it ab initio} calculations such as DFT. The validity of these approximations in realistic adsorbate-surface systems has not been examined rigorously. The lack of a rigorous procedure to construct a multi-orbital AN Hamiltonian with nuclear degrees of freedom from DFT calculations has hindered the application of the AN Hamiltonian to study nonadiabatic dynamics in adsorbate-surface systems, which is the challenge we address in this work (see \cref{Tab:compare_previous_work}).

\par
In this work, we first give an overview of the multi-orbital AN theory, including orbital overlaps. We then show a fit-free mapping from DFT to the multi-orbital AN Hamiltonian, including the bare nuclear potential $\hat{V}_N(R)$. 
We derive analytical expressions for hybridization energy, electronic dynamics, and electronic friction for the multi-orbital AN model. 
Different from recently developed projection-operator diabatization (POD2) approaches \cite{ghan2020improved,ghan2023interpreting,hertl2026first} which compute the single-orbital $\Delta(E,R)$ through a series of orbital orthogonalizations, our method makes use of a Kramers-Kronig relation between the projected density of states (PDOS) and the Green's function. 
Starting from the adsorbate PDOS in a non-orthogonal basis (where adsorbate orbitals overlap with surface orbitals), we can directly compute $\mathbf{\Delta}(E,R)$ in an orthogonalized basis (where adsorbate orbitals are orthogonal to surface orbitals), without explicit change-of-basis.
Because every parameter of the AN Hamiltonian ($\hat{V}_N(R)$, $\mathbf{\Delta}(E,R)$, $\hat{H}_{ad}(R)$) is obtained directly from DFT without fitting or an assumed functional form, the approximations underlying previous AN studies, such as the single-orbital, energy-independent, or semi-elliptic forms of $\mathbf{\Delta}(E,R)$, can be tested rather than assumed.
\begin{table*}
\begin{tabular}{|c|c|c|c|c|c|}
     \hline
     Context &Nonadiabatic dynamics  & Catalysis & Kondo & POD2 & This work \\
     & \cite{gardner2023efficient, gardner2023assessing,preston2025nonadiabatic, dan2026nonadiabatic, shenvi2009nonadiabatic,dou2017born, dou2015frictional,brandbyge1995electronically,mizielinski2005electronic}& \cite{greiner2018free, tiwari2021reactivity,vijay2022limits,vojvodic2014electronic, norskov2014fundamental} & physics \cite{shick2022spin,valli2020kondo,jacob2010dynamical} &  \cite{ghan2023interpreting,hertl2026first}& \\
     \hline
     single vs. & single & single & multi & single & multi \\
     multi-orbital & &&& & \\
     \hline
     $\Delta(E,R)$ & separable $f(E)g(R)$,  & semi-ellipse & self-consistency & from DFT & from DFT \\
     & $f(E)$ usually constant && or from $G(E)$ &($H_{KS}$)&(PDOS)\\
     \hline
     $V_N(R)$ & harmonic or fit to Morse & N/A & N/A & N/A & from DFT \\
     \hline
\end{tabular}
\caption{\textbf{Comparison with previous studies that use the AN Hamiltonian.} N/A: not available. POD: projection operator diabatization.~\cite{hertl2026first}}
\label{Tab:compare_previous_work}
\end{table*}

\section{Mapping between DFT and AN Hamiltonian}

We first summarize the notation used throughout the paper. A symbol with a hat (e.g., $\hat{A}$) denotes an operator, which is independent of the basis. A symbol without a hat (e.g., $\mathbf{A}$ or $\overline{\mathbf{A}}$) denotes the matrix representation of the corresponding operator in a certain orbital basis. An overbar (e.g., $\overline{\mathbf{A}}$ or $|\overline{\phi}_\lambda\rangle$) emphasizes that the underlying basis is orthonormal, e.g., $\overline{H}_{\lambda\tau}=\langle \overline{\phi}_\lambda|\hat{H}|\overline{\phi}_\tau\rangle$. A symbol without an overbar (e.g., $\mathbf{A}$ or $|\phi_\lambda\rangle$) refers to a basis that is in general non-orthonormal. In particular, the fermionic operators $a_\lambda$ must be associated with orthonormal orbitals $|\overline{\phi}_\lambda\rangle$ in order to satisfy the canonical anti-commutation relations. This is why $\overline{H}$ and $\overline{V}$ in \cref{Eq:H_ad_second_quantized,Eq:H_coup} carry overbars. The Greek letters $\mu$ and $\nu$ are reserved for labeling adsorbate orbitals, $\sigma$ is reserved for labeling surface orbitals, and $\lambda$ and $\tau$ are used for generic orbitals of either type.

\subsection{Electronic Hamiltonian}
\label{Sec:adsorbate_hamiltonian}
Since our electronic part $\hat{H}_{el}=\hat{H}_\text{ad}+\hat{H}_\text{bath}+\hat{H}_\text{coup}$ consists of non-interacting fermions, we can treat it as a single-particle Hamiltonian, i.e.,
\begin{align}
\begin{split}
    \hat{H}_{el}(R) = & \sum_{\lambda,\tau}\overline{H}_{\lambda\tau}(R)|\overline{\phi}_\lambda\rangle\langle \overline{\phi}_\tau| ,
\end{split}
\end{align}
where $\lambda$ and $\tau$ sum over all adsorbate and surface orbitals. The orbitals $\{|\overline{\phi}_\lambda\rangle\}$ are orthonormal. 
In this section, we will focus on a fixed nuclear coordinate $R$ and omit it for notational simplicity.

In Gaussian-based DFT calculations, the atomic orbitals (AO) are usually not orthonormal, and the adsorbate orbitals can overlap with the surface orbitals. Working in a non-orthogonal basis $\{|\phi_\lambda\rangle\}$ with the metric, $\langle \phi_\lambda|\phi_\tau\rangle = S_{\lambda\tau}$, $\hat{H}_{el}$ can be expressed as
\begin{align}
\begin{split}
    \hat{H}_{el} = & \sum_{\lambda,\tau}H_{\lambda\tau}|\phi^\lambda\rangle\langle \phi^\tau| ,
\end{split}
\end{align}
where the covariant matrix elements are $H_{\lambda\tau}=\langle \phi_\lambda|\hat{H}_{el}|\phi_\tau\rangle$. We use tensor notation for non-orthogonal orbitals.\cite{head2000tensors} $|\phi^\lambda\rangle$ and $|\phi_\lambda\rangle$ denote contravariant and covariant orbitals, respectively. See \cref{App:tensor_notation} for a brief review of the tensor notation.
\par
We partition the basis orbitals $|\phi_\lambda\rangle$ into two parts: adsorbate and surface, denoted as $a$ and $s$, respectively. For an atom-centered basis, this partition is natural. AO-centered at adsorbate (surface) atoms are grouped into $a$ ($s$). We denote the submatrices of a general covariant matrix $\mathbf{A}$ as $\mathbf{A}_{aa}$, $\mathbf{A}_{as}$, $\mathbf{A}_{sa}$, and $\mathbf{A}_{ss}$.
If $\mathbf{A}$ is a contravariant matrix, then notations such as $\mathbf{A}^{aa}$ will be used. When we work in an orthonormal basis, there is no difference between covariant and contravariant orbitals, and we will only use lower indices.

\subsection{Adsorbate-projected Green's function}
\label{Sec:adsorbate_projected_G}
The adsorbate submatrix $\mathbf{G}^{aa}(E)$ of the contravariant retarded Green's function matrix, 
\begin{equation}
    \mathbf{G}(E) = \lim_{\epsilon\rightarrow0^+}((E\mathbf{I}+i\epsilon)\mathbf{S}-\mathbf{H})^{-1},
\label{Eq:Greens_matrix_with_overlap}
\end{equation}
fully characterizes $\hat{H}_\text{el}$. 
We will see that $\mathbf{G}^{aa}(E)$ connects DFT results to $\hat{H}_\text{el}$, and physical observables regarding $\hat{H}_\text{el}$ are derived from quantities closely related to $\mathbf{G}^{aa}(E)$.
\par
Since changing the basis of the surface orbitals among themselves does not affect $\mathbf{G}^{aa}(E)$, we work in a basis where the surface orbitals have already been diagonalized such that $\mathbf{H}_{ss}$ is diagonal and $\mathbf{S}_{ss}=\mathbf I$. In general, the adsorbate orbitals remain non-orthogonal (i.e., $\mathbf{S}_{aa}\neq \mathbf I$) and overlap with the surface orbitals (i.e., $\mathbf{S}_{as}\neq 0$).
In \cref{App:Deriving_Gaa_multi_orb_AN}, we show that the matrix $\mathbf{G}^{aa}(E)$ is expressed as,
\begin{equation}
    \mathbf{G}^{aa}(E) = (\mathbf{Q}(E))^{-1},
\label{Eq:multi_orb_G_phi_inv}
\end{equation}
with 
\begin{equation}
    \mathbf{Q}(E) = E \mathbf{S}_\text{eff} - \mathbf{H}_\text{eff} - \mathbf{\Lambda}(E) + i\mathbf{\Delta}(E).
\label{Eq:phi_GS1}
\end{equation}
The effective adsorbate Hamiltonian $\mathbf{H}_\text{eff}$ and effective adsorbate overlap $\mathbf{S}_\text{eff}$ matrices are
\begin{align}
\begin{split}
    H_{\text{eff},\mu\nu} &= H_{\mu\nu} 
    +\sum_{\sigma\in s} \left(S_{\mu \sigma} E_\sigma S_{\sigma \nu} - H_{\mu \sigma}S_{\sigma\nu} - S_{\mu \sigma}H_{\sigma\nu}\right)
\label{Eq:H_GS1_from_PDOS}
\end{split}
\end{align}
and
\begin{equation}
    S_{\text{eff},\mu\nu}= S_{\mu\nu} - \sum_{\sigma\in s}S_{\mu \sigma} S_{\sigma \nu}.
\label{Eq:S_GS1_from_PDOS}
\end{equation}
The matrix element of the hybridization function matrix $\mathbf{\Delta}(E)$ is 
\begin{align}
\begin{split}
     \Delta_{\mu\nu}(E) = \pi \sum_{\sigma\in s} &(  H_{\mu \sigma} - E_\sigma  S_{\mu \sigma}) \\
    &(H_{\sigma\nu} - E_\sigma  S_{\sigma\nu})\delta(E-E_\sigma),
\label{Eq:multi_orb_AN_Delta}
\end{split}
\end{align}
where $\mu$ and $\nu$ are adsorbate orbital indices, while $\sigma$ indexes the surface orbitals. 
$\mathbf{\Lambda}(E)$ is the Hilbert transform of $\mathbf{\Delta}(E)$ (via the Kramers-Kronig relation\cite{nitzan2024chemical}), i.e.,
\begin{align}
\begin{split}
    & \Lambda_{\mu\nu}(E) = \frac{1}{\pi}\mathcal{P}\int \frac{ \Delta_{\mu\nu}(E')}{E-E'
    } dE'.
\label{Eq:multi_orb_Lambda}
\end{split}
\end{align}
The Kramers-Kronig relation holds because $\mathbf{\Lambda}(E)$ and $\mathbf{\Delta}(E)$ are the Hermitian and anti-Hermitian parts of the retarded self-energy matrix induced by the surface.

\par
As we show in \cref{Sec:constructing_Hel_from_DFT}, \cref{Eq:phi_GS1} contains an implicit orthogonalization in which the adsorbate orbitals become orthogonal to the surface orbitals; this is why $\mathbf{H}_\text{eff}$ and $\mathbf{S}_\text{eff}$ appear in place of $\mathbf{H}_{aa}$ and $\mathbf{S}_{aa}$. Consequently, even though $\mathbf{\Delta}(E)$ in \cref{Eq:multi_orb_AN_Delta} is computed using the adsorbate-surface overlap $\mathbf{S}_{as}$ of the original AO basis, it can be viewed as the hybridization in an orthogonalized basis where $\mathbf{S}_{as}=0$. We classify the different implicit orthogonalization schemes at the end of \cref{Sec:constructing_Hel_from_DFT}.

\subsection{Constructing $\hat{H}_{el}$ from DFT}
\label{Sec:constructing_Hel_from_DFT}
A key quantity that connects DFT and the multi-orbital AN Hamiltonian is the adsorbate projected density of states (PDOS). 
In the multi-orbital picture, the PDOS is a matrix-valued function, defined (in contravariant form) as
\begin{align}
\begin{split}
    \text{PDOS}^{\mu\nu}(E) &= \sum_{i} \langle \phi^\mu|\psi_i\rangle\langle\psi_i|\phi^\nu\rangle \delta(E-E_i) \\
    &= \sum_i C^\mu_{\cdot i}(C^{\nu}_{\cdot i})^*\delta(E-E_i),
\label{Eq:multi_orb_PDOS_def}
\end{split}
\end{align}
where $|\psi_i\rangle$ and $E_i$ are the molecular orbitals (MO) and MO energies of $\mathbf{H}$, with overlap $\mathbf{S}$. We note again that $\mu$ and $\nu$ index the adsorbate AO. The second line is written in terms of the coefficient matrix $C^\mu_{\cdot i} = \langle \phi^\mu|\psi_i\rangle$. Thus, the PDOS matrix can be computed directly from the Kohn-Sham (KS) DFT eigenvalues and the coefficient matrix. 
\par
In practice, the KS eigenvalues are discretized, making $\text{PDOS}(E)$ highly singular. 
To obtain a continuous PDOS, we replace the $\delta$ function in \cref{Eq:multi_orb_PDOS_def} with the normalized Lorentzian smearing function
\begin{equation}
    \delta_\epsilon(E) = \frac{1}{\pi} \frac{\epsilon}{E^2+\epsilon^2}.
\label{Eq:Lorentzian}
\end{equation}
We show in \cref{Sec:app_construct_Hel_from_PDOS} that the use of Lorentzian smearing is crucial for constructing $\hat{H}_{el}$ from DFT, since Lorentzian smearing preserves the analytical relationships in the multi-orbital AN theory. The smearing correction is simply a constant shift of $\epsilon \mathbf{S}_\text{eff}$ in $\mathbf{\Delta}(E)$. Therefore, Lorentzian smearing allows one to obtain accurate results using finite $\epsilon$, without the need to check for convergence in the limit $\epsilon\rightarrow0^+$.
Other smearing methods, such as Gaussian smearing, do not preserve the analytical relationships in AN theory, and can fail to produce the correct $\mathbf{\Delta}(E)$, even qualitatively. \cref{App:model_Hamiltonians} compares Lorentzian and Gaussian smearing.
\par
To obtain $\mathbf{G}^{aa}(E)$ from PDOS$(E)$, we first note that in the AN model, PDOS is equal to the anti-Hermitian part of $\mathbf{G}^{aa}$ times $-1/\pi$, i.e.,
\begin{equation}
    \text{PDOS}(E) = -\frac{1}{\pi} \frac{\mathbf{G}^{aa}(E)-\mathbf{G}^{aa\dagger}(E)}{2i} .
\label{Eq:multi_orb_AN_PDOS_from_G}
\end{equation}
Next, we notice that the Hermitian and anti-Hermitian parts of $\mathbf{G}^{aa}(E)$ obey the Kramers-Kronig relation because in the time domain, $\mathbf{G}^{aa}(t)$ is nonzero only when $t>0$.\cite{nitzan2024chemical}
Therefore, 
\begin{equation}
    \mathbf{G}^{aa}(E) = -i\pi\text{PDOS}(E) + \mathcal{P}\int \frac{\text{PDOS}(E')}{E-E'} dE'
\label{Eq:Gaa_from_PDOS}
\end{equation}
\par
The Lorentzian-smeared $\mathbf{\Delta}(E)$ is obtained as the anti-Hermitian part of $\mathbf{Q}(E)=(\mathbf{G}^{aa}(E))^{-1}$ (see \cref{Eq:phi_GS1}), i.e.,
\begin{equation}
    \mathbf{\Delta}(E) = \frac{\mathbf{Q}(E)-\mathbf{Q}^\dagger(E)}{2i} - \epsilon \mathbf{S}_\text{eff}.
\label{Eq:multi_orb_Delta_from_phi_smeared}
\end{equation}
The origin of the $\epsilon \mathbf{S}_\text{eff}$ constant shift due to Lorentzian smearing is derived in \cref{Sec:app_construct_Hel_from_PDOS}. Since $\mathbf{\Delta}(E)$ usually decays to $0$ as $E\rightarrow \pm\infty$, $\mathbf{\Delta}(E)$ can be obtained by applying a constant matrix shift to the anti-Hermitian part of $\mathbf{Q}(E)$ such that $\mathbf{\Delta}(\pm\infty)=0$, so prior knowledge of $\mathbf{S}_\text{eff}$ is not needed.
Performing a Hilbert transform, we can obtain $\mathbf{\Lambda}(E)$ from $\mathbf{\Delta}(E)$ (see \cref{Eq:multi_orb_Lambda}).
Taking the Hermitian part of $\mathbf{Q}(E)$ in \cref{Eq:phi_GS1}, we see that
\begin{equation}
    E \mathbf{S}_\text{eff} - \mathbf{H}_\text{eff} = \frac{\mathbf{Q}(E)+\mathbf{Q^\dagger(E)}}{2} + \mathbf{\Lambda}(E).
\label{Eq:multi_orbital_phi_hermitian}
\end{equation}
Therefore, $\mathbf{H}_\text{eff}$ and $\mathbf{S}_\text{eff}$ can be obtained by identifying the slope and intercept from the right-hand side of \cref{Eq:multi_orbital_phi_hermitian}.
\par
The construction above involves two choices: the contravariant basis defining the PDOS matrix and, in periodic DFT, the stage at which the Brillouin-zone average is performed. As we show in \cref{App:implicit_orthogonalization_schemes}, these choices correspond to different implicit orthogonalization schemes. The PDOS in the contravariant basis derived from the full adsorbate+surface AO, denoted $\text{PDOS}^{\mu\nu}$, yields the $\hat{H}_{el}$ of a Gram-Schmidt (GS) orthogonalization in which the adsorbate orbitals are projected to the orthogonal subspace of the surface orbitals (denoted GS-fixS). The PDOS in the contravariant basis restricted to the adsorbate AO, denoted $\widetilde{\text{PDOS}}^{\mu\nu}$, corresponds to the reverse scheme (denoted GS-fixA), in which the surface orbitals are projected to the orthogonal subspace of the adsorbate orbitals. GS-fixA corresponds to the POD2GS approach used in Refs.~\onlinecite{ghan2023interpreting}, where the surface states are Gram-Schmidt orthogonalized against adsorbate orbitals.
Furthermore, averaging the $\hat{H}_{el}$ constructed at each $\mathbf{k}$ point over the Brillouin zone excludes lateral interactions with adsorbates in neighboring unit cells, whereas constructing $\hat{H}_{el}$ from the $\mathbf{k}$-averaged PDOS includes lateral interactions. 

The resulting four variations of the AN Hamiltonian (summarized in \cref{fig:PDOS_orthogonalization} of \cref{App:implicit_orthogonalization_schemes}) are useful for studying different physical properties and understanding the effect of lateral interactions. For example, to simulate the adsorbate electronic dynamics after an optical pump excitation (see \cref{Sec:electronic_dynamics}), we use GS-fixA to better capture the optical excitation between adsorbate orbitals. In the study of vibrational relaxation (see \cref{Sec:vib_relax}), we use GS-fixS to ensure that surface orbitals are independent of the adsorbate coordinate $R$.

\subsection{Numerical verification}

We verify our method for constructing $\hat{H}_{el}$, the electronic part of the AN Hamiltonian, in model Hamiltonians and realistic chemical systems. The $\mathbf{\Delta}(E)$, $\mathbf{H}_\text{eff}$, and $\mathbf{S}_\text{eff}$ obtained from the PDOS agree quantitatively with those directly computed from the full adsorbate+surface Hamiltonian.
We first validate the procedure on two analytically solvable models: a three-level Hamiltonian with randomly generated matrix elements, and the single-orbital AN Hamiltonian with a semi-elliptical $\Delta(E)$,\cite{newns1969} widely used in the catalysis community to rationalize binding energy trends.\cite{greiner2018free, tiwari2021reactivity,vijay2022limits,vojvodic2014electronic, norskov2014fundamental} In both cases, the $\Delta(E)$ constructed numerically from the PDOS reproduces the analytical results (see \cref{App:model_Hamiltonians}).
\par
\begin{figure}
    \centering
    \includegraphics[scale=0.8]{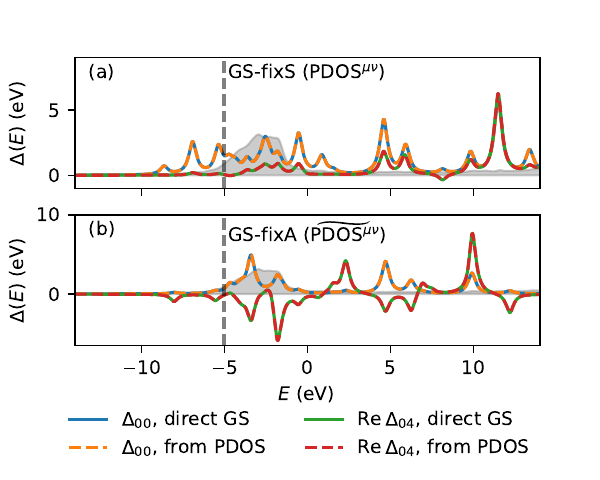}
    \caption{\textbf{Numerical validation of the hybridization function.} (a) $\mathbf{\Delta}(E)$ obtained from the non-orthogonal $\text{PDOS}^{\mu\nu}$ agrees with that obtained via direct GS-fixS orthogonalization. (b) $\mathbf{\Delta}(E)$ obtained from the non-orthogonal $\widetilde{\text{PDOS}}^{\mu\nu}$ agrees with that obtained via direct GS-fixA orthogonalization.
    The gray shaded area is proportional to the Cu bulk density of states. Vertical dashed lines represent the lowest orbital energy (1s) of $\mathbf{H}_{aa}$. We have shifted the energy so that the Fermi energy $\mu$ is set to $0$. In constructing the PDOS, we apply a Lorentzian smearing width of $0.27$ eV. We sample the energy at a spacing of $dE=2.7\times 10^{-3}$ eV, from -216 to 219 eV.  }
    \label{fig:Cu_H_check_Delta}
\end{figure}

Next, we perform a periodic DFT calculation of H on the fcc site of Cu (111) surface and take the alpha-spin component of the KS Hamiltonian at $\mathbf{k}=\Gamma$ point. Five atomic orbitals (AO), indexed from 0 to 4, are centered at the adsorbate H atom. They correspond to the 1s, 2s, and 2p shells. 
\cref{fig:Cu_H_check_Delta} shows that the $\mathbf{\Delta}(E)$ obtained from the non-orthogonal $\text{PDOS}^{\mu\nu}$ and $\widetilde{\text{PDOS}}^{\mu\nu}$ are equal to the $\mathbf{\Delta}(E)$ obtained from direct GS-fixS and GS-fixA orthogonalization. Quantitative agreement is obtained in both the diagonal and off-diagonal elements of $\mathbf{\Delta}(E)$. 
\par
\begin{figure*}
    \centering
    \includegraphics[scale=0.85]{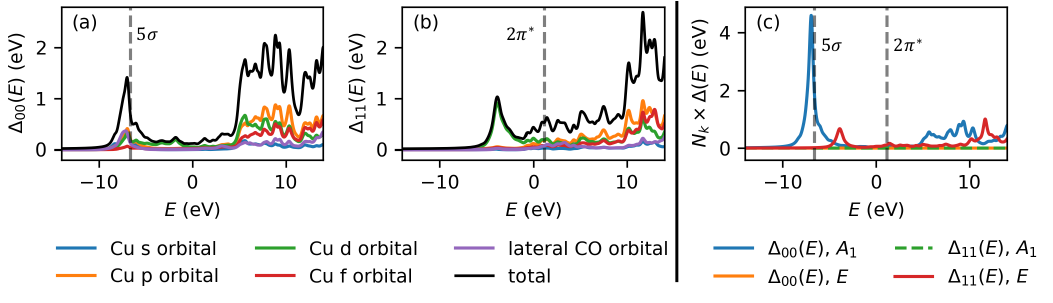}
    \caption{\textbf{Decomposition of the hybridization function $\mathbf{\Delta}(E)$ into different contributions.} Orbital 0 denotes the $5\sigma$ MO of the CO. Orbitals 1 and 2 denote the doubly degenerate $2\pi^*$ MO of the CO. (a)-(b) Contributions to $\Delta_{00}(E)$ and $\Delta_{11}(E)$ from different Cu orbitals and lateral CO molecules. (c) Decomposing $\mathbf{\Delta}(E)$ due to the Cu atom adjacent to the C atom into different symmetry sectors. Orbitals 0 and 1 selectively couple to symmetry sectors $A_1$ and $E$, respectively. The value of $\mathbf{\Delta}(E)$ is amplified by $N_k$, the number of unit cells in the supercell, so that the single-atom contribution is insensitive to the size of the supercell.}
    \label{fig:CO_Cu_Delta_decomposition}
\end{figure*}

Finally, we consider CO adsorbed on the atop site of the Cu (111) surface. The PDOS is chosen so that the resulting $\hat{H}_{el}$ corresponds to GS-fixS orthogonalization with lateral interaction. Once $\mathbf{\Delta}(E)$, $\mathbf{H}_\text{eff}$, and $\mathbf{S}_\text{eff}$ in the AO basis of the CO molecule are obtained, we project these quantities onto three of the CO frontier molecular orbitals (MOs). The $5\sigma$ HOMO is indexed as the 0\textsuperscript{th} adsorbate orbital, and the two degenerate $2\pi^*$ LUMOs are indexed as the first and second orbitals (see \cref{fig:single_vs_multi_orbital_schematics}). The coefficients of the MO are obtained by diagonalizing the adsorbate submatrix $\mathbf{H}_{aa}$ (with overlap $\mathbf{S}_{aa}$). $\Delta_{00}(E)$ and $\Delta_{11}(E)$ are shown as the curves labeled ``total" in \cref{fig:CO_Cu_Delta_decomposition} (a)-(b).
The $\mathbf{\Delta}(E)$ computed directly through GS-fixS orthogonalization on the supercell agrees quantitatively with the $\mathbf{\Delta}(E)$ constructed from PDOS. 

\section{Structure of the hybridization function $\mathbf{\Delta}(E,R)$}
\label{Sec:structure_Delta}
Having established a first-principles method to construct the hybridization function $\mathbf{\Delta}(E,R)$, we can now examine some common assumptions about $\mathbf{\Delta}(E,R)$ in the literature. Here is a list of common assumptions:
\begin{itemize}
    \item Assumption 1 (wide-band limit): $\mathbf{\Delta}(E,R)$ is constant in $E$.\cite{gardner2023efficient, gardner2023assessing, preston2025nonadiabatic,shenvi2009nonadiabatic,dou2017born, dou2015frictional}
    \item Assumption 2 (semi-elliptical form): $\mathbf{\Delta}(E,R)$ can be qualitatively modeled as a semi-ellipse.\cite{greiner2018free, tiwari2021reactivity,vijay2022limits,vojvodic2014electronic, norskov2014fundamental}
    \item Assumption 3 (single-orbital representation): The behavior of the adsorbate-surface system can be captured using only one adsorbate orbital, corresponding to two states (occupied and empty). 
    \item Assumption 4 (d-band dominance): $\mathbf{\Delta}(E,R)$ has a small constant (in $E$) contribution from the s and p metal bands. The metal d band has a large but narrow (in $E$) contribution to $\mathbf{\Delta}(E,R)$.\cite{norskov2014fundamental, santos2009model, nilsson2005electronic, norskov2011density}
    \item Assumption 5 (bulk-DOS proportionality): $\mathbf{\Delta}(E,R)$ is proportional to the bulk density of states $D(E)$.\cite{santos2009model}
    \item Assumption 6 (separability): $\mathbf{\Delta}(E,R)$ can be factorized as $\mathbf{f}(E)g(R)$. \cite{dan2026nonadiabatic, dan2023theoretical} This assumption includes Assumptions 1 and 5 as special cases.
\end{itemize}
From \cref{fig:Cu_H_check_Delta} and \cref{fig:CO_Cu_Delta_decomposition} (a)-(b), we find that Assumptions 1, 2, and 5 are usually not satisfied. Assumption 1 (wide-band limit) could be reasonable for $\Delta_{00}(E)$ in CO on Cu (111) (see \cref{fig:CO_Cu_Delta_decomposition} (a)), since $\Delta_{00}(E)$ is roughly constant around the Fermi energy (i.e., $-5$ to $5$ eV). We show below that, by symmetry, Assumption 5 (bulk-DOS proportionality) fails in highly symmetric configurations. Each of Assumptions 3--6 is examined in a dedicated subsection below.

\subsection{Hybridization ratio (Assumption 3: single-orbital representation)}
We characterize how well a single orbital representation captures the full adsorbate-surface interaction by the hybridization ratio (h.r.), defined as
\begin{equation}
    \text{h.r.} = \frac{\int^\mu_{-\infty} \Delta_{00}(E)/S_{\text{eff},00} \,dE}{\int^\mu_{-\infty} \text{Tr}(\mathbf{S}_\text{eff}^{-1}\mathbf{\Delta}(E))\,dE},
\label{Eq:representation_ratio}
\end{equation}
where orbital $0$ is the representative single orbital. h.r. is the ratio of the integrated hybridization strength below the Fermi energy between orbital 0 and the sum of all orbitals.
The factor of $\mathbf{S}_\text{eff}^{-1}$ ensures that the quantity is invariant under arbitrary rotations of adsorbate orbitals. h.r. ranges from 0 to 1. An h.r. value of 1 indicates that the single orbital perfectly captures all interactions between the adsorbate and surface below the Fermi level.
\par
Using the GS-fixS orthogonalization with lateral interaction, the basis set 1s orbital of H in H/Cu has an h.r. of 0.80. By optimizing the coefficients of the representative orbital, we can achieve a maximum h.r. of 0.94. This means that Assumption 3 (single-orbital representation) is reasonable for H/Cu. For the CO/Cu system, the $5\sigma$ HOMO has an h.r. of 0.47, and each of the $2\pi^*$ LUMOs has an h.r. of 0.27. Therefore, Assumption 3 fails for CO/Cu. Mixing the HOMO with the LUMOs will only make the h.r. smaller than 0.47, since the $\mathbf{\Delta}(E)$ is diagonal due to symmetry (see \cref{Sec:Decompose_Delta_symmetry}).

\subsection{Decomposition of $\mathbf{\Delta}(E)$ by surface orbital angular momentum (Assumption 4: d-band dominance)}
Considering CO adsorbed on Cu at the equilibrium geometry $R=R_0$, we discuss a procedure to decompose $\mathbf{\Delta}(E)$ into different surface contributions.
First, we partition the surface orbitals into $M$ different groups $s_1, s_2,\cdots,s_M$. Note that the ``surface" orbitals may contain other adsorbates if lateral interaction is included (see \cref{fig:PDOS_orthogonalization}).
$\mathbf{\Delta}(E)$ can be decomposed into a sum of different orbital group contributions, i.e.,
\begin{equation}
    \mathbf{\Delta}(E) = \sum_{j=1}^M \mathbf{\Delta}^{(s_j)}(E).
\label{Eq:Delta_decomposition}
\end{equation}
Each orbital group contribution $\mathbf{\Delta}^{(s_j)}(E)$ is defined as
\begin{align}
\begin{split}
    \Delta^{(s_j)}_{\mu\nu}(E) = \pi \sum_{\sigma\in s} &( H_{\mu \sigma} - E_\sigma S_{\mu \sigma}) \\
    &(H_{\sigma\nu} - E_\sigma S_{\sigma\nu})\delta(E-E_\sigma) P_\sigma^{(s_j)},
\end{split}
\end{align}
where $\sigma$ indexes the eigenstates of the surface block. $P_\sigma^{s_j}$ is the total L\"owdin population of the surface MO $\psi_\sigma$ on the orbital group $s_j$.\cite{lowdin1950non} 
Since $\sum_{j=1}^M P_\sigma^{(s_j)} =1$, the sum of $\mathbf{\Delta}^{(s_j)}(E)$ is equal to $\mathbf{\Delta}(E)$ (i.e., \cref{Eq:Delta_decomposition} holds). We compute the orbital-group contributions by explicitly diagonalizing the supercell surface Hamiltonian.
\par
In \cref{fig:CO_Cu_Delta_decomposition} (a)-(b), we decompose the surface orbitals in the supercell into Cu orbitals and lateral CO orbitals. We further group the Cu orbitals by angular momentum number $l$.
We see that in both $\Delta_{00}(E)$ and $\Delta_{11}(E)$, the Cu d-band contribution dominates between around $-5$ and $0$ eV. From $0$ to $10$ eV, Cu p, d, and f orbitals tend to hybridize, and they have roughly equal contributions to $\mathbf{\Delta}(E)$.
Our analysis shows that Assumption 4 (d-band dominance) may need closer reconsideration. While the metal d band is still the dominating contribution near the Fermi level, the d band contribution is not always large and narrow in $E$, and the contribution from s, p, and f bands may not be negligible near the Fermi level ($\approx -10$ to $10$ eV). 

\subsection{Decomposition of $\mathbf{\Delta}(E)$ using point group symmetry (Assumption 5: bulk-DOS proportionality)}
\label{Sec:Decompose_Delta_symmetry}
The atomic environment of the CO adsorbate on the atop site of Cu (111) possesses a $C_{3v}$ point group symmetry.
The $5\sigma$ HOMO (orbital 0) transforms as the $A_1$ irreducible representation (irrep), and the $2\pi^*$ LUMOs (orbitals 1 and 2) transform as the two-dimensional irrep $E$.
Using Schur's lemma,\cite{Hall2015LieGroups} we show in \cref{App:representation_group} that the $3\times 3$ matrix $\mathbf{\Delta}(E)$ is diagonal, with $\Delta_{11}(E)=\Delta_{22}(E)$.
\cref{fig:CO_Cu_Delta_decomposition} (c) decomposes the contribution of $\mathbf{\Delta}(E)$ from the Cu atom adjacent to the C atom into the irreps of $C_{3v}$. The orbitals that transform as $A_1$ are $s$, $p_z$, $d_{z^2}$, $f_{y(3x^2-y^2)}$, and $f_{z^3}$. The only orbital that transforms as $A_2$ is $f_{x(x^2-3y^2)}$. The orbitals that transform as $E$ are $p_x$, $p_y$, $d_{xy}$, $d_{x^2-y^2}$, $d_{xz}$, $d_{yz}$, $f_{xyz}$, $f_{z(x^2-y^2)}$, $f_{xz^2}$, and $f_{yz^2}$. \cref{App:representation_group} provides a general method for determining which linear combinations of spherical harmonics belong to a point-group irrep. 
Since we focus on a single Cu atom in the supercell (containing $N_k$ unit cells), the contribution of $\mathbf{\Delta}(E)$ from the Cu atom scales as $1/N_k$. To make sure the single Cu atom contribution to $\mathbf{\Delta}(E)$ is insensitive to the k-mesh density, we amplify the values of $\mathbf{\Delta}(E)$ in \cref{fig:CO_Cu_Delta_decomposition} (c) by $N_k$. We see that adsorbate orbitals do not couple to all surface states. Instead, they selectively couple to different sets of surface states due to symmetry. Orbital 0 (i.e., $5\sigma$) only couples to the $A_1$ symmetry sector, and orbitals 1 and 2 (i.e., $2\pi^*$) only couple to the $E$ symmetry sector. Therefore, when there is a high degree of symmetry in the adsorbate-surface system, taking $\mathbf{\Delta}(E)$ to be proportional to the bulk density of states (i.e., Assumption 5) may be a poor approximation.

\subsection{Dependence on nuclear coordinate $R$ (Assumption 6: separability)}
\begin{figure}
    \centering
    \includegraphics[width=0.9\linewidth]{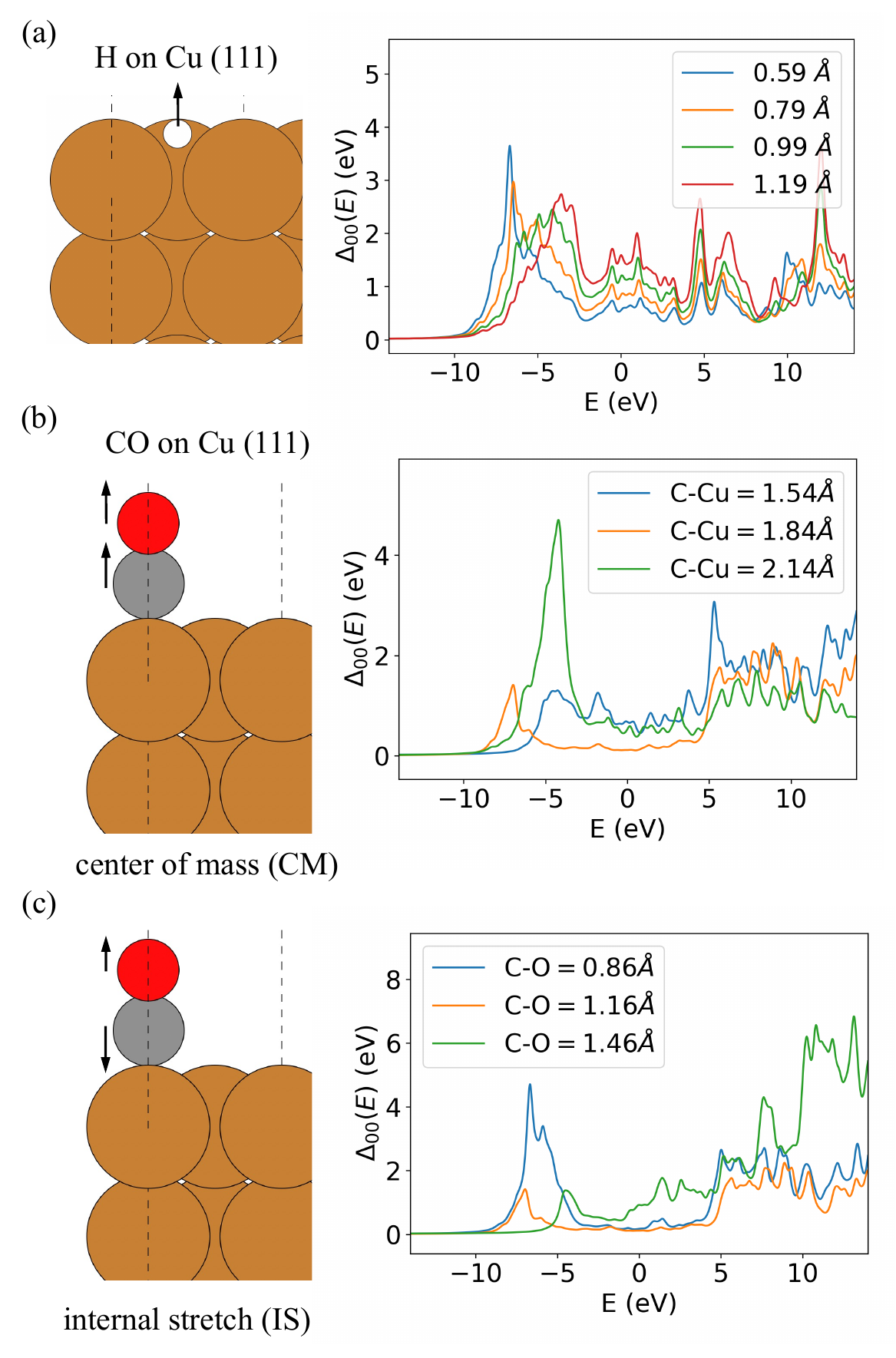}
    \caption{\textbf{Hybridization functions of realistic surface problems.} (a) Dependence of $\Delta_{00}(E)$ (H 1s orbital) on the H atom height in H-on-Cu system. (b) Dependence of $\Delta_{00}(E)$ (CO $5\sigma$ HOMO) on the center-of-mass (CM) motion in the CO-on-Cu system. (c) Dependence of $\Delta_{00}(E)$ (CO $5\sigma$ HOMO) on the internal stretch (IS) motion in the CO-on-Cu system.}
    \label{fig:R_dependence_Delta}
\end{figure}

We now examine how $\mathbf{\Delta}(E,R)$ depends on the nuclear coordinate $R$. Note that separability (i.e., $\mathbf{\Delta}(E,R)=\mathbf{f}(E)g(R)$) requires the fixed-$R$ curves in \cref{fig:R_dependence_Delta} to be scalar multiples of a single lineshape $\mathbf{f}(E)$: the peak positions must be independent of $R$, and the ratio between any two curves must be independent of $E$. 
\par
\cref{fig:R_dependence_Delta} (a) shows the $\mathbf{\Delta}(E,R)$ for H/Cu at different H atom heights. Between $E=-5$ and 5 eV, $\mathbf{\Delta}(E,R)$ approximately satisfies this condition and can be well-approximated as a separable function, so Assumption 6 (separability) is reasonable in this case. However, farther away from the Fermi energy (say $-10$ to $10$ eV), $\mathbf{\Delta}(E,R)$ is no longer separable, as the main peak below $E=0$ shifts with the H atom height and the curves cross. As the H atom height increases, $\mathbf{\Delta}(E)$ appears to increase, contrary to intuition. This is because the H 1s orbital strongly overlaps with the Cu orbitals for the heights considered. Therefore, in the GS-fixS orthogonalization picture, increasing the H atom height decreases the overlap, thereby increasing the effective norm of the adsorbate orbital (i.e., $S_{\text{eff},00}$), which leads to larger $\mathbf{\Delta}(E)$. Increasing the height further beyond the strong-overlap regime would decrease $\mathbf{\Delta}(E)$.
\par
\cref{fig:R_dependence_Delta} (b)-(c) examine the dependence of $\mathbf{\Delta}(E,R)$ on two nuclear degrees of freedom along the z-axis in the CO/Cu system. The first degree of freedom is the center of mass (CM) mode, where the C-O bond length is fixed, while the entire CO molecule moves along the z-axis (see \cref{fig:R_dependence_Delta} (b)). The second degree of freedom is the internal stretching (IS) mode of CO, where the C-O bond stretches while the center of mass of CO is fixed (see \cref{fig:R_dependence_Delta} (c)). We see that Assumption 6 breaks down, since $\Delta_{00}(E, R)$ is not separable for both the CM and IS degrees of freedom: the peak positions shift with $R$, and the curves cross rather than rescale. Furthermore, $\mathbf{\Delta}(E,R)$ behaves non-monotonically in both the CM mode coordinate and IS mode coordinate. 
\par

\section{Hybridization energy and the bare nuclear potential $V_N(R)$}
\label{Sec:hybridization_energy_and_V_N}
So far, we have shown how to construct $\hat{H}_{el}(R)$ (see \cref{Eq:H_tot}), the electronic part of the AN Hamiltonian, at a fixed $R$. To construct the full AN Hamiltonian $\hat{H}_{tot}$, we need to know the bare nuclear potential $V_N(R)$. We will construct $V_N(R)$ by requiring consistency between the total energy from DFT and the total energy of $\hat{H}_{tot}$ with classical nuclear degrees of freedom. Once the nuclear potential $V_N(R)$ is obtained, one can re-interpret the nuclear degrees of freedom to be quantum mechanical, allowing for the investigation of nuclear quantum effects, which is difficult using conventional DFT.
\subsection{Connection between DFT and the classical-nuclei AN Hamiltonian}
\par
In the AN Hamiltonian with classical nuclei at a fixed chemical potential $\mu$ and zero temperature, the grand potential (i.e., $\Omega=E-\mu N$) is equal to
\begin{equation}
    \Omega_{\text{tot}}(R) = V_N(R) + \Omega(\hat{H}_{el}(R),\mu).
\end{equation}
In \cref{App:hybridization_energy}, we show that the hybridization energy $E_{\text{hyb}}(R)$, defined as the difference between the grand potential of $\hat{H}_{el}$ and $\hat{H}_\text{bath}$, is given by
\begin{align}
\begin{split}
     E_\text{hyb}(R) &=  \Omega(\hat{H}_{\text{el}}(R),\mu) - \Omega(\hat{H}_{\text{bath}}, \mu) \\
    &=  \int^\mu_{-\infty} \eta(E,R)\, dE ,
\label{Eq:single_orbital_bindingE}
\end{split}
\end{align}
where 
\begin{equation}
    \eta(E) = \frac{1}{\pi}\Big(\arg(\det \mathbf{Q}(E)) - \arg(\det \mathbf{Q}(-\infty))\Big)
\label{Eq:eta_def}
\end{equation}
at a fixed $R$.
The function $\arg(z)$ computes the argument, or angle, of the complex number $z$. The angle is defined only up to integer multiples of $2\pi$. However, one must pick the angle consistently to make sure the function $\arg(\det \mathbf{Q}(E))$ is continuous in $E$. $\eta(E)$ is uniquely defined because shifting the entire $\arg(\det \mathbf{Q}(E))$ by a multiple of $2\pi$ has no effect on $\eta(E)$, which is defined as the difference between two $\arg(\det \mathbf{Q}(E))$. 
From \cref{Eq:single_orbital_bindingE}, we see that $\eta(E)$ can be interpreted as the contribution to $E_\text{hyb}$ due to the hybridized MO at energy $E$. Therefore, we shall call $\eta(E)$ the hybridization energy density.
\cref{Eq:eta_def} has a useful property: it is invariant under any basis transformation (including non-orthogonal transformations) of the matrix $\mathbf{Q}(E)$. 
Therefore, we can compute $\eta(E)$ conveniently in the non-orthogonal AO basis.
\par
In the special case of a single adsorbate orbital, \cref{Eq:single_orbital_bindingE} reduces to a well-known expression for the hybridization energy, which is commonly used in the d-band theory for catalysis,\cite{newns1969, norskov2014fundamental,vojvodic2014electronic,vijay2022limits, wang2020bayesian} (see \cref{App:hybridization_energy}). The binding energy in d-band theory is sometimes decomposed into a hybridization energy and an orthogonalization energy. We show in \cref{App:bypassing_ortho_E} that, by choosing the GS-fixS orthogonalization scheme, one can bypass an orthogonalization energy contribution.
\par
In the limit of an infinitely thick slab, the change in grand potential at a fixed $\mu$ is equal to the change in energy at a fixed $N$, relative to some adsorbate configuration $R_0$.
Requiring $E_{\text{DFT}}(R)-E_{\text{DFT}}(R_0)$ to equal $\Omega_\text{tot}(R)-\Omega_\text{tot}(R_0)$, we can obtain the bare nuclear potential as
\begin{align}
\begin{split}
    V_N(R) - V_N(R_0) &= (E_\text{DFT}(R)- E_\text{DFT}(R_0))  \\
    &- ( E_\text{hyb}(R) - E_{\text{hyb}}(R_0)).
\label{Eq:V_N_construction}
\end{split}
\end{align}
\cref{App:hybridization_energy} provides further details, including the proof of this grand-potential--energy equivalence and the finite-temperature extension of \cref{Eq:single_orbital_bindingE}.

\begin{figure*}
    \centering
    \includegraphics[width=0.9\linewidth]{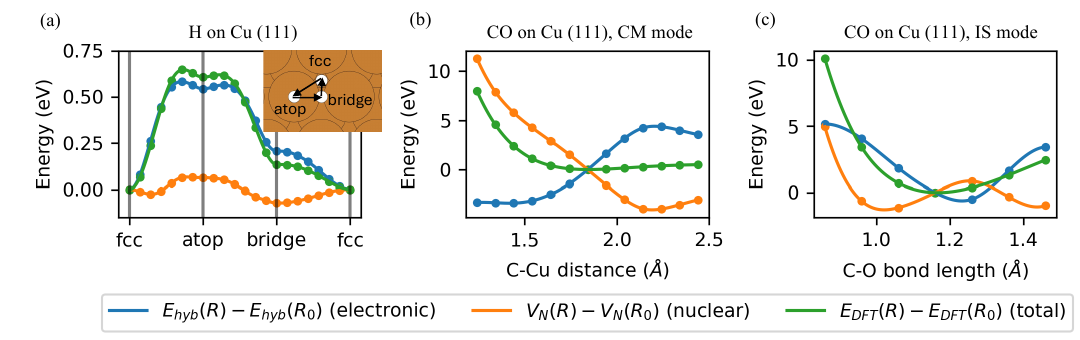}
    \caption{\textbf{Decomposition of the total energy from DFT into the electronic hybridization energy and the bare nuclear potential in the classical-nuclei AN Hamiltonian.} (a) H atom traversing across the fcc, atop, and bridge binding sites of the Cu (111) surface. (b) CM mode of CO on Cu (111). (c) IS mode of CO on Cu (111).}
    \label{fig:V_N_all}
\end{figure*}

\subsection{Validation of the AN Hamiltonian with nuclear dependence}
First, we investigate the energy landscape as a hydrogen atom (H) traverses across the fcc, atop, and bridge binding sites on Cu (111) surface (see \cref{fig:V_N_all} (a)).
The H atom height at each binding site is determined by minimizing $E_\text{DFT}$. We move the H atom along straight line segments connecting these binding sites and decompose the total energy into $E_\text{hyb}(R)$ and $V_N(R)$.
The reference coordinate $R_0$ is taken to be the fcc binding site.
We find that the total energy is dominated by the electronic contribution $E_\text{hyb}$, while the nuclear energy landscape $V_N$ is relatively flat. Interestingly, compared to the fcc site, the bridge site has a lower $V_N$, even though it has a higher total energy and $E_\text{hyb}$. Between the fcc and atop sites, the total energy and $E_\text{hyb}$ exhibit an energy barrier, while $V_N$ exhibits a shallow local minimum.

\par
\cref{fig:V_N_all} (b) shows the energy decomposition as we move along the CM mode of CO/Cu (see \cref{fig:R_dependence_Delta}). We choose the Cu-C distance as the coordinate for the vertical motion of the CO molecule. The energy decomposition for the CM mode in CO/Cu is shown . We have set the optimized geometry $R_0$ to be the reference state. 
The total energy has a shallow minimum at a C-Cu distance of around 1.84 $\AA$. 
When approaching the surface from far away, $E_\text{hyb}$, the electronic contribution, shows an energy barrier of around 1 eV. This barrier is offset by the nuclear contribution $V_N$, so that the total energy curve $E_\text{DFT}$ is barrier-less. As the C-Cu distance decreases further, the total energy is dominated by the sharp increase of $V_N$. 
\par
\cref{fig:V_N_all} (c) shows the energy decomposition along the IS mode in CO on Cu (see \cref{fig:R_dependence_Delta}). We use the C-O distance as the coordinate for the IS motion.
The bare nuclear potential $V_N$ exhibits a double-minimum structure. This can be rationalized by comparing it to the energy decomposition of an isolated CO molecule, where $\mathbf{\Delta}(E) = 0$ (see \cref{Sec:isolated_CO}). The trend resembles that of the isolated CO for C-O bond lengths $< \approx 1.2 \AA$. At larger C-O bond lengths, the C atom overlaps strongly with the Cu atom, resulting in the second minimum of $V_N$.

\section{Electronic dynamics}
\label{Sec:electronic_dynamics}
 
To study the electronic dynamics in the second-quantized picture, we will work in an orthonormal basis. Due to the correspondence described in \cref{App:implicit_orthogonalization_schemes}, the $\hat{H}_{el}$ constructed from the PDOS will already have the adsorbate orbitals orthogonal to the surface orbitals. In this section, we will work in a fully orthonormal basis by further diagonalizing the adsorbate orbitals with respect to $\mathbf{H}_\text{eff}$ and $\mathbf{S}_\text{eff}$. 
\par
Working at a fixed $R$, the electron tunneling rate from the adsorbate orbital $|\overline{\phi}_\mu\rangle$ to the surface is commonly obtained using the Fermi golden rule,\cite{hertl2026first, ghan2023interpreting} i.e.,
\begin{align}
\begin{split}
    \Gamma_{el} 
    &= \frac{2}{\hbar}\Delta_{\mu\mu}(E_\mu),
\label{Eq:electron_tunneling_rate_Fermi}
\end{split}
\end{align}
where $E_\mu$ is the energy of the adsorbate orbital $|\overline{\phi}_\mu\rangle$.
\cref{Eq:electron_tunneling_rate_Fermi} has been shown to yield reasonable agreement with experiments.\cite{ghan2023interpreting} For typical adsorbate-surface systems, the tunneling time $1/\Gamma_{el}$ is usually less than 10 fs.\cite{hertl2026first, ghan2023interpreting} Tunneling time has been experimentally measured indirectly using the core hole clock method.\cite{wurth2000ultrafast,fohlisch2003energy} However, due to the short time scale, direct time-resolved experimental measurements are still lacking. Therefore, a comprehensive understanding of the fs-timescale electronic dynamics based on experiments is still unavailable.
\par
\begin{figure}
    \centering
    \includegraphics[width=0.9\linewidth]{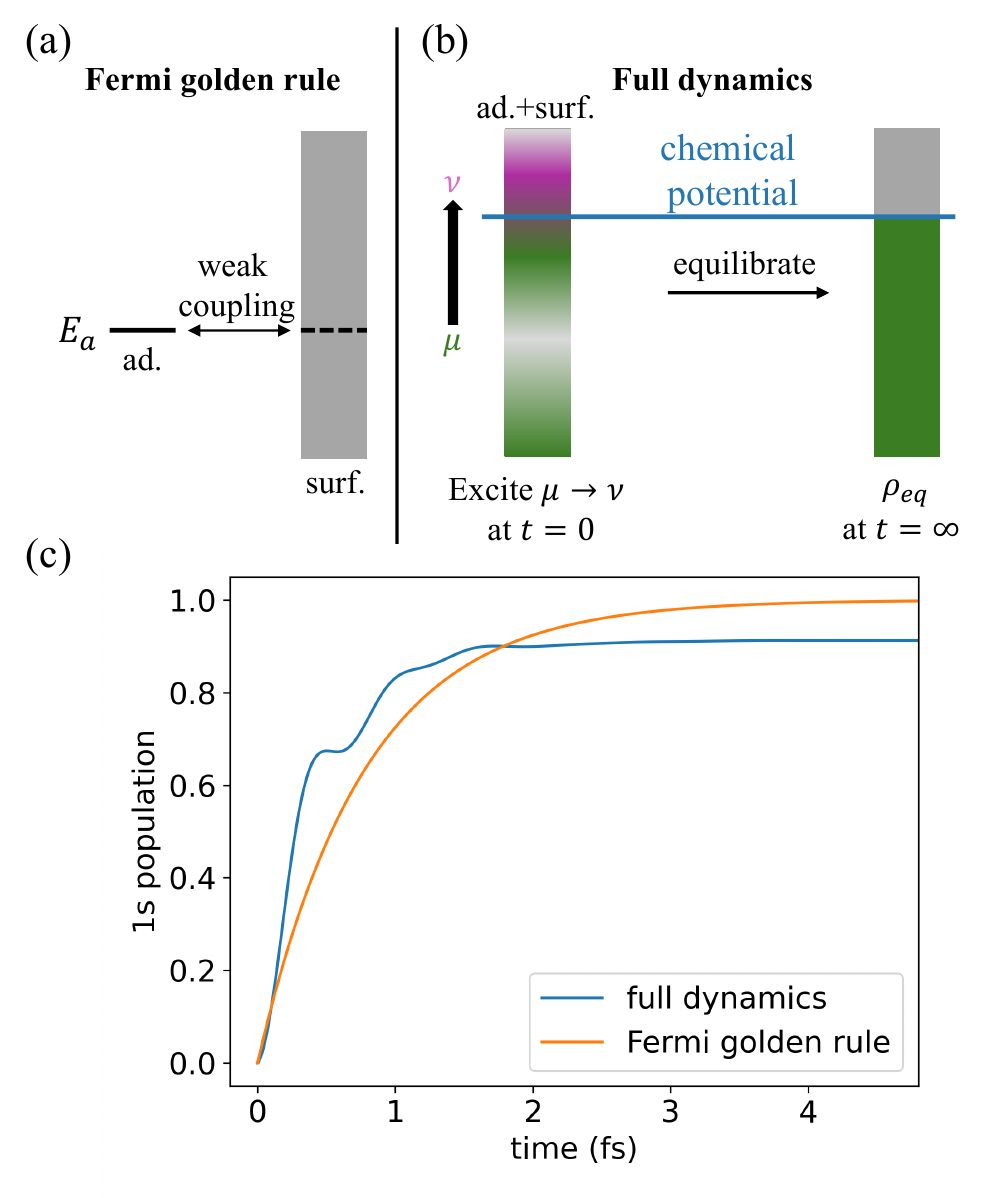}
    \caption{\textbf{Application to electronic dynamics.} (a) Fermi's golden rule assumes an initially uncoupled state and weak resonant adsorbate-surface coupling. (b) In the full electronic dynamics, the initial state is prepared from the coupled equilibrium state $\rho_{eq}$ by exciting an electron from orbital $\mu$ to orbital $\nu$. The population $p_\sigma(t)$ of orbital $\sigma$ is measured as the system returns to equilibrium. (c) Population dynamics of the H 1s orbital. At $t=0$, electron is excited from 1s to 2p\textsubscript{x}.
    }
    \label{fig:Cu_H_electronic_dynamics}
\end{figure}

Taking a step beyond the perturbative Fermi golden rule, we show that if we assume a time-independent AN Hamiltonian, then the full electronic dynamics can be computed exactly using Wick's theorem. \cite{rammer2011quantum}
Different from Fermi's golden rule, which starts from the decoupled initial state, we let the initial state be the many-electron ground state of the coupled adsorbate-surface system (see \cref{fig:Cu_H_electronic_dynamics}a-b). 
Exciting an electron from adsorbate orbital $\mu$ to adsorbate orbital $\nu$ at time $t=0$, we derive a closed-form expression for the population of orbital $\sigma$ as a function of time $t$ (see \cref{App:electronic_dynamics}).
Applying this closed-form solution to the model Hamiltonian with semi-elliptical $\Delta(E)$, we find that the exact dynamics can exhibit fast oscillations and an additional slow decay component, which differ significantly from the Fermi golden rule (see \cref{App:electronic_dynamics} for this model demonstration).
\par
Our assumption of time-independent AN Hamiltonian corresponds to assuming the electron density does not change significantly during the time evolution.
We note that Ref.~\onlinecite{mizielinski2005electronic} has shown how to compute the electronic dynamics under a time-dependent AN Hamiltonian, which has the time-dependence originating from the time-dependent mean-field. However, $\mathbf{\Delta}(E)$ is taken to be constant in $E$ in Ref.~\onlinecite{mizielinski2005electronic}.
\par
To simulate the electronic dynamics for H/Cu (111), we use the GS-fixA orthogonalization scheme with lateral interaction to preserve the shape of H AO. We first label the diagonalized adsorbate orbital as 1s, 2s, 2p\textsubscript{x}, 2p\textsubscript{y}, and 2p\textsubscript{z}. Focusing on the alpha spin component, we consider an electronic excitation from 1s to 2p\textsubscript{x} at $t=0$ and compute the population of 1s as a function of time.
The population dynamics under the time-independent $\hat{H}_{el}$ and under Fermi's golden rule (\cref{Eq:electron_tunneling_rate_Fermi}) are shown in \cref{fig:Cu_H_electronic_dynamics}. We see that Fermi's golden rule agrees with the population dynamics at short times. However, Fermi's golden rule cannot capture the oscillatory part of the population dynamics and the long time equilibrium population.

\par
Deviations from Fermi's golden rule may not be measurable using the core hole clock method \cite{wurth2000ultrafast,fohlisch2003energy}, which infers the tunneling rate by measuring the spectral linewidth. The fast decay due to Fermi's golden rule is likely to dominate the spectral broadening effect, thereby overwhelming the signal from other dynamical effects. Direct, time-resolved measurements would help uncover more nuanced electron transfer dynamics.

\section{Vibrational relaxation and friction}
\label{Sec:vib_relax}

Since metal surfaces contain low energy electronic excitations that couple to nuclear degrees of freedom, adsorbate-metal systems are highly nonadiabatic. The energy exchange between nuclear motion and electronic excitations leads to enhanced vibrational relaxation and frictional effects. These effects have been demonstrated via spectral linewidth measurements,\cite{lamont1995dynamics} time-resolved spectroscopy,\cite{harris1990vibrational,omiya2014coverage,germer1994picosecond,saalfrank2006quantum} and atomic/molecular scattering.\cite{bunermann2015electron}
\par
Quantizing the vibrational degree of freedom and applying Fermi's golden rule to the multi-orbital AN Hamltonian, we show that the friction coefficient (with a dimension of 1/[Time]) can be expressed in terms of the hybridization energy density $\eta$, i.e., 
\begin{equation}
    \gamma_f = \frac{\pi\hbar}{m}\Big( \frac{\partial \eta(\mu,R)}{\partial R}\Big)^2.
\label{Eq:electronic_friction_no_smear}
\end{equation}
$\gamma_f$ is also equal to the relaxation rate of the vibrational energy. 
In deriving \cref{Eq:electronic_friction_no_smear}, we have made two assumptions. First, we have assumed that the density operator does not vary significantly around $\mu$, where electronic transitions take place. Second, we assume that the density operator at $\mu$ is a pure state. These assumptions are explained in detail in \cref{App:single_orbital_vib_relaxation_rate}.
We note that Ref.~\onlinecite{brandbyge1995electronically} has derived the electronic friction of the single-orbital AN Hamiltonian using path integral. The result is consistent with \cref{Eq:electronic_friction_no_smear} in the single-orbital special case.
\par
We compute the vibrational energy relaxation rate $\gamma_f$ for three nuclear degrees of freedom: H atom height in H/Cu, CM mode in CO/Cu, and IS mode in CO/Cu. The derivative $\partial \eta/\partial R$ in \cref{Eq:electronic_friction_no_smear} is computed via finite difference with $0.01\,\AA$ increments in $R$. The mass of the IS mode in CO/Cu is taken to be the reduced mass of a diatomic molecule, i.e., $M_C M_O/(M_C+M_O)$. 
The results, with comparison to literature results, are tabulated in \cref{tab:CO_Cu_electronic_friction}. 
We note that in the infrared reflectivity experiment reported in Ref.~\onlinecite{lamont1995dynamics}, the nuclear degree of freedom is inferred to be the parallel motion of a H atom on the three-fold symmetric site of Cu (111), which includes the fcc and hcp sites. The transient infrared spectroscopy study of Ref.~\onlinecite{owrutsky1992femtosecond} reports a vibrational dephasing time $T_2$ of 2 ps. Since $T_2$ is related to the population relaxation time $T_1$ and pure dephasing time $T_2^*$ by $(T_2)^{-1} = (2T_1)^{-1} + (T_2^*)^{-1}$, we infer that the vibrational relaxation time $T_1$ is $\geq 1$ ps.\cite{nitzan2024chemical}
\par
Most previous theoretical calculations of $\gamma_f$ \cite{maurer2016ab,forsblom2007vibrational} are derived directly from DFT, which integrates out electronic degrees of freedom and treats the nonadiabatic effects simply as Langevin friction on classical nuclei. In order to model the coupled nuclear-electronic dynamics, one could use AN Hamiltonians. 
We use our calculations of $\gamma_f$ based on the AN Hamiltonian to test whether our AN Hamiltonian captures nonadiabatic effects correctly.
$\gamma_f$ derived from the multi-orbital AN model is within the same order of magnitude as literature results. To compare with the results derived from single-orbital AN models, we take the H 1s orbital to be the single orbital in H/Cu and take the $5\sigma$ HOMO of CO to be the single orbital in CO/Cu. The $\gamma_f$ of H/Cu is well approximated using the single-orbital AN model. However, the $\gamma_f$ of CO/Cu derived from the single-orbital AN model is likely wrong by orders of magnitude. Therefore, the conventional single-orbital AN Hamiltonian fails to capture the nonadiabatic dynamics of CO/Cu, and possibly a large class of other adsorbate-surface systems. For these systems, accounting for multiple adsorbate orbitals in the AN Hamiltonian is crucial to understanding the nonadiabatic dynamics, even just qualitatively.

\begin{table}[]
    \centering
    Vibrational relaxation time $1/\gamma_f$ (ps)
    \begin{tabular}{|c|c|c|c|c|c|}
        \hline
        & \cref{Eq:electronic_friction_no_smear} & \cref{Eq:electronic_friction_no_smear} & Lit. & Lit.  \\
        & multi-orb. & single-orb. & (calc.) & (expt.) \\
         \hline
         H/Cu & 0.66 & 0.57 & & 0.7\cite{lamont1995dynamics} \\
         \hline
         CM, CO/Cu & 46.2 & 272 & 30.1\cite{maurer2016ab} & \\
         \hline
         IS, CO/Cu & 1.5 & 1764 & 5.9,\cite{maurer2016ab} 1.8\cite{forsblom2007vibrational} & $\geq$1\cite{owrutsky1992femtosecond} \\
         \hline
    \end{tabular}
    \caption{\textbf{Vibrational relaxation times of H/Cu (111) and CO/Cu (111).} calc: calculation. expt: experiment. Literature calculation values are computed by applying Fermi's golden rule \cite{forsblom2007vibrational} or first-order time-dependent perturbation theory \cite{maurer2016ab} on DFT KS orbitals. }
    \label{tab:CO_Cu_electronic_friction}
\end{table}

\section{Conclusion}
For more than five decades, the AN Hamiltonian has been the standard effective model for chemisorption and nonadiabatic dynamics at metal surfaces, underpinning frameworks that range from the d-band model of catalysis to electronic friction theories of vibrational relaxation. Yet in studies of nonadiabatic dynamics and catalysis trends, its central ingredient, the hybridization function $\mathbf{\Delta}(E,R)$, has almost always been assumed, simplified, or fitted rather than derived; first-principles constructions have so far been restricted to fixed nuclear geometries or to a single adsorbate orbital. This work removes that limitation. Using exact analytic relationships, we establish a fit-free mapping from periodic DFT to the multi-orbital AN Hamiltonian through the PDOS.
\par
With this, the AN Hamiltonian for nonadiabatic dynamics becomes a first-principles effective theory: it combines the \textit{ab initio} rigor of strong-correlation approaches with explicit nuclear degrees of freedom, while its non-interacting form keeps key observables available in closed form. Every parameter ($\mathbf{\Delta}(E,R)$, $\hat{H}_\text{ad}(R)$, and $\hat{V}_N(R)$) is now derivable from DFT, and the approximations that previous studies made are no longer necessary and can be rigorously tested. Our calculations on H/Cu and CO/Cu show that this approach is essential to study realistic systems. Widely used assumptions, including the wide-band limit, bulk-DOS proportionality, and separability of $\mathbf{\Delta}(E,R)$ in $E$ and $R$, fail in these realistic systems, and the single-orbital representation itself breaks down dramatically for CO/Cu.
\par
Within the multi-orbital AN theory, we derived closed-form expressions for the hybridization energy density, electronic dynamics following an excitation, and the electronic friction. 
Our improved electronic-dynamics calculation reveals oscillations and potential slow-decay components that are invisible to Fermi's golden rule.
The vibrational relaxation time calculations reveal that modeling nonadiabatic dynamics with a single adsorbate orbital can result in drastic failures, and that multiple adsorbate orbitals are required to capture nonadiabatic effects realistically. 

The first-principles AN Hamiltonian constructed here provides a rigorous starting point for several directions: quantum and nonadiabatic nuclear dynamics on the bare nuclear potential $\hat{V}_N(R)$ coupled to the first-principles $\mathbf{\Delta}(E,R)$, correlated extensions that reintroduce electron-electron interactions on the adsorbate as in the original Anderson model, and a re-examination of descriptor-based reactivity trends in catalysis with hybridization functions that are computed rather than assumed. We anticipate that this framework will enable microscopic modeling of reaction dynamics at metal surfaces.

\section*{Acknowledgments}
This work was supported by Harvard University's
startup funds and the DOE Office of Fusion Energy Sciences
``Foundations for quantum simulation of warm dense matter'' project.
This research used the FAS Research Computing cluster supported by the Faculty of Arts and Sciences (FAS) Division of Science Research Computing Group at Harvard. 
The authors thank Anton Ni for assistance with using the periodic boundary condition extension of Q-Chem.~\cite{lee2021approaching,lee2022faster,rettig2023even,ni2025gaussian,dinh2026efficient}

\section*{AI Statement}
We used GPT-5 to help search for literature results and discuss ideas. GitHub Copilot assisted with computational tasks. We used Anthropic's Opus 5 and Fable 5 in the preparation of the manuscript. All uses were with supervision and verification by the authors.

\section*{References}
\bibliographystyle{unsrt}
\typeout{}
\bibliography{references.bib}

@article{ghan2020improved,
  title={Improved projection-operator diabatization schemes for the calculation of electronic coupling values},
  author={Ghan, Simiam and Kunkel, Christian and Reuter, Karsten and Oberhofer, Harald},
  journal={Journal of Chemical Theory and Computation},
  volume={16},
  number={12},
  pages={7431--7443},
  year={2020},
  publisher={ACS Publications}
}

@article{jacob2010dynamical,
  title={Dynamical mean-field theory for molecular electronics: Electronic structure and transport properties},
  author={Jacob, D and Haule, K and Kotliar, G},
  journal={Physical Review B—Condensed Matter and Materials Physics},
  volume={82},
  number={19},
  pages={195115},
  year={2010},
  publisher={APS}
}

@article{valli2020kondo,
  title={Kondo screening in {C}o adatoms with full Coulomb interaction},
  author={Valli, Angelo and Bahlke, Marc Philipp and Kowalski, Alexander and Karolak, Michael and Herrmann, Carmen and Sangiovanni, Giorgio},
  journal={Physical Review Research},
  volume={2},
  number={3},
  pages={033432},
  year={2020},
  publisher={APS}
}

@article{shick2022spin,
  title={Spin-orbit coupling and Kondo resonance in the {C}o adatom on the {C}u (100) surface: {DFT} plus exact diagonalization study},
  author={Shick, AB and Tchaplianka, M and Lichtenstein, AI},
  journal={Physical Review B},
  volume={106},
  number={24},
  pages={245115},
  year={2022},
  publisher={APS}
}

@article{dan2023theoretical,
  title={Theoretical study of nonadiabatic hydrogen atom scattering dynamics on metal surfaces using the hierarchical equations of motion method},
  author={Dan, Xiaohan and Shi, Qiang},
  journal={The Journal of Chemical Physics},
  volume={159},
  number={4},
  year={2023},
  publisher={AIP Publishing}
}

@article{forsblom2007vibrational,
  title={Vibrational lifetimes of cyanide and carbon monoxide on noble and transition metal surfaces},
  author={Forsblom, Mattias and Persson, Mats},
  journal={The Journal of chemical physics},
  volume={127},
  number={15},
  year={2007},
  publisher={AIP Publishing}
}

@article{maurer2016ab,
  title={Ab initio tensorial electronic friction for molecules on metal surfaces: Nonadiabatic vibrational relaxation},
  author={Maurer, Reinhard J and Askerka, Mikhail and Batista, Victor S and Tully, John C},
  journal={Physical Review B},
  volume={94},
  number={11},
  pages={115432},
  year={2016},
  publisher={APS}
}

@article{owrutsky1992femtosecond,
  title={Femtosecond coherent transient infrared spectroscopy of {CO} on {C}u (111)},
  author={Owrutsky, JC and Culver, JP and Li, M and Kim, YR and Sarisky, MJ and Yeganeh, MS and Yodh, AG and Hochstrasser, RM},
  journal={The Journal of chemical physics},
  volume={97},
  number={6},
  pages={4421--4427},
  year={1992},
  publisher={American Institute of Physics}
}

@book{zee2016group,
  title={Group theory in a nutshell for physicists},
  author={Zee, Anthony},
  year={2016},
  publisher={Princeton University Press}
}

@book{Hall2015LieGroups,
  author    = {Hall, Brian C.},
  title     = {Lie Groups, Lie Algebras, and Representations: An Elementary Introduction},
  series    = {Graduate Texts in Mathematics},
  edition   = {2},
  publisher = {Springer Cham},
  year      = {2015},
  doi       = {10.1007/978-3-319-13467-3},
  isbn      = {978-3-319-13467-3},
  pages     = {449}
}

@article{norskov2011density,
  title={Density functional theory in surface chemistry and catalysis},
  author={N{\o}rskov, Jens K and Abild-Pedersen, Frank and Studt, Felix and Bligaard, Thomas},
  journal={Proceedings of the National Academy of Sciences},
  volume={108},
  number={3},
  pages={937--943},
  year={2011},
  publisher={National Academy of Sciences}
}

@article{nilsson2005electronic,
  title={The electronic structure effect in heterogeneous catalysis},
  author={Nilsson, Anders and Pettersson, Lars GM and Hammer, Bj{\o}rk and Bligaard, Thomas and Christensen, Claus H and N{\o}rskov, Jens Kehlet},
  journal={Catalysis letters},
  volume={100},
  number={3},
  pages={111--114},
  year={2005},
  publisher={Springer}
}

@article{lowdin1950non,
  title={On the non-orthogonality problem connected with the use of atomic wave functions in the theory of molecules and crystals},
  author={L{\"o}wdin, Per-Olov},
  journal={The Journal of Chemical Physics},
  volume={18},
  number={3},
  pages={365--375},
  year={1950},
  publisher={American Institute of Physics}
}

@misc{nist_CO,
    author = {P.J. Linstrom and W.G. Mallard},
    title={{NIST} Chemistry WebBook, {NIST} Standard Reference Database Number 69},
    note = {National Institute of Standards and Technology, Gaithersburg MD, 20899, https://doi.org/10.18434/T4D303, (retrieved July 14, 2026).}
}

@article{hollins1979interactions,
  title={Interactions of {CO} molecules adsorbed on {C}u (111)},
  author={Hollins, P and Pritchard, J},
  journal={Surface Science},
  volume={89},
  number={1-3},
  pages={486--495},
  year={1979},
  publisher={Elsevier}
}

@article{chen2023accurate,
  title={Accurate descriptions of molecule-surface interactions in electrocatalytic {CO$_2$} reduction on the copper surfaces},
  author={Chen, Zheng and Liu, Zhangyun and Xu, Xin},
  journal={Nature Communications},
  volume={14},
  number={1},
  pages={936},
  year={2023},
  publisher={Nature Publishing Group UK London}
}

@article{xu2018atomic,
  title={Atomic and molecular adsorption on {C}u (111)},
  author={Xu, Lang and Lin, Joshua and Bai, Yunhai and Mavrikakis, Manos},
  journal={Topics in Catalysis},
  volume={61},
  number={9},
  pages={736--750},
  year={2018},
  publisher={Springer}
}

@article{pang2007adsorption,
  title={Adsorption of atoms on {C}u surfaces: a density functional theory study},
  author={Pang, Xian-Yong and Xue, Li-Qin and Wang, Gui-Chang},
  journal={Langmuir},
  volume={23},
  number={9},
  pages={4910--4917},
  year={2007},
  publisher={ACS Publications}
}

@article{mudiyanselage2013adsorption,
  title={Adsorption of hydrogen on the surface and sub-surface of {C}u (111)},
  author={Mudiyanselage, Kumudu and Yang, Yixiong and Hoffmann, Friedrich M and Furlong, Octavio J and Hrbek, Jan and White, Michael G and Liu, Ping and Stacchiola, Dar{\'\i}o J},
  journal={The Journal of Chemical Physics},
  volume={139},
  number={4},
  year={2013},
  publisher={AIP Publishing}
}

@article{sakong2003dissociative,
  title={Dissociative adsorption of hydrogen on strained {C}u surfaces},
  author={Sakong, Sung and Gro{\ss}, Axel},
  journal={Surface science},
  volume={525},
  number={1-3},
  pages={107--118},
  year={2003},
  publisher={Elsevier}
}

@article{mizielinski2005electronic,
  title={Electronic nonadiabatic effects in the adsorption of hydrogen atoms on metals},
  author={Mizielinski, Matthew S and Bird, David M and Persson, Mats and Holloway, Stephen},
  journal={The Journal of chemical physics},
  volume={122},
  number={8},
  year={2005},
  publisher={AIP Publishing}
}

@article{epifanovsky2021software,
  title={Software for the frontiers of quantum chemistry: An overview of developments in the Q-Chem 5 package},
  author={Epifanovsky, Evgeny and Gilbert, Andrew TB and Feng, Xintian and Lee, Joonho and Mao, Yuezhi and Mardirossian, Narbe and Pokhilko, Pavel and White, Alec F and Coons, Marc P and Dempwolff, Adrian L and others},
  journal={The Journal of chemical physics},
  volume={155},
  number={8},
  year={2021},
  publisher={AIP Publishing}
}

@article{kresse1999ultrasoft,
  title={From ultrasoft pseudopotentials to the projector augmented-wave method},
  author={Kresse, Georg and Joubert, Daniel},
  journal={Physical review b},
  volume={59},
  number={3},
  pages={1758},
  year={1999},
  publisher={APS}
}

@article{kresse1996efficiency,
  title={Efficiency of ab-initio total energy calculations for metals and semiconductors using a plane-wave basis set},
  author={Kresse, Georg and Furthm{\"u}ller, J{\"u}rgen},
  journal={Computational materials science},
  volume={6},
  number={1},
  pages={15--50},
  year={1996},
  publisher={Elsevier}
}

@article{kresse1996efficient,
  title={Efficient iterative schemes for ab initio total-energy calculations using a plane-wave basis set},
  author={Kresse, Georg and Furthm{\"u}ller, J{\"u}rgen},
  journal={Physical review B},
  volume={54},
  number={16},
  pages={11169},
  year={1996},
  publisher={APS}
}

@article{grimme2011effect,
  title={Effect of the damping function in dispersion corrected density functional theory},
  author={Grimme, Stefan and Ehrlich, Stephan and Goerigk, Lars},
  journal={Journal of computational chemistry},
  volume={32},
  number={7},
  pages={1456--1465},
  year={2011},
  publisher={Wiley Online Library}
}

@article{vandevondele2007gaussian,
  title={Gaussian basis sets for accurate calculations on molecular systems in gas and condensed phases},
  author={VandeVondele, Joost and Hutter, J{\"u}rg},
  journal={The Journal of chemical physics},
  volume={127},
  number={11},
  year={2007},
  publisher={AIP Publishing}
}

@article{hartwigsen1998relativistic,
  title={Relativistic separable dual-space Gaussian pseudopotentials from {H} to {R}n},
  author={Hartwigsen, Christian and G{\oe}decker, Sephen and Hutter, J{\"u}rg},
  journal={Physical Review B},
  volume={58},
  number={7},
  pages={3641},
  year={1998},
  publisher={APS}
}

@article{goedecker1996separable,
  title={Separable dual-space Gaussian pseudopotentials},
  author={Goedecker, Stefan and Teter, Michael and Hutter, J{\"u}rg},
  journal={Physical Review B},
  volume={54},
  number={3},
  pages={1703},
  year={1996},
  publisher={APS}
}

@article{perdew1996generalized,
  title={Generalized gradient approximation made simple},
  author={Perdew, John P and Burke, Kieron and Ernzerhof, Matthias},
  journal={Physical review letters},
  volume={77},
  number={18},
  pages={3865},
  year={1996},
  publisher={APS}
}

@article{aizawa1998first,
  title={First-principles study of {CO} bonding to {P}t (111): validity of the Blyholder model},
  author={Aizawa, Hideaki and Tsuneyuki, Shinji},
  journal={Surface science},
  volume={399},
  number={2-3},
  pages={L364--L370},
  year={1998},
  publisher={Elsevier}
}

@article{blyholder1964molecular,
  title={Molecular orbital view of chemisorbed carbon monoxide},
  author={Blyholder, George},
  journal={The Journal of Physical Chemistry},
  volume={68},
  number={10},
  pages={2772--2777},
  year={1964},
  publisher={ACS Publications}
}

@article{anderson1961localized,
  title={Localized magnetic states in metals},
  author={Anderson, Philip Warren},
  journal={Physical Review},
  volume={124},
  number={1},
  pages={41},
  year={1961},
  publisher={APS}
}

@article{nitzan2001electron,
  title={Electron transmission through molecules and molecular interfaces},
  author={Nitzan, Abraham},
  journal={Annual review of physical chemistry},
  volume={52},
  number={1},
  pages={681--750},
  year={2001},
  publisher={Annual Reviews 4139 El Camino Way, PO Box 10139, Palo Alto, CA 94303-0139, USA}
}

@book{eliaz2019physical,
  title={Physical electrochemistry: fundamentals, techniques, and applications},
  author={Eliaz, Noam and Gileadi, Eliezer},
  year={2019},
  publisher={John Wiley \& Sons}
}

@book{schmickler2010interfacial,
  title={Interfacial electrochemistry},
  author={Schmickler, Wolfgang and Santos, Elizabeth},
  year={2010},
  publisher={Springer Science \& Business Media}
}

@article{sanna2025steam,
  title={Steam reforming of methane: state of the art and novel technologies},
  author={Sanna, Aimaro and Openshaw, Dillon and Oghotomo, Princess and Bagnato, Giuseppe},
  journal={Reaction Chemistry \& Engineering},
  volume={10},
  number={9},
  pages={1963--1977},
  year={2025},
  publisher={Royal Society of Chemistry}
}

@article{ertl1980surface,
  title={Surface science and catalysis—studies on the mechanism of ammonia synthesis: the PH Emmett award address},
  author={Ertl, G},
  journal={Catalysis Reviews Science and Engineering},
  volume={21},
  number={2},
  pages={201--223},
  year={1980},
  publisher={Taylor \& Francis}
}

@article{rammer2011quantum,
  title={Quantum field theory of non-equilibrium states},
  author={Rammer, J{\o}rgen},
  journal={Quantum Field Theory of Non-equilibrium States},
  year={2011}
}

@article{hammer1996co,
  title={{CO} chemisorption at metal surfaces and overlayers},
  author={Hammer, Bj{\o}rk and Morikawa, Y and N{\o}rskov, Jens Kehlet},
  journal={Physical review letters},
  volume={76},
  number={12},
  pages={2141},
  year={1996},
  publisher={APS}
}

@book{bellman1997introduction,
  title={Introduction to matrix analysis},
  author={Bellman, Richard},
  year={1997},
  publisher={SIAM}
}

@article{brandbyge1995electronically,
  title={Electronically driven adsorbate excitation mechanism in femtosecond-pulse laser desorption},
  author={Brandbyge, Mads and Hedeg{\aa}rd, Per and Heinz, TF and Misewich, JA and Newns, DM},
  journal={Physical Review B},
  volume={52},
  number={8},
  pages={6042},
  year={1995},
  publisher={APS}
}

@article{saalfrank2006quantum,
  title={Quantum dynamical approach to ultrafast molecular desorption from surfaces},
  author={Saalfrank, Peter},
  journal={Chemical reviews},
  volume={106},
  number={10},
  pages={4116--4159},
  year={2006},
  publisher={ACS Publications}
}

@article{omiya2014coverage,
  title={Coverage dependent non-adiabaticity of {CO} on a copper surface},
  author={Omiya, Takuma and Arnolds, Heike},
  journal={The Journal of Chemical Physics},
  volume={141},
  number={21},
  year={2014},
  publisher={AIP Publishing}
}

@article{germer1994picosecond,
  title={Picosecond time-resolved adsorbate response to substrate heating: Spectroscopy and dynamics of {CO}/{C}u (100)},
  author={Germer, Thomas A and Stephenson, John C and Heilweil, Edwin J and Cavanagh, Richard R},
  journal={The Journal of chemical physics},
  volume={101},
  number={2},
  pages={1704--1716},
  year={1994},
  publisher={American Institute of Physics}
}

@article{harris1990vibrational,
  title={Vibrational energy transfer to metal surfaces probed by sum generation: {CO}/{C}u (100) and {CH$_3$S}/{A}g (111)},
  author={Harris, AL and Levinos, NJ and Rothberg, L and Dubois, LH and Dhar, L and Shane, SF and Morin, M},
  journal={Journal of Electron Spectroscopy and Related Phenomena},
  volume={54},
  pages={5--16},
  year={1990},
  publisher={Elsevier}
}

@article{lamont1995dynamics,
  title={Dynamics of atomic adsorbates: hydrogen on {C}u (111)},
  author={Lamont, CLA and Persson, BNJ and Williams, GP},
  journal={Chemical physics letters},
  volume={243},
  number={5-6},
  pages={429--434},
  year={1995},
  publisher={Elsevier}
}

@article{bunermann2015electron,
  title={Electron-hole pair excitation determines the mechanism of hydrogen atom adsorption},
  author={B{\"u}nermann, Oliver and Jiang, Hongyan and Dorenkamp, Yvonne and Kandratsenka, Alexander and Janke, Svenja M and Auerbach, Daniel J and Wodtke, Alec M},
  journal={Science},
  volume={350},
  number={6266},
  pages={1346--1349},
  year={2015},
  publisher={American Association for the Advancement of Science}
}

@article{wurth2000ultrafast,
  title={Ultrafast electron dynamics at surfaces probed by resonant Auger spectroscopy},
  author={Wurth, W and Menzel, D},
  journal={Chemical Physics},
  volume={251},
  number={1-3},
  pages={141--149},
  year={2000},
  publisher={Elsevier}
}

@article{fohlisch2003energy,
  title={Energy dependence of resonant charge transfer from adsorbates to metal substrates},
  author={F{\"o}hlisch, A and Menzel, D and Feulner, P and Ecker, M and Weimar, R and Kostov, KL and Tyuliev, G and Lizzit, S and Larciprete, R and Hennies, F and others},
  journal={Chemical physics},
  volume={289},
  number={1},
  pages={107--115},
  year={2003},
  publisher={Elsevier}
}

@article{hertl2026first,
    author = {Hertl, Nils and Koczor-Benda, Zsuzsanna and Maurer, Reinhard J.},
    title = {First-principles Newns–Anderson Hamiltonian construction for chemisorbed hydrogen at metal surfaces},
    journal = {The Journal of Chemical Physics},
    volume = {165},
    number = {6},
    pages = {064105},
    year = {2026},
    month = {08},
    issn = {0021-9606},
    doi = {10.1063/5.0334786},
    url = {https://doi.org/10.1063/5.0334786},
    eprint = {https://pubs.aip.org/aip/jcp/article-pdf/doi/10.1063/5.0334786/21099830/064105_1_5.0334786.pdf},
}

@article{wang2020bayesian,
  title={Bayesian learning of chemisorption for bridging the complexity of electronic descriptors},
  author={Wang, Siwen and Pillai, Hemanth Somarajan and Xin, Hongliang},
  journal={Nature communications},
  volume={11},
  number={1},
  pages={6132},
  year={2020},
  publisher={Nature Publishing Group UK London}
}

@article{head2000tensors,
  title={Tensors in electronic structure theory: basic concepts and applications to electron correlation models},
  author={Head-Gordon, Martin and Lee, Michael and Maslen, Paul and van Voorhis, Troy and Gwaltney, Steven and others},
  journal={Modern Methods and Algorithms of Quantum Chemistry},
  volume={3},
  pages={593--638},
  year={2000},
  publisher={John von Neumann Institute for Computing J{\"u}lich}
}

@article{santos2006model,
  title={A model for bond-breaking electron transfer at metal electrodes},
  author={Santos, E and Koper, MTM and Schmickler, W},
  journal={Chemical physics letters},
  volume={419},
  number={4-6},
  pages={421--425},
  year={2006},
  publisher={Elsevier}
}

@article{vojvodic2014electronic,
  title={Electronic structure effects in transition metal surface chemistry},
  author={Vojvodic, A and N{\o}rskov, JK and Abild-Pedersen, F},
  journal={Topics in catalysis},
  volume={57},
  number={1},
  pages={25--32},
  year={2014},
  publisher={Springer}
}

@article{vijay2022limits,
  title={Limits to scaling relations between adsorption energies?},
  author={Vijay, Sudarshan and Kastlunger, Georg and Chan, Karen and N{\o}rskov, Jens K},
  journal={The Journal of Chemical Physics},
  volume={156},
  number={23},
  year={2022},
  publisher={AIP Publishing}
}

@book{nitzan2024chemical,
  title={Chemical dynamics in condensed phases: relaxation, transfer, and reactions in condensed molecular systems},
  author={Nitzan, Abraham},
  year={2024},
  publisher={Oxford university press}
}

@book{norskov2014fundamental,
  title={Fundamental concepts in heterogeneous catalysis},
  author={N{\o}rskov, Jens K and Studt, Felix and Abild-Pedersen, Frank and Bligaard, Thomas},
  year={2014},
  publisher={John Wiley \& Sons}
}

@article{tiwari2021reactivity,
  title={Reactivity of transition-metal alloys to oxygen and sulfur},
  author={Tiwari, Rajarshi and Nelson, James and Xu, Chen and Sanvito, Stefano},
  journal={Physical Review Materials},
  volume={5},
  number={8},
  pages={083801},
  year={2021},
  publisher={APS}
}

@article{greiner2018free,
  title={Free-atom-like d states in single-atom alloy catalysts},
  author={Greiner, Mark T and Jones, TE and Beeg, Sebastian and Zwiener, Leon and Scherzer, Michael and Girgsdies, Frank and Piccinin, S and Armbr{\"u}ster, Marc and Knop-Gericke, Axel and Schl{\"o}gl, Robert},
  journal={Nature chemistry},
  volume={10},
  number={10},
  pages={1008--1015},
  year={2018},
  publisher={Nature Publishing Group UK London}
}

@article{persson1980vibrational,
  title={Vibrational lifetime for CO adsorbed on Cu (100)},
  author={Persson, BNJ and Persson, M},
  journal={Solid State Communications},
  volume={36},
  number={2},
  pages={175--179},
  year={1980},
  publisher={Elsevier}
}

@article{dou2015frictional,
  title={Frictional effects near a metal surface},
  author={Dou, Wenjie and Nitzan, Abraham and Subotnik, Joseph E},
  journal={The Journal of chemical physics},
  volume={143},
  number={5},
  year={2015},
  publisher={AIP Publishing}
}

@article{dou2017born,
  title={Born-Oppenheimer dynamics, electronic friction, and the inclusion of electron-electron interactions},
  author={Dou, Wenjie and Miao, Gaohan and Subotnik, Joseph E},
  journal={Physical review letters},
  volume={119},
  number={4},
  pages={046001},
  year={2017},
  publisher={APS}
}

@article{shenvi2009nonadiabatic,
  title={Nonadiabatic dynamics at metal surfaces: Independent-electron surface hopping},
  author={Shenvi, Neil and Roy, Sharani and Tully, John C},
  journal={The Journal of chemical physics},
  volume={130},
  number={17},
  year={2009},
  publisher={AIP Publishing}
}

@article{gardner2023assessing,
  title={Assessing mixed quantum-classical molecular dynamics methods for nonadiabatic dynamics of molecules on metal surfaces},
  author={Gardner, James and Habershon, Scott and Maurer, Reinhard J},
  journal={The Journal of Physical Chemistry C},
  volume={127},
  number={31},
  pages={15257--15270},
  year={2023},
  publisher={ACS Publications}
}

@article{gardner2023efficient,
  title={Efficient implementation and performance analysis of the independent electron surface hopping method for dynamics at metal surfaces},
  author={Gardner, James and Corken, Daniel and Janke, Svenja M and Habershon, Scott and Maurer, Reinhard J},
  journal={The Journal of Chemical Physics},
  volume={158},
  number={6},
  year={2023},
  publisher={AIP Publishing}
}

@article{dan2026nonadiabatic,
  title={Nonadiabatic H-atom scattering channels on {G}e (111) elucidated by the hierarchical equations of motion},
  author={Dan, Xiaohan and Long, Zhuoran and Qiu, Tianyin and Menzel, Jan Paul and Shi, Qiang and S Batista, Victor},
  journal={The Journal of Chemical Physics},
  volume={164},
  number={2},
  year={2026},
  publisher={AIP Publishing}
}

@article{preston2025nonadiabatic,
  title={Nonadiabatic quantum dynamics of molecules scattering from metal surfaces},
  author={Preston, Riley J and Ke, Yaling and Rudge, Samuel L and Hertl, Nils and Borrelli, Raffaele and Maurer, Reinhard J and Thoss, Michael},
  journal={Journal of Chemical Theory and Computation},
  volume={21},
  number={3},
  pages={1054--1063},
  year={2025},
  publisher={ACS Publications}
}

@article{santos2009model,
  title={Model for the electrocatalysis of hydrogen evolution},
  author={Santos, Elizabeth and Lundin, Angelica and P{\"o}tting, Kay and Quaino, Paola and Schmickler, Wolfgang},
  journal={Physical Review B—Condensed Matter and Materials Physics},
  volume={79},
  number={23},
  pages={235436},
  year={2009},
  publisher={APS}
}

@article{ghan2023interpreting,
  title={Interpreting ultrafast electron transfer on surfaces with a converged first-principles Newns--Anderson chemisorption function},
  author={Ghan, Simiam and Diesen, Elias and Kunkel, Christian and Reuter, Karsten and Oberhofer, Harald},
  journal={The Journal of Chemical Physics},
  volume={158},
  number={23},
  year={2023},
  publisher={AIP Publishing}
}

@article{newns1969,
  title={Self-consistent model of hydrogen chemisorption},
  author={Newns, DM},
  journal={Physical Review},
  volume={178},
  number={3},
  pages={1123},
  year={1969},
  publisher={APS}
}

@book{brown2009complex,
  title={Complex variables and applications},
  author={Brown, James Ward and Churchill, Ruel V},
  year={2009},
  publisher={McGraw-Hill,}
}

@article{lee2021approaching,
  title={Approaching the basis set limit in Gaussian-orbital-based periodic calculations with transferability: Performance of pure density functionals for simple semiconductors},
  author={Lee, Joonho and Feng, Xintian and Cunha, Leonardo A and Gonthier, J{\'e}r{\^o}me F and Epifanovsky, Evgeny and Head-Gordon, Martin},
  journal={The Journal of Chemical Physics},
  volume={155},
  number={16},
  year={2021},
  publisher={AIP Publishing}
}

@article{lee2022faster,
  title={Faster exact exchange for solids via occ-RI-K: Application to combinatorially optimized range-separated hybrid functionals for simple solids with pseudopotentials near the basis set limit},
  author={Lee, Joonho and Rettig, Adam and Feng, Xintian and Epifanovsky, Evgeny and Head-Gordon, Martin},
  journal={Journal of chemical theory and computation},
  volume={18},
  number={12},
  pages={7336},
  year={2022}
}

@article{rettig2023even,
  title={Even faster exact exchange for solids via tensor hypercontraction},
  author={Rettig, Adam and Lee, Joonho and Head-Gordon, Martin},
  journal={Journal of Chemical Theory and Computation},
  volume={19},
  number={17},
  pages={5773--5784},
  year={2023},
  publisher={ACS Publications}
}

@article{ni2025gaussian,
  title={Gaussian-based periodic grand canonical density functional theory with implicit solvation for computational electrochemistry},
  author={Ni, Anton Z and Rettig, Adam and Lee, Joonho},
  journal={Journal of Chemical Theory and Computation},
  volume={21},
  number={21},
  pages={10961--10970},
  year={2025},
  publisher={ACS Publications}
}

@article{dinh2026efficient,
  title={Efficient all-electron periodic Fourier-transformed Coulomb method},
  author={Dinh, Hieu Q and Rettig, Adam and Feng, Xintian and Lee, Joonho},
  journal={The Journal of Chemical Physics},
  volume={164},
  number={5},
  year={2026},
  publisher={AIP Publishing}
}

\clearpage
\appendix
\numberwithin{equation}{section}
\setcounter{figure}{0}
\renewcommand{\thefigure}{A\arabic{figure}}

\section{DFT calculation}
\subsection{H on Cu(111)}
For H atoms adsorbed on Cu (111) surface, the unit cell contains a $2\times 2\times 3$ Cu slab, with 3 layers and $2\times 2=4$ atoms per layer. The Cu slab is taken from an FCC lattice with a lattice constant of $3.615\,\text{\AA}$. A single H atom is placed in the unit cell, corresponding to a $1/4$ coverage. Unless otherwise noted, the H atom is placed at the fcc site, at a vertical distance of $0.888 \,\text{\AA}$ above the top Cu atom. The equilibrium geometry is obtained by geometry optimization using VASP\cite{kresse1996efficient, kresse1996efficiency, kresse1999ultrasoft} with PBE exhange-correlation functional.\cite{perdew1996generalized} We then perform periodic DFT with atom-centered basis functions using QCPBC,~\cite{lee2021approaching,lee2022faster,rettig2023even,ni2025gaussian,dinh2026efficient} an in-house periodic extension based on Q-Chem\cite{epifanovsky2021software} that is developed in our group. We use the PBE exchange-correlation functional,\cite{perdew1996generalized} GTH-PBE pseudopotential,\cite{goedecker1996separable,hartwigsen1998relativistic} DZVP-MOLOPT-PBE-GTH basis set,\cite{vandevondele2007gaussian} and D3-BJ dispersion correction.\cite{grimme2011effect} The Brillouin zone is sampled using a $5\times 5\times 1$ $\Gamma$-centered grid. The self-consistent field (SCF) procedure is performed with unrestricted spin, and a Fermi smearing width of $0.001$ a.u. (atomic units). The kinetic energy cutoff for the real-space grid is set to 1500 eV.
For each H atom, five AO basis functions are used, corresponding to the 1s, 2s, and 2p shells. 
\par
The adsorption energy
\begin{equation}
    E_{ads} = E_{\text{Cu-H}} - E_{\text{Cu slab}} - \frac{1}{2}E_{\text{H}_2}
\label{Eq:H_adsorption_energy}
\end{equation}
is found to be $-0.25$ eV. The difference to literature values ($-0.18$ eV\cite{sakong2003dissociative} and $-0.85/3=-0.28$ eV\cite{mudiyanselage2013adsorption}) is less than $0.1$ eV.
If $E_{\text{H}}$, the atomic hydrogen energy, is used instead of $\frac{1}{2}E_{\text{H}_2}$ in \cref{Eq:H_adsorption_energy}, then our calculation gives $E_{\text{ads}}=-2.54$ eV. Previous reported values are $-2.43$ eV,\cite{pang2007adsorption} $-2.45$ eV, and $-2.27$ eV.\cite{xu2018atomic}

\subsection{CO on Cu(111)}
We consider CO molecule adsorbed on the atop site of Cu (111) surface, with the C-O axis perpendicular to the surface. We use the same $2\times2\times3$ Cu slab as in the H/Cu example. The C atom in CO faces downward, closer to the Cu surface. We perform geometry optimizations and single-point calculations with the same settings as in the H/Cu example. 
The only differences are that the SCF is performed with restricted spin, and dispersion correction is not used. It has been reported that, for CO adsorption on Cu, adding the D3BJ dispersion correction to the PBE functional leads to over-binding.\cite{chen2023accurate}
\par
The adsorption energy
\begin{equation}
    E_{ads} = E_{\text{Cu-CO}} - E_{\text{Cu slab}} - E_{\text{CO}}
\end{equation}
is found to be -0.71 eV. This compares with the experimental value of -0.52 eV, determined by temperature-dependent desorption.\cite{hollins1979interactions} Another study using DFT with the PBE functional obtained a value of -0.75 eV.\cite{chen2023accurate}

\section{Tensor notation for non-orthogonal basis}
\label{App:tensor_notation}
In treating non-orthogonal orbital basis functions, we will adopt the tensor notation as in.\cite{head2000tensors} The basis atomic orbitals (AO) are denoted as $|\phi_\lambda\rangle$, with a lower index (or a covariant index). $\lambda$ indexes over the different orbitals centered on different atoms. We denote the dual basis as $\langle\phi^\lambda|$, with an upper index (or a contravariant index). The dual basis satisfies the biorthogonal relation: $\langle \phi^\lambda|\phi_\tau\rangle = \delta^\lambda_{\cdot\tau}$. The Hermitian conjugates of $|\phi_\lambda\rangle$ and $\langle \phi^\lambda|$ are $\langle \phi_\lambda|$ and $|\phi^\lambda\rangle$, respectively. The overlaps between different orbitals are denoted as $S_{\lambda\tau} = \langle \phi_\lambda|\phi_\tau\rangle$ .
Expanding $\langle\phi^\lambda|$ as the linear combination $\sum_\tau A^{\lambda\tau}\langle\phi_\tau|$ and using the bi-orthogonal relation, we see that $\mathbf{A} = \mathbf{S}^{-1}$. Therefore, the covariant orbitals $|\phi_\lambda\rangle$ and the contravariant orbitals $|\phi^\lambda\rangle$ are related by $|\phi^\lambda\rangle = \sum_\tau|\phi_\tau\rangle (\mathbf{S}^{-1})^{\tau\lambda}$ and $|\phi_\lambda\rangle=\sum_\tau|\phi^\tau\rangle S_{\tau\lambda}$, where we have used the property $S_{\lambda\tau}=S^*_{\tau\lambda}$. 
Change of basis from $|\phi_\lambda\rangle$ to $|\tilde{\phi}_\lambda\rangle$ is expressed as $|\tilde{\phi}_\lambda\rangle=\sum_\tau |\phi_\tau\rangle T^\tau_{\cdot \lambda}$. This implies that $\langle \tilde{\phi}^\lambda| = \sum_\tau (\mathbf{T}^{-1})^\lambda_{\cdot\tau} \langle \phi^\tau|$. Taking the Hermitian conjugates, we have $\langle\tilde{\phi}_\lambda| = \sum_\tau (\mathbf{T}^\dagger)_\lambda^{\cdot \tau}\langle \phi_\tau|$ and $|\tilde{\phi}^\lambda\rangle = \sum_\tau |\phi^\tau\rangle (\mathbf{T}^{-1 \dagger})_\tau^{\cdot \lambda}$.
\par
We use the hat notation (e.g. $\hat{A}$) to denote an operator without reference to any basis. When a basis is chosen, the matrix representation of an operator $\hat{A}$ is denoted as $A$, without the hat. The matrix elements are defined with respect to the covariant or contravariant basis. For example, $A_{\lambda\tau} = \langle \phi_\lambda|\hat{A}|\phi_\tau\rangle$, $A^{\lambda\tau} = \langle \phi^\lambda|\hat{A}|\phi^\tau\rangle$, and $A^\lambda_{\cdot\tau} = \langle \phi^\lambda|\hat{A}|\phi_\tau\rangle$. In particular, the identity operator $\hat{I}$ has the matrix elements $I^\lambda_{\cdot\tau}= \delta^\lambda_{\cdot\tau}$, $I_{\lambda\tau} = S_{\lambda\tau}$, and $I^{\lambda\tau}=(\mathbf{S}^{-1})^{\lambda\tau}$. Using the completeness relation $\hat{I} = \sum_\lambda | \phi^\lambda\rangle\langle \phi_\lambda|$, we can express an operator $\hat{A}$ in terms of the matrix elements. For example,
$\hat{A} = \sum_{\lambda,\tau}|\phi^\lambda\rangle A_{\lambda\tau}\langle\phi^\tau|$.

\section{Deriving the adsorbate-projected Green's function in the multi-orbital AN model with overlap}
\label{App:Deriving_Gaa_multi_orb_AN}

Working with the single-particle Hamiltonain $\hat{H}$, the Green's operator $\hat{G}(E)$ is defined according to
\begin{equation}
    \hat{G}(E) \lim_{\epsilon\rightarrow 0^+}((E+i\epsilon)\hat{I}-\hat{H}) = \hat{I}.
\label{Eq:Green_operator_identity}
\end{equation}
We will take the matrix of $\hat{G}(E)$ to be contravariant, and take the matrix of $E\hat{I}-\hat{H}$ to be covariant, so that
\begin{equation}
    \sum_{\lambda} G^{\rho\lambda}(E)\lim_{\epsilon\rightarrow 0^+}((E+i\epsilon)S_{\lambda\tau}-H_{\lambda\tau}) = \delta^\rho_{\cdot \tau},
\label{Eq:Green_matrix_identity_explicit_tensor_index}
\end{equation}
where $S_{\lambda\tau} = \langle\phi_\lambda|\phi_\tau\rangle$ is the overlap matrix.
Keeping in mind the contravariant nature of $\mathbf{G}$ and the covariant nature of $\mathbf{S}$ and $\mathbf{H}$, we can rewrite the matrix equation of \cref{Eq:Green_matrix_identity_explicit_tensor_index} compactly as
\begin{equation}
    \mathbf{G}(E) = \lim_{\epsilon\rightarrow0^+}((E+i\epsilon)\mathbf{S}-\mathbf{H})^{-1}.
\label{Eq:app_Greens_matrix_with_overlap}
\end{equation}
\par
We define the uncoupled Hamiltonian $\mathbf{H}_0$ by setting the off-diagonal blocks $\mathbf{H}_{as}$ and $\mathbf{H}_{sa}$ in $\mathbf{H}$ to zero. Similarly, the uncoupled overlap matrix $\mathbf{S}_0$ is defined by setting $\mathbf{S}_{as}$ and $\mathbf{S}_{sa}$ to zero. We write the differences as $\Delta \mathbf{H} = \mathbf{H}-\mathbf{H}_0$ and $\Delta \mathbf{S} = \mathbf{S}-\mathbf{S}_0$. 
The Green's functions $\mathbf{G}$ and $\mathbf{G}_0$ are defined as 
\begin{equation}
    \mathbf{G}(E) = (E\mathbf{S}-\mathbf{H})^{-1}
\label{Eq:app_G_full_def}
\end{equation}
and
\begin{equation}
    \mathbf{G}_0(E) = (E\mathbf{S}_0 - \mathbf{H}_0)^{-1}.
\end{equation}
We have omitted the limit $\lim_{\epsilon\rightarrow 0^+}$ for notational simplicity.
Inverting the equations, we see that 
\begin{equation}
    \mathbf{G}_0^{-1}(E) = \mathbf{G}^{-1}(E) + \Delta \mathbf{H} - E\Delta \mathbf{S}.
\end{equation}
Left multiplying by $\mathbf{G}_0(E)$ and right multiplying by $\mathbf{G}(E)$, we obtain the Dyson equation with overlap
\begin{equation}
    \mathbf{G}(E) = \mathbf{G}_0(E) + \mathbf{G}_0(E)(\Delta \mathbf{H} - E\Delta \mathbf{S})\mathbf{G}(E).
\label{Eq:App_modified_Dyson}
\end{equation}
Since $\mathbf{H}_0$ and $\mathbf{S}_0$ are block-diagonal, $\mathbf{G}_0$ is also block-diagonal.
Projecting \cref{Eq:App_modified_Dyson} to the adsorbate block, we have
\begin{equation}
    \mathbf{G}^{aa} = \mathbf{G}^{aa}_{0} + \mathbf{G}^{aa}_{0} (\mathbf{H} - E\mathbf{S})_{as} \mathbf{G}^{sa}.
\label{Eq:Gaa_Dyson}
\end{equation}
We have used the fact that the off-diagonal block $\mathbf{G}^{as}_{0}$ is zero.
Similarly,
\begin{equation}
    \mathbf{G}^{sa} = \mathbf{G}^{ss}_{0} ( \mathbf{H} - E\mathbf{S})_{sa}\mathbf{G}^{aa}.
\label{Eq:Gka_Dyson}
\end{equation}
Substituting \cref{Eq:Gka_Dyson} into \cref{Eq:Gaa_Dyson} and defining $\mathbf{Q}=(\mathbf{G}^{aa})^{-1}$, we see that 
\begin{align}
\begin{split}
    \mathbf{Q} = (\mathbf{G}_{0}^{aa})^{-1}-(\mathbf{H} - E\mathbf{S})_{as}\mathbf{G}_{0}^{ss}(\mathbf{H} - E\mathbf{S})_{sa}.
\label{Eq:app_multi_orb_phi0}
\end{split}
\end{align}
\par
To further simplify $\mathbf{Q}$, we notice that changing the basis of the surface orbitals among themselves does not affect the adsorbate submatrices $\mathbf{G}^{aa}$ or $\mathbf{Q}$. This is because the change of basis matrix $\mathbf{T}$ is block diagonal with $\mathbf{T}^a_{\cdot a} = 1$, so the inverse $\mathbf{T}^{-1}$ is also block diagonal with $(\mathbf{T}^{-1})^a_{\cdot a} = 1$. Therefore, the adsorbate orbitals $|\phi_\mu\rangle$ and $|\phi^\mu\rangle$ are invariant under the change of basis among surface orbitals.
Taking advantage of this invariance, we will work in the diagonalized surface basis, where $\mathbf{S}_{ss} = 1$ and $\mathbf{H}_{ss}$ is diagonal. The diagonal elements of $\mathbf{H}_{ss}$ are the diagonalized surface orbital energies $E_\sigma$. The adsorbate orbitals remain unchanged and are non-orthogonal (i.e., $\mathbf{S}_{aa}\neq 1$) in general. $\mathbf{S}_{as}$ and $\mathbf{H}_{as}$ also remain nonzero in general. 
\par
Under this partially diagonalized basis, $\mathbf{G}_{0}^{ss}$ is diagonal, with the diagonal element equal to
\begin{equation}
    G_{0}^{\sigma\sigma} = \lim_{\epsilon\rightarrow 0^+} \frac{1}{E-E_\sigma+i\epsilon}.
\label{Eq:app_G0kk_element}
\end{equation}
Substituting \cref{Eq:app_G0kk_element} into \cref{Eq:app_multi_orb_phi0} and using the identity
\begin{equation}
    \lim_{\epsilon\rightarrow0^+} \frac{1}{E-E_0+i\epsilon} = \mathcal{P}\frac{1}{E-E_0} -i\pi\delta(E-E_0),
\end{equation}
we obtain
\begin{align}
\begin{split}
    \mathbf{Q}(E) 
    &= \lim_{\epsilon\rightarrow0^+} (E+i\epsilon)\mathbf{S}_{aa}  - \mathbf{H}_{aa} - \mathbf{K}(E) + i\mathbf{\Delta}(E),
\label{Eq:app_phi_0}
\end{split}
\end{align}
where
\begin{align}
\begin{split}
    \Delta_{\mu\nu}(E) = \pi \sum_{\sigma \in s} &(H_{\mu \sigma} - E_\sigma S_{\mu \sigma}) \\
    &(H_{\sigma\nu} - E_\sigma S_{\sigma\nu})\delta(E-E_\sigma)
\label{Eq:app_multi_orb_AN_Delta}
\end{split}
\end{align}
and
\begin{equation}
    K_{\mu\nu}(E) = \sum_{\sigma \in s}\mathcal{P}\frac{(H_{\mu \sigma} - E S_{\mu \sigma})(H_{\sigma\nu} - E S_{\sigma\nu})}{E-E_\sigma}.
\label{Eq:app_multi_orb_AN_K}
\end{equation}
We remind the reader that $\mu$ and $\nu$ in \cref{Eq:app_multi_orb_AN_Delta,Eq:app_multi_orb_AN_K} are adsorbate orbital indices, while $\sigma$ is a surface orbital index.
\par
Different from the conventional single-orbital AN theory without overlap, in the presence of adsorbate-surface overlap (i.e., $\mathbf{S}_{as}\neq 0$), $\mathbf{K}(E)$ is not the Hilbert transform of $\mathbf{\Delta}(E)$. Instead, the Hilbert transform of $\mathbf{\Delta}(E)$ is
\begin{align}
\begin{split}
    &\Lambda_{\mu\nu}(E) = \frac{1}{\pi}\mathcal{P}\int \frac{\Delta_{\mu\nu}(E')}{E-E'
    } dE' \\
    &= \sum_{\sigma\in s}\mathcal{P}\frac{( H_{\mu \sigma} - E_\sigma S_{\mu \sigma})(H_{\sigma\nu} - E_\sigma S_{\sigma\nu})}{E-E_\sigma},
\label{Eq:app_multi_orb_Lambda}
\end{split}
\end{align}
where the $E$ in the numerator of \cref{Eq:app_multi_orb_AN_K} is replaced with $E_\sigma$.
The difference between $\mathbf{K}(E)$ and $\mathbf{\Lambda}(E)$ is obtained from the following calculation:
\begin{widetext}
\begin{align}
\begin{split}
    K_{\mu\nu}(E) &= \sum_{\sigma\in s}\mathcal{P}\frac{( H_{\mu \sigma} - E_\sigma S_{\mu \sigma} - (E-E_\sigma)S_{\mu \sigma})( H_{\sigma\nu} - E_\sigma S_{\sigma\nu}- (E-E_\sigma)S_{\sigma\nu})}{E-E_\sigma} \\
    &=\sum_{\sigma\in s}\mathcal{P}\frac{( H_{\mu \sigma} - E_\sigma S_{\mu \sigma} )( H_{\sigma\nu} - E_\sigma S_{\sigma\nu})}{E-E_\sigma} \\
    &\qquad\quad + \mathcal{P}\frac{E-E_\sigma}{E-E_\sigma}\Big( -S_{\mu \sigma}H_{\sigma\nu} - H_{\mu \sigma}S_{\sigma\nu} + 2S_{\mu \sigma}E_\sigma S_{\sigma\nu} \Big) \\
    & \qquad\quad + \mathcal{P}\frac{E-E_\sigma}{E-E_\sigma} S_{\mu \sigma}(E-E_\sigma)S_{\sigma\nu} \\
    &= \Lambda_{\mu\nu}(E) + \sum_{\sigma\in s}(E+E_\sigma) S_{\mu \sigma}S_{\sigma\nu} - S_{\mu \sigma}H_{\sigma\nu} - H_{\mu \sigma}S_{\sigma\nu}.
\label{Eq:app_multi_orb_K_1}
\end{split}
\end{align}
\end{widetext}
Using \cref{Eq:app_multi_orb_K_1}, we can express $\mathbf{Q}(E)$ in \cref{Eq:app_phi_0} in an alternative form as
\begin{equation}
    \mathbf{Q}(E) = E \mathbf{S}_\text{eff} - \mathbf{H}_\text{eff} - \mathbf{\Lambda}(E) + i\mathbf{\Delta}(E),
\label{Eq:app_phi_GS1}
\end{equation}
where the effective Hamiltonian and overlap matrices are
\begin{align}
\begin{split}
    H_{\text{eff},\mu\nu} &= H_{\mu\nu} +\sum_{\sigma\in s} S_{\mu \sigma} E_\sigma S_{\sigma \nu} - H_{\mu \sigma}S_{\sigma\nu} - S_{\mu \sigma}H_{\sigma\nu}
\label{Eq:H_GS1_from_PDOS_app}
\end{split}
\end{align}
and
\begin{equation}
    S_{\text{eff},\mu\nu}= S_{\mu\nu} - \sum_{\sigma \in s}S_{\mu \sigma} S_{\sigma \nu}.
\label{Eq:S_GS1_from_PDOS_app}
\end{equation}

\subsection{Evaluating Hilbert transform through Fourier transform}
We note that the principal value integral in the Hilbert transform does not pose a significant issue in numerical evaluation. This is because the Hilbert transform can be computed efficiently via Fourier transform, as implemented in the open-source SciPy routine scipy.signal.hilbert. We describe the connection between the Hilbert transform and Fourier transform below for completeness.
\par
Since the Hilbert transform of a matrix is performed element-wise, we only need to consider the Hilbert transform of a scalar function. Furthermore, the Hilbert transform of a complex-valued function can be decomposed as a sum of the Hilbert transform of the real part and the imaginary part, so it suffices to consider the Hilbert transform of a real-valued scalar function.
The Hilbert transform of a real-valued scalar function $\Delta(E)$ is defined as
\begin{align}
\begin{split}
    \Lambda(E) &= \frac{1}{\pi} \mathcal{P}\int^\infty_{-\infty} \frac{\Delta(E')}{E-E'} dE' \\
    &=\lim_{\epsilon\rightarrow0^+} \Big(\int^{E-\epsilon}_{-\infty} \frac{\Delta(E')}{E-E'} dE' + \int^{\infty}_{E+\epsilon} \frac{\Delta(E')}{E-E'} dE' \Big).
\label{Eq:app_hilbert_def}
\end{split}
\end{align}
Instead of computing the integral in the limit of $\epsilon\rightarrow 0$, we can compute $\Lambda$ using Fourier transforms.
\par
First, we note that the Hilbert transform implies that 
\begin{equation}
    \Lambda(E) - i\Delta(E) = \frac{1}{\pi}\lim_{\epsilon\rightarrow 0^+} \int \frac{\Delta(E')}{E-E'+i\epsilon} \,dE'.
\end{equation}
Defining the complex-valued function $\chi(E)$ as $\Lambda(E)-i\Delta(E)$, we notice that $\chi(E)$ has no pole in the upper complex plane. Assuming $\chi(E)$ decays fast enough as $|E|\rightarrow\infty$ in the upper plane, then the Fourier transform 
\begin{equation}
    \chi(t) = \frac{1}{2\pi} \int dE\, \chi(E) e^{-iEt}
\end{equation}
is nonzero only when $t>0$. When $t<0$, $\chi(t) = 0$.
\par
Next, we can write $\chi(t)$ as a sum of a Hermitian function $\chi_e$ and an anti-Hermitian function $\chi_o$, i.e.,
\begin{equation}
    \chi(t) = \chi_e(t) + \chi_o(t).
\end{equation}
The Hermitian part $\chi_e$ is
\begin{equation}
    \chi_e(t) =
    \begin{cases}
        \chi(t)/2 \quad\,\qquad, t>0 \\
        \chi^*(-t)/2 \qquad , t<0.
    \end{cases}
\end{equation}
Hermiticity of $\chi_e$ (i.e., $\chi_e(-t) = \chi^*_e(t)$) implies that it is the Fourier transform of a purely real function.
The anti-Hermitian part $\chi_o$ is
\begin{equation}
    \chi_o(t) = 
    \begin{cases}
        \chi(t)/2 \qquad\qquad, t>0 \\
        -\chi^*(-t)/2 \qquad , t<0.
    \end{cases}
\end{equation}
Anti-Hermiticity of $\chi_o$ (i.e., $\chi_o(-t) = -\chi^*_o(t)$) implies that it is the Fourier transform of a purely imaginary function. Combining these facts, we see that the Fourier transform of $-i\Delta(E)$ is $\chi_o(t)$, and the Fourier transform of $\Lambda(E)$ is $\chi_e(t)$.
\par
Therefore, a step-by-step procedure for computing $\Lambda$ from $\Delta$ is given below:
\begin{enumerate}
    \item Take the Fourier transform of $-i\Delta(E)$ 
    \begin{equation}
        \chi_o(t) = \frac{-i}{2\pi} \int dE\,\Delta(E)e^{-iEt}.
    \end{equation}

    \item Obtain $\chi(t)$ by taking
    \begin{equation}
        \chi(t) = 
        \begin{cases}
            2\chi_o(t) \qquad , t>0 \\
            0 \qquad\qquad, t<0
        \end{cases}
    \end{equation}

    \item $\Lambda(E)$ is the real part of the inverse Fourier transform of $\chi(t)$
    \begin{equation}
        \Lambda(E) = \text{Re} \,\chi(E) = \text{Re} \, \int dt \, \chi(t) e^{iEt}.
    \end{equation}
\end{enumerate}

\section{Constructing $\hat{H}_{el}$ from PDOS}
\label{Sec:app_construct_Hel_from_PDOS}
To show the detailed derivation for the procedure of constructing $\hat{H}_{el}$ from PDOS, we will first show the derivation without smearing. Then, we introduce Lorentzian smearing and show that the procedure remains valid under Lorentzian smearing, with a constant correction to $\mathbf{\Delta}(E)$. 

\subsection{Without smearing}
First, we show that PDOS is proportional to the anti-Hermitian part of $\mathbf{G}^{aa}$ (see \cref{Eq:multi_orb_AN_PDOS_from_G}).
We denote $|\psi_i\rangle$ and $E_i$ as the normalized MO states and energies of $\hat{H}$.
Using the completeness relation of MOs
\begin{equation}
    \hat{I} = \sum_i |\psi_i\rangle\langle\psi_i|,
\label{Eq:app_completeness_relation}
\end{equation}
we can write the Green's function matrix as
\begin{align}
\begin{split}
    G^{\mu\nu} &= \lim_{\epsilon\rightarrow 0^+} \sum_i \langle \phi^\mu|\psi_i\rangle\langle\psi_i|\frac{1}{(E+i\epsilon)\hat{I}-\hat{H}}|\psi_i\rangle\langle\psi_i|\phi^\nu\rangle \\
    &= \sum_i C^\mu_{\cdot i} (C^\nu_{\cdot i})^* (\mathcal{P}\frac{1}{E-E_i}-i\pi\delta(E-E_i)).
\end{split}
\end{align}
The Hermitian conjugate of the matrix $\mathbf{G}^{aa}$ is
\begin{align}
\begin{split}
    (\mathbf{G}^{\dagger})^{\mu\nu} &= (G^{\nu\mu})^* \\
    &= \sum_i C^\mu_{\cdot i} (C^\nu_{\cdot i})^* (\mathcal{P}\frac{1}{E-E_i}+i\pi\delta(E-E_i)).
\end{split}
\end{align}
Therefore, 
\begin{align}
\begin{split}
    -\frac{1}{\pi}\Big(\frac{\mathbf{G} - \mathbf{G}^{\dagger}}{2i}\Big)^{\mu\nu} &= \sum_i C^\mu_{\cdot i} (C^\nu_{\cdot i})^*\delta(E-E_i) \\
    &= \text{PDOS}^{\mu\nu},
\label{Eq:app_G_to_PDOS_no_smear}
\end{split}
\end{align}
and PDOS is equal to $-1/\pi$ times the anti-Hermitian part of $\mathbf{G}^{aa}$.
\par
Next, we show that the Hermitian and anti-Hermitian parts of $\mathbf{G}^{aa}$ satisfy the Kramers-Kronig relation, so that PDOS can be used to reconstruct the full $\mathbf{G}^{aa}$.
Since $\mathbf{G}^{aa}(E)$ has no pole in the upper complex plane and decays to $0$ as $|E|\rightarrow\infty$, the real and imaginary parts of $\mathbf{G}^{aa}(E)$ satisfy the Kramers-Kronig relation
\begin{equation}
    \text{Re } \mathbf{G}^{aa}(E) = -\frac{1}{\pi} \mathcal{P} \int \frac{\text{Im }\mathbf{G}^{aa}(E')}{E-E'}dE'.
\label{Eq:app_KK_relation_matrix}
\end{equation}
The real and imaginary parts are taken element-wise.
We then decompose $\mathbf{G}_{aa}(E)$ into the sum of a Hermitian part $\mathbf{A}(E)$ and an anti-Hermitian part $i\mathbf{B}(E)$ (with Hermitian $\mathbf{B}(E)$), i.e.,
\begin{equation}
    \mathbf{G}^{aa}(E) = \mathbf{A}(E) + i\mathbf{B}(E).
\end{equation}
Since $\mathbf{A}(E)$ and $\mathbf{B}(E)$ are not necessarily real-valued, we cannot identify $\mathbf{A}(E)$ as $\text{Re }\mathbf{G}_{aa}(E)$ and $\mathbf{B}(E)$ as $\text{Im }\mathbf{G}_{aa}(E)$.
$\mathbf{A}$ and $\mathbf{B}$ are given by
\begin{equation}
    \mathbf{A}(E) = \frac{\mathbf{G}^{aa}(E) + \mathbf{G}^{aa\dagger}(E)}{2}
\end{equation}
and
\begin{equation}
    \mathbf{B}(E) = \frac{\mathbf{G}^{aa}(E)-\mathbf{G}^{aa\dagger}(E)}{2i}.
\end{equation}
We further decompose $\mathbf{A}$ and $\mathbf{B}$ into real and imaginary parts, such that 
\begin{equation}
    \mathbf{A}(E) = \mathbf{A}_R(E) + i\mathbf{A}_I(E)
\end{equation}
and 
\begin{equation}
    \mathbf{B}(E) = \mathbf{B}_R(E) + i\mathbf{B}_I(E).
\end{equation}
Therefore, $\mathbf{A}_R$ and $\mathbf{B}_R$ are Hermitian (H.). $\mathbf{A}_I$ and $\mathbf{B}_I$ are anti-Hermitian (A.H.).
Now we can separate $\mathbf{G}^{aa}$ into real and imaginary parts, i.e.,
\begin{equation}
    \mathbf{G}^{aa}(E) = \underbrace{(\overbrace{\mathbf{A}_R(E)}^{\text{H.}}\overbrace{-\mathbf{B}_I(E)}^{\text{A.H.}})}_{\text{Re }\mathbf{G}^{aa}} + i\underbrace{(\overbrace{\mathbf{B}_R(E)}^{\text{H.}}+\overbrace{\mathbf{A}_I(E)}^{\text{A.H.}})}_{\text{Im }\mathbf{G}^{aa}}.
\end{equation}
Since the Kramers-Kronig relation (\cref{Eq:app_KK_relation_matrix}) applies to the matrix element-wise, the Hilbert transform of a Hermitian matrix is Hermitian, and the Hilbert transform of an anti-Hermitian matrix is anti-Hermitian. Therefore, $\mathbf{A}_R$ and $\mathbf{B}_R$ satisfy the Kramers-Kronig relation
\begin{equation}
    \mathbf{A}_R(E) = -\frac{1}{\pi}\mathcal{P}\int \frac{\mathbf{B}_R(E')}{E-E'} dE'.
\label{Eq:app_KK_AR_BR}
\end{equation}
Similarly,
\begin{equation}
    \mathbf{B}_I(E) = \frac{1}{\pi}\mathcal{P}\int \frac{\mathbf{A}_I(E')}{E-E'} dE'.
\end{equation}
Inverting the equation, we have
\begin{equation}
    \mathbf{A}_I(E) = -\frac{1}{\pi}\mathcal{P}\int \frac{\mathbf{B}_I(E')}{E-E'} dE'.
\label{Eq:app_KK_AI_BI}
\end{equation}
Combining \cref{Eq:app_KK_AR_BR,Eq:app_KK_AI_BI}, we have
\begin{equation}
    \mathbf{A}(E) = -\frac{1}{\pi}\mathcal{P}\int \frac{\mathbf{B}(E')}{E-E'} dE'.
\end{equation}
Note the similarity to \cref{Eq:app_KK_relation_matrix}.
Now, $\mathbf{G}^{aa}$ can be constructed from the PDOS using
\begin{equation}
    \mathbf{G}^{aa}(E) = i\mathbf{B}(E) -\frac{1}{\pi} \mathcal{P}\int \frac{\mathbf{B}(E')}{E-E'} dE',
\label{Eq:app_PDOS_to_G_no_smear}
\end{equation}
where 
\begin{equation}
    \mathbf{B}(E) = -\pi \text{PDOS}(E).
\end{equation}
\par
Once $\mathbf{Q}(E) = (\mathbf{G}^{aa}(E))^{-1}$ is obtained, we can compute $\mathbf{\Delta}(E)$, $\mathbf{H}_\text{eff}$, and $\mathbf{S}_\text{eff}$ by taking the Hermitian and anti-Hermitian parts of \cref{Eq:phi_GS1}, so that 
\begin{equation}
    \mathbf{\Delta}(E) = \frac{\mathbf{Q}(E)-\mathbf{Q}^\dagger(E)}{2i}
\end{equation}
and
\begin{equation}
    E \mathbf{S}_\text{eff} - \mathbf{H}_\text{eff} = \frac{\mathbf{Q}(E) + \mathbf{Q}^\dagger(E)}{2} + \mathbf{\Lambda}(E),
\end{equation}
where $\mathbf{\Lambda}(E)$ is the Hilbert transform of $\mathbf{\Delta}(E)$.

\subsection{With Lorentzian smearing}
The Lorentzian smearing function is 
\begin{equation}
    \delta_\epsilon(E) = \frac{1}{\pi} \frac{\epsilon}{E^2+\epsilon^2}.
\label{Eq:app_Lorentzian}
\end{equation}
We will first show that the relationship between PDOS and $\mathbf{G}^{aa}$ is preserved under Lorentzian smearing. Then we show that the Hilbert transform between $\mathbf{\Delta}(E)$ and $\mathbf{\Lambda}(E)$ is preserved under Lorentzian smearing. Finally, we show that the relationship between $\mathbf{Q} = (\mathbf{G}^{aa})^{-1}$ and ($\mathbf{\Delta}(E),\mathbf{H}_{eff},\mathbf{S}_{eff}$) is preserved under Lorentzian smearing, up to a constant offset in $\mathbf{\Delta}(E)$. In this section, we will denote smeared quantities with a subscript $\epsilon$, e.g., $\text{PDOS}_\epsilon$.

\subsubsection{Relationship between $\text{PDOS}_\epsilon$ and $\mathbf{G}^{aa}_\epsilon$}

The smeared PDOS is defined as 
\begin{align}
\begin{split}
    \text{PDOS}_\epsilon^{\mu\nu}(E) &= \sum_{i} \langle \phi^\mu|\psi_i\rangle\langle\psi_i|\phi^\nu\rangle \delta_\epsilon(E-E_i) \\
    &= \sum_i C^\mu_{\cdot i}(C^{\nu}_{\cdot i})^*\delta_\epsilon(E-E_i).
\label{Eq:app_smeared_PDOS_def}
\end{split}
\end{align}
This is related to the unsmeared PDOS by
\begin{equation}
    \text{PDOS}_\epsilon(E) = \text{PDOS}(E) * \delta_\epsilon(E),
\label{Eq:smeared_PDOS_def}
\end{equation}
where $*$ denotes convolution, i.e.,
\begin{equation}
    f(E)*g(E) = \int dE'\, f(E')g(E-E').
\end{equation}
The convolution is applied element-wise. 
\par
The smeared Green's function operator $\hat{G}_\epsilon(E)$ is defined as
\begin{equation}
    \hat{G}_\epsilon = \frac{1}{(E+i\epsilon)\hat{I} - \hat{H}},
\label{Eq:multi_orbital_G_operator_smeared}
\end{equation}
where $\epsilon$ is a nonzero positive number.
Expressed as a contravariant matrix $\mathbf{G}_\epsilon(E)$,
\begin{equation}
    \mathbf{G}_\epsilon(E) = ((E+i\epsilon)\mathbf{S}-\mathbf{H})^{-1}.
\end{equation}
$\mathbf{G}^{aa}_\epsilon(E)$ is the adsorbate submatrix of $\mathbf{G}_\epsilon(E)$. We claim that $\mathbf{G}_\epsilon$ is also equal to
\begin{equation}
    \mathbf{G}_\epsilon(E) = \mathbf{G}(E) * \delta_\epsilon(E).
\label{Eq:app_smeared_G_equal_to_convolution}
\end{equation}
To prove \cref{Eq:app_smeared_G_equal_to_convolution}, we show that the $\lambda\tau$-th elements on both sides are equal.
The $\lambda\tau$-th element of the right hand side of \cref{Eq:app_smeared_G_equal_to_convolution} is
\begin{align}
\begin{split}
    \int dE' \sum_i \langle\phi^\lambda|\psi_i\rangle\frac{1}{E'+i\sigma^{0+}-E_i}\langle\psi_i|\psi^\tau\rangle \delta_\epsilon(E-E').
\label{Eq:app_expand_smeared_G_RHS}
\end{split}
\end{align}
$\{|\psi_i\rangle\}$ are the MO that diagonalize $\mathbf{H}$ with overlap $\mathbf{S}$. The notation $\sigma^{0+}$ means the limit $\lim_{\sigma\rightarrow 0^+}$ is taken.
Evaluating the integral using contour integration, \cref{Eq:app_expand_smeared_G_RHS} becomes
\begin{equation}
    \sum_i \langle\phi^\lambda|\psi_i\rangle\frac{1}{E+i\epsilon-E_i}\langle \psi_i|\phi^\tau\rangle = \langle\phi^\lambda|\hat{G}_\epsilon(E)|\phi^\tau\rangle,
\label{Eq:app_smeared_G_2}
\end{equation}
which is the $\lambda\tau$-th element of the left hand side of \cref{Eq:app_smeared_G_equal_to_convolution}.
\par
One can show that
\begin{equation}
    \text{PDOS}_\epsilon(E) = -\frac{1}{\pi} \frac{\mathbf{G}_{\epsilon}^{aa}(E)-\mathbf{G}_{\epsilon}^{aa\dagger}(E)}{2i}.
\label{Eq:multi_orb_AN_PDOS_from_G_smeared}
\end{equation}
using \cref{Eq:app_smeared_G_2}.
Since $\mathbf{G}_\epsilon^{aa}(E)$ also has no pole in the upper half-plane and decays to zero fast enough as $|E| \rightarrow \infty$. Therefore, $\mathbf{G}_{\epsilon,aa}(E)$ satisfies the same Kramers-Kronig relation, so that 
\begin{equation}
    \mathbf{G}^{aa}_\epsilon(E) = -i\pi\text{PDOS}_\epsilon(E) + \mathcal{P}\int \frac{\text{PDOS}_\epsilon(E')}{E-E'} dE'.
\label{Eq:app_G_from_PDOS_smeared}
\end{equation}
To summarize, \cref{Eq:multi_orb_AN_PDOS_from_G_smeared,Eq:app_G_from_PDOS_smeared} are the Lorentzian-smeared generalization of \cref{Eq:app_G_to_PDOS_no_smear,Eq:app_PDOS_to_G_no_smear}. These relationships hold exactly under Lorentzian smearing with finite width.

\subsubsection{Relationship between $\mathbf{\Delta}_\epsilon$ and $\mathbf{\Lambda}_\epsilon$}
The smeared $\mathbf{\Delta}(E)$ is defined as 
\begin{align}
\begin{split}
    \Delta_{\epsilon,\mu\nu}(E) = \pi \sum_{\sigma\in s} &( H_{\mu \sigma} - E_\sigma S_{\mu \sigma}) \\
    &(H_{\sigma\nu} - E_\sigma S_{\sigma\nu})\delta_\epsilon(E-E_\sigma).
\label{Eq:multi_orb_AN_Delta_smeared}
\end{split}
\end{align}
$\mathbf{\Delta}_\epsilon(E)$ is also equal to the convolution 
\begin{equation}
    \mathbf{\Delta}_\epsilon(E) = \mathbf{\Delta}(E) * \delta_\epsilon(E).
\label{Eq:smeared_Delta_def}
\end{equation}
We define the smeared $\mathbf{\Lambda}(E)$ as the convolution
\begin{equation}
    \mathbf{\Lambda}_\epsilon(E) = \mathbf{\Lambda}(E) * \delta_\epsilon(E),
\end{equation}
where
\begin{equation}
    \mathbf{\Lambda}(E) = \frac{1}{\pi}\mathcal{P}\int dE' \frac{\mathbf{\Delta}(E')}{E-E'}
\end{equation}
(copied from \cref{Eq:multi_orb_Lambda}).
\par
We claim that the smeared $\mathbf{\Lambda}_\epsilon(E)$ and $\mathbf{\Delta}_\epsilon(E)$ are related by the same Hilbert transform, i.e.,
\begin{equation}
    \mathbf{\Lambda}_\epsilon(E) = \frac{1}{\pi}\mathcal{P}\int dE' \frac{\mathbf{\Delta}_\epsilon(E')}{E-E'}.
\end{equation}
To prove this, we expand $\mathbf{\Delta}_\epsilon$ as a convolution, i.e.,
\begin{align}
\begin{split}
    \mathbf{\Lambda}_\epsilon(E) &= \frac{1}{\pi}\mathcal{P}\int dE'dE''\,\frac{\mathbf{\Delta}(E'-E'')\delta_\epsilon(E'')}{E-E'}.
\end{split}
\end{align}
Changing the integration order and making the change of variable $(E',E'')\rightarrow (\tilde{E},E'')$ where $\tilde{E} = E'-E''$, we have
\begin{align}
\begin{split}
    \mathbf{\Lambda}_\epsilon(E) &= \frac{1}{\pi}\mathcal{P}\int dE'' d\tilde{E} \frac{\mathbf{\Delta}(\tilde{E})}{(E-E'')-\tilde{E}} \delta_\epsilon(E'') \\
    &=\int dE''\, \mathbf{\Lambda}(E-E'')\delta_\epsilon(E'') \\
    &= \mathbf{\Lambda}(E) * \delta_\epsilon(E).
\end{split}
\end{align}

\subsubsection{Relationship between $\mathbf{Q}_\epsilon$ and ($\mathbf{\Delta}_\epsilon,\mathbf{H}_{eff},\mathbf{S}_{eff}$)}
We define the smeared quantity $\mathbf{Q}_\epsilon(E)$ as
\begin{equation}
    \mathbf{Q}_\epsilon(E) = (\mathbf{G}_\epsilon^{aa}(E))^{-1} .
\label{Eq:multi_smeared_G_inv}
\end{equation}
We will show that 
\begin{equation}
    \mathbf{Q}_\epsilon(E) = E\mathbf{S}_\text{eff} - \mathbf{H}_\text{eff} -\mathbf{\Lambda}_\epsilon(E) + i(\mathbf{\Delta}_\epsilon(E)+\epsilon \mathbf{S}_\text{eff}).
\label{Eq:app_smeared_phi}
\end{equation}
Note that, unlike other smeared quantities,
\begin{equation}
    \mathbf{Q}_\epsilon(E) \neq \mathbf{Q}(E) * \delta_\epsilon(E).
\end{equation}
\cref{Eq:app_smeared_phi} implies that \cref{Eq:phi_GS1} is almost exactly preserved under Lorentzian smearing, with the constant smearing correction of $\epsilon \mathbf{S}_\text{eff}$ on $\mathbf{\Delta}(E)$.
\par
We now prove \cref{Eq:app_smeared_phi}.
Following the steps of \crefrange{Eq:app_G_full_def}{Eq:App_modified_Dyson} using the smeared $\mathbf{G}_\epsilon(E)$, we obtain the smeared Dyson equation
\begin{equation}
    \mathbf{G}_\epsilon(E) = \mathbf{G}_{0,\epsilon}(E) + \mathbf{G}_{0,\epsilon}(E) (\Delta \mathbf{H} - (E+i\epsilon)\Delta \mathbf{S}) \mathbf{G}_\epsilon(E).
\label{Eq:multi_orbital_smeared_Dyson_equation}
\end{equation}
Following the derivation of \cref{Eq:app_phi_0} using Lorentzian-smeared quantities, we see that 
\begin{equation}
    \mathbf{Q}_\epsilon(E) = (\mathbf{G}_{\epsilon}^{aa}(E))^{-1} = (E+i\epsilon)\mathbf{S}_{aa} - \mathbf{H}_{aa} - \mathbf{L}(E),
\end{equation}
where
\begin{equation}
    L_{\mu\nu}(E) = \sum_{\sigma \in s} \frac{(\mathbf{H}-(E+i\epsilon)\mathbf{S})_{\mu \sigma}(\mathbf{H}-(E+i\epsilon)\mathbf{S})_{\sigma \nu}}{E-E_\sigma+i\epsilon}.
\end{equation}
Following the decomposition method in \cref{Eq:app_multi_orb_K_1} to simplify $\mathbf{Q}_\epsilon(E)$, we obtain \cref{Eq:app_smeared_phi}.

\section{implicit orthogonalization schemes}
\label{App:implicit_orthogonalization_schemes}
\begin{figure*}
    \centering
    \includegraphics[scale=0.35]{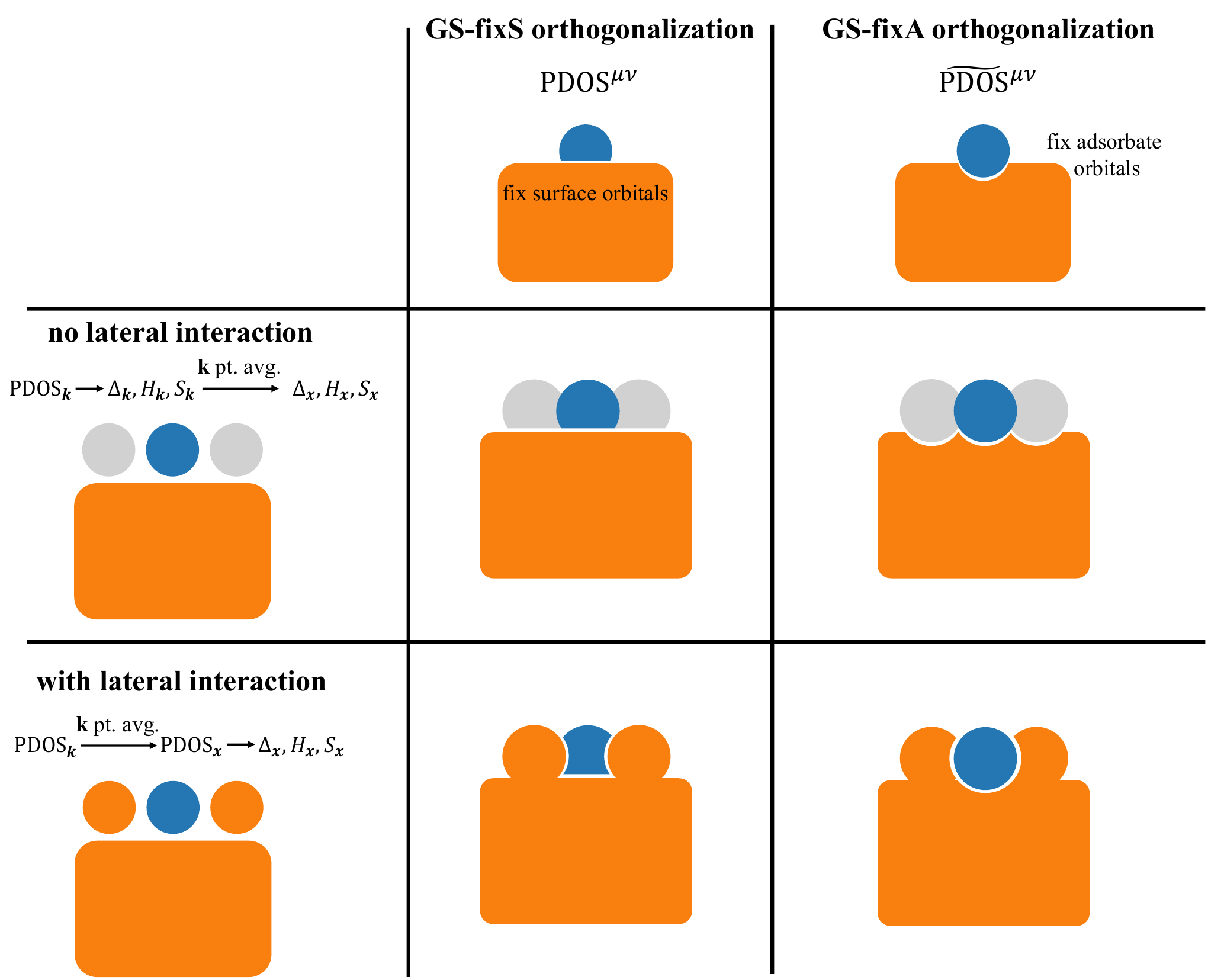}
    \caption{Columns: The $\hat{H}_{el}$ constructed from two different PDOS matrices ($\text{PDOS}^{\mu\nu}$ and $\widetilde{\text{PDOS}}^{\mu\nu}$) in non-orthogonal basis are equal to the $\hat{H}_{el}$ obtained from two different GS orthogonalization schemes. Rows: The order of $\mathbf{k}$-point averaging determines whether lateral interaction between adsorbates is included in $\hat{H}_{el}$. The combination of the choice of PDOS matrices and the order of $\mathbf{k}$-point averaging results in four different variations of $\hat{H}_{el}$.}
    \label{fig:PDOS_orthogonalization}
\end{figure*}

We first show that PDOS and $\widetilde{\text{PDOS}}$ correspond to the GS-fixS and GS-fixA orthogonalization schemes. Then we show that different $\mathbf{k}$-point averaging procedures correspond to whether lateral interactions are included. \cref{fig:PDOS_orthogonalization} summarizes the different implicit orthogonalization schemes.

\par
Two choices exist for the contravariant basis of the PDOS.
First, taking the contravariant adsorbate orbitals derived from the full adsorbate+surface basis, we have
\begin{equation}
    \text{PDOS}^{\mu\nu}(E) = \langle \phi^\mu|\delta(E-\hat{H})|\phi^\nu\rangle,
\end{equation}
where
\begin{equation}
    |\phi^\mu\rangle = \sum_{\tau\in \text{all AO}} |\phi_\tau\rangle \big(\mathbf{S}^{-1} \big)^{\tau\mu}.
\label{Eq:contravariant_basis}
\end{equation}
$\mu$ and $\nu$ are adsorbate indices. This corresponds to applying the theory of \cref{Sec:adsorbate_projected_G} to the basis set AO, without modification.
Second, we can also use a different PDOS matrix by choosing the basis to be the contravariant adsorbate orbitals derived from only the adsorbate AO, i.e.,
\begin{equation}
    \widetilde{\text{PDOS}}^{\mu\nu}(E) = \langle \widetilde{\phi}^\mu|\delta(E-\hat{H})|\widetilde{\phi}^\nu\rangle,
\end{equation}
where
\begin{equation}
    |\widetilde{\phi}^\mu\rangle = \sum_{\tau\in a} |\phi_\tau\rangle \big((\mathbf{S}_{aa})^{-1}\big)^{\tau\mu} .
\label{Eq:adsorbated_restricted_contravariant_basis}
\end{equation}
As we show below, this corresponds to applying the theory of \cref{Sec:adsorbate_projected_G} to a modified AO basis where the surface orbitals are projected to the orthogonal subspace of adsorbate AO.

\subsection{PDOS-to-GS-orthogonalization correspondences}
\label{App:PDOS_orthogonalization_correspondence}

We consider two types of GS orthogonalization such that the adsorbate orbitals and the surface orbitals become orthogonal (i.e., $\mathbf{S}_{as}= 0$).
The first type of GS orthogonalization, denoted as GS-fixS (or GSS for short), fixes the surface orbitals, while the adsorbate orbitals are projected to the orthogonal subspace of the surface orbitals. Mathematically, an adsorbate orbital $|\phi_\mu\rangle$ transforms into 
\begin{equation}
    |\phi_\mu\rangle_{GSS} = (\hat{I}-\hat{\Pi}_{surf})|\phi_\mu\rangle,
\end{equation}
where
\begin{equation}
    \hat{\Pi}_{surf} = \sum_{\sigma_1, \sigma_2\in s} |\phi_{\sigma_1}\rangle \big((\mathbf{S}_{ss})^{-1}\big)^{\sigma_1 \sigma_2} \langle \phi_{\sigma_2}|
\label{Eq:surface_projector}
\end{equation}
is the projector onto the subspace spanned by the surface orbitals. The GS-fixS orthogonalized adsorbate orbitals remain non-orthogonal in general, i.e., $\langle \phi_\mu|\phi_\nu\rangle_{GSS} \neq \delta_{\mu\nu}$.
The second type of GS orthogonalization, denoted as GS-fixA (or GSA for short), fixes the adsorbate orbitals, while the surface orbitals are projected to the orthogonal subspace of the adsorbate orbitals. Mathematically, a surface orbital $|\phi_\sigma\rangle$ transforms into
\begin{equation}
    |\phi_\sigma\rangle_{GSA} = (\hat{I}-\hat{\Pi}_{ad})|\phi_\sigma\rangle,
\end{equation}
where
\begin{equation}
    \hat{\Pi}_{ad} = \sum_{\mu,\nu \in \text{ad.}} |\phi_\mu\rangle \big( (\mathbf{S}_{aa})^{-1}\big)^{\mu\nu}\langle \phi_\nu|
\label{Eq:adsorbate_projector}
\end{equation}
is the projector onto the subspace spanned by the adsorbate orbitals.
The top row of \cref{fig:PDOS_orthogonalization} illustrates the two orthogonalization schemes.
\par
To show why $\text{PDOS}^{\mu\nu}$ and $\widetilde{\text{PDOS}}^{\mu\nu}$ correspond to GS-fixS and GS-fixA, we will make use of different bases. To keep track of the different bases, we list the notations for the bases here:
\begin{itemize}
    \item Basis 0: The original AO basis from electronic structure calculations. The overlap matrix
    \begin{equation}
        \raisebox{0\height}{\includegraphics[scale=0.5]{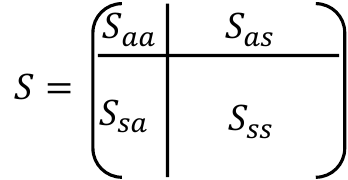}}
    \end{equation}
    has no special structures.

    \item Basis GS-fixS: The basis is obtained by fixing the surface orbitals and projecting the adsorbate orbitals according to $|\phi_\mu\rangle \rightarrow (\hat{I}-\hat{\Pi}_{surf})|\phi_\mu\rangle$. The overlap matrix in this basis is
    \begin{equation}
        \mathbf{S}^{(GSS)} = \mathbf{V}_{GSS}^\dagger \mathbf{S} \mathbf{V}_{GSS},
    \label{Eq:app_GS1_S_0}
    \end{equation}
    where the change of basis matrix $\mathbf{V}_{GSS}$ is
    \begin{equation}
        \raisebox{0\height}{\includegraphics[scale=0.5]{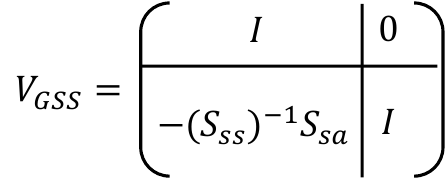}}.
    \label{Eq:app_V_GS1}
    \end{equation}
    Substituting into \cref{Eq:app_GS1_S_0}, we obtain the explicit form of $\mathbf{S}^{(GSS)}$ as
    \begin{equation}
        \raisebox{0\height}{\includegraphics[scale=0.5]{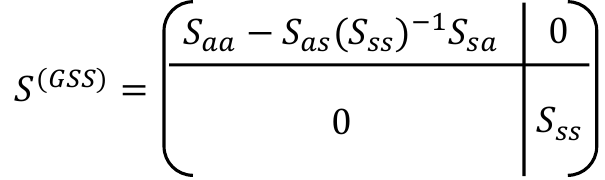}}.
    \end{equation}

    \item Basis GS-fixA: The basis is obtained by fixing the adsorbate orbitals and projecting the surface orbitals according to $|\phi_\sigma\rangle \rightarrow (\hat{I}-\hat{\Pi}_{ad})|\phi_\sigma\rangle$. The change of basis matrix is
    \begin{equation}
        \raisebox{0\height}{\includegraphics[scale=0.5]{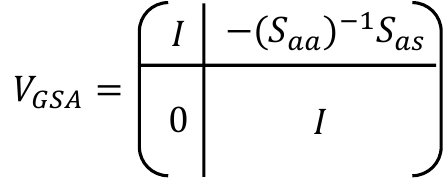}}.
    \end{equation}
    The overlap matrix in this basis is
    \begin{equation}
        \raisebox{0\height}{\includegraphics[scale=0.5]{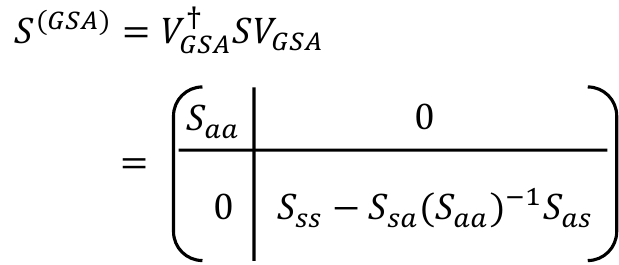}}.
    \label{Eq:app_S_GS2}
    \end{equation}
\end{itemize}

\par
We summarize the correspondence between the two PDOS types and the two GS orthogonalization schemes in two theorems. 
\par
\textit{Theorem 1:} The ($\mathbf{\Delta}(E)$, $\mathbf{H}_\text{eff}$, $\mathbf{S}_\text{eff}$) derived from $\text{PDOS}^{\mu\nu}$ in Basis 0 is equal to the ($\mathbf{\Delta}(E)$, $\mathbf{H}_{aa}$, $\mathbf{S}_{aa}$) in Basis GS-fixS.
\par
\textit{Theorem 2:} The ($\mathbf{\Delta}(E)$, $\mathbf{H}_\text{eff}$, $\mathbf{S}_\text{eff}$) derived from $\widetilde{\text{PDOS}}^{\mu\nu}$ in Basis 0 is equal to the ($\mathbf{\Delta}(E)$, $\mathbf{H}_{aa}$, $\mathbf{S}_{aa}$) in Basis GS-fixA.
\par
Note that the three quantities $\mathbf{\Delta}(E)$, $\mathbf{H}_\text{eff}$, and $\mathbf{S}_\text{eff}$ completely characterize $\hat{H}_{el}$, the electronic part of the AN Hamiltonian. If the adsorbate-surface overlap $\mathbf{S}_{as}=0$, then $\mathbf{H}_\text{eff}=\mathbf{H}_{aa}$ and $\mathbf{S}_\text{eff}=\mathbf{S}_{aa}$.
\par
\textit{Proof of Theorem 1:}
It suffices to show that the adsorbate contravariant orbitals remain unchanged from Basis 0 to Basis GS-fixS, since this implies that $\text{PDOS}^{\mu\nu}$ in Basis 0 is equal to the $\text{PDOS}^{\mu\nu}$ in Basis GS-fixS. This means that the ($\mathbf{\Delta}(E)$, $\mathbf{H}_\text{eff}$, $\mathbf{S}_\text{eff}$) derived from $\text{PDOS}^{\mu\nu}$ in Basis 0 is equal to the ($\mathbf{\Delta}(E)$, $\mathbf{H}_\text{eff}$, $\mathbf{S}_\text{eff}$) derived from $\text{PDOS}^{\mu\nu}$ in Basis GS-fixS, which is equal to the ($\mathbf{\Delta}(E)$, $\mathbf{H}_{aa}$, $\mathbf{S}_{aa}$) in Basis GS-fixS, since $\mathbf{S}_{as}=0$ in Basis GS-fixS.
\par
The contravariant orbitals transform as
\begin{equation}
    | \phi^\lambda\rangle_{GSS} = \sum_{\tau}|\phi^\tau\rangle\big((\mathbf{V}_{GSS})^{-1\dagger}\big)_{\tau}^{\cdot\lambda}
\end{equation}
(see \cref{App:tensor_notation}).
From \cref{Eq:app_V_GS1}, we see that 
\begin{equation}
        \raisebox{0\height}{\includegraphics[scale=0.5]{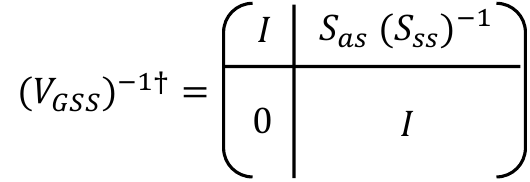}}.
\end{equation}
Therefore, the contravariant orbitals transform in an opposite way from the covariant orbitals. In GS-fixS orthogonalization, the contravariant adsorbate orbitals are fixed, while the contravariant surface orbitals are projected to an orthogonal subspace.
Since the contravariant adsorbate orbitals in Basis 0 are the same as those in Basis GS-fixS, Theorem 1 is proved. $\blacksquare$

\par
\textit{Proof of Theorem 2:}
Similar to the proof of Theorem 1, to prove Theorem 2, it suffices to show that the contravariant adsorbate orbitals in GS-fixA are equal to the adsorbate-restricted contravariant adsorbate orbitals in Basis 0
(see \cref{Eq:adsorbated_restricted_contravariant_basis}). Since $\mathbf{S}^{(GSA)}$ is block-diagonal (see \cref{Eq:app_S_GS2}), with the same $\mathbf{S}_{aa}$ in the adsorbate block and the same covariant adsorbate orbitals, the contravariant adsorbate basis in Basis GS-fixA is equal to \cref{Eq:adsorbated_restricted_contravariant_basis}. Hence, Theorem 2 is proved. $\blacksquare$ 

\subsection{Lateral interaction}
\label{App:lateral_interaction}
In periodic DFT, the adsorbate-surface unit cell is repeated $N$ times to form a supercell. $N$ is equal to the number of $\mathbf{k}$ point samples. This means an adsorbate in a unit cell is also repeated in the supercell, and adsorbates interact laterally across unit cells. The $\hat{H}_{el}$ for a local adsorbate is obtained by averaging over $\mathbf{k}$-points.
We now show that, by performing the $\mathbf{k}$-point averaging at different steps, one obtains the $\hat{H}_{el}$ with or without lateral interactions. Combining this with the two orthogonalization schemes using two different PDOS matrices, we obtain four variations of $\hat{H}_{el}$ (see \cref{fig:PDOS_orthogonalization}). The different $\hat{H}_{el}$ can be useful in studying different physical properties and understanding the effect of lateral interactions.

\subsubsection{Average over $\hat{H}_{el,\mathbf{k}}$: no lateral interaction}
First, we consider the case where the PDOS at different $\mathbf{k}$-point (denoted $\text{PDOS}_\mathbf{k}$) are used to construct $\hat{H}_{el}$ at $\mathbf{k}$ (denoted $\hat{H}_{el,\mathbf{k}}$). Then we take the $\mathbf{k}$-point average of the de-localized $\hat{H}_{el,\mathbf{k}}$ to obtain the local Hamiltonian, denoted as $\hat{H}_{el,\mathbf{x}}$.
Since the surface part of $\hat{H}_{el,\mathbf{k}}$ at each $\mathbf{k}$ includes only surface AO, the surface part of the local AN Hamiltonian $\hat{H}_{el,\mathbf{x}} = \frac{1}{N}\sum_\mathbf{k}H_{el,\mathbf{k}}$ will also include only the surface AO. Therefore, the final AN Hamiltonian contains no lateral interaction with other adsorbates. Since $\hat{H}_{el}$ is fully characterized by $(\mathbf{\Delta}(E),\mathbf{H}_{aa},\mathbf{S}_{aa})$ in an orthogonalized basis (i.e., $\mathbf{S}_{as}=0$), we will show explicitly that the local $(\mathbf{\Delta}(E),\mathbf{H}_{aa},\mathbf{S}_{aa})$ without lateral interaction is equal to the average of $(\mathbf{\Delta}_\mathbf{k}(E),\mathbf{H}_{aa,\mathbf{k}},\mathbf{S}_{aa,\mathbf{k}})$ over $\mathbf{k}$-points, both in GS-fixS and GS-fixA orthogonalizations.
\par
We will denote an orbital $\lambda$ in unit cell $\mathbf{x}$ as $|\phi_{\lambda,\mathbf{x}}\rangle$. It is related to $|\phi_{\lambda,\mathbf{k}}\rangle$, the corresponding orbital at wavevector $\mathbf{k}$, by
\begin{equation}
    |\phi_{\lambda,\mathbf{k}}\rangle = \frac{1}{\sqrt{N}}\sum_{\mathbf{x}} e^{i\mathbf{k}\cdot\mathbf{x}}|\phi_{\lambda,\mathbf{x}}\rangle
\label{Eq:phi_k_as_sum_phi_R}
\end{equation}
and
\begin{equation}
    |\phi_{\lambda,\mathbf{x}}\rangle = \frac{1}{\sqrt{N}}\sum_{\mathbf{x}} e^{-i\mathbf{k}\cdot\mathbf{x}}|\phi_{\lambda,\mathbf{k}}\rangle.
\label{Eq:phi_R_as_sum_phi_k}
\end{equation}
Let us define the projection operators $\hat{\Pi}_{surf}$ and $\hat{\Pi}_{ad}$, which act on the orbital space of the supercell. $\hat{\Pi}_{surf}$ projects onto the subspace spanned by all surface orbitals in the supercell (see \cref{Eq:surface_projector}). Similarly, $\hat{\Pi}_{ad}$ projects onto the subspace spanned by all adsorbate orbitals in the supercell (see \cref{Eq:adsorbate_projector}).
By lattice translational symmetry, $\hat{\Pi}_{surf}$, $\hat{\Pi}_{ad}$, and $\hat{H}$ are block-diagonal in $\mathbf{k}$.
In the GS-fixS orthogonalized basis, we define the surface density of states operator as
\begin{equation}
    \hat{\rho}_{s1}(E) = \delta(E-\hat{\Pi}_{surf}\hat{H}\hat{\Pi}_{surf}).
\end{equation}
In the GS-fixA orthogonalized basis, we define the surface density of states operator as
\begin{equation}
    \hat{\rho}_{s2}(E) = \delta(E-(1-\hat{\Pi}_{ad})\hat{H}(1-\hat{\Pi}_{ad})).
\end{equation}
Since $\hat{\Pi}_{surf}$, $\hat{\Pi}_{ad}$, and $\hat{H}$ are all block-diagonal in $\mathbf{k}$, $\hat{\rho}_{s1}$ and $\hat{\rho}_{s2}$ are also block-diagonal in $\mathbf{k}$.

\par
Having defined the notations, we now show that the $\mathbf{k}$-point average of the GS-fixS $(\mathbf{\Delta}_\mathbf{k}(E),\mathbf{H}_{aa,\mathbf{k}},\mathbf{S}_{aa,\mathbf{k}})$ is equal to $(\mathbf{\Delta}(E),\mathbf{H}_{aa},\mathbf{S}_{aa})$ for a localized adsorbate in the supercell, without lateral interaction and under the GS-fixS orthogonalization.
$(\mathbf{\Delta}_\mathbf{k}(E),\mathbf{H}_{aa,\mathbf{k}},\mathbf{S}_{aa,\mathbf{k}})$ can be expressed as
\begin{align}
\begin{split}
        \mathbf{\Delta}_{\mathbf{k},\mu\nu}(E) = \pi \langle \phi_{\mu,\mathbf{k}}|&(1-\hat{\Pi}_{surf})\hat{H}\hat{\rho}_{s1}(E) \\
    &\qquad\qquad\hat{H} (1-\hat{\Pi}_{surf})|\phi_{\nu,\mathbf{k}}\rangle,
\label{Eq:app_Delta_GS1_k}
\end{split}
\end{align}
\begin{equation}
    \big(\mathbf{H}_{aa,\mathbf{k}}\big)_{\mu\nu} = \langle \phi_{\mu,\mathbf{k}}|(1-\hat{\Pi}_{surf}) \hat{H} (1-\hat{\Pi}_{surf})|\phi_{\nu,\mathbf{k}}\rangle,
\label{Eq:app_Haa_GS1_k}
\end{equation}
and
\begin{equation}
    \big(\mathbf{S}_{aa,\mathbf{k}}\big)_{\mu\nu} = \langle \phi_{\mu,\mathbf{k}}|(1-\hat{\Pi}_{surf}) (1-\hat{\Pi}_{surf})|\phi_{\nu,\mathbf{k}}\rangle.
\label{Eq:app_Saa_GS1_k}
\end{equation}
Notice that the operators in \crefrange{Eq:app_Delta_GS1_k}{Eq:app_Saa_GS1_k} are all block-diagonal in $\mathbf{k}$.
Using \cref{Eq:phi_R_as_sum_phi_k}, we see that the local $(\mathbf{\Delta}(E),\mathbf{H}_{aa}, \mathbf{S}_{aa})$ in the unit cell $\mathbf{x}$ are given by
\begin{align}
\begin{split}
    \Delta_{\mu\nu}(E) &= \pi \langle \phi_{\mu,\mathbf{x}}|(1-\hat{\Pi}_{surf})\hat{H}\hat{\rho}_{s1}(E) \\
    &\qquad\qquad\qquad\qquad\hat{H} (1-\hat{\Pi}_{surf})|\phi_{\nu,\mathbf{x}}\rangle \\
    &=\frac{1}{N}\sum_\mathbf{k} \Delta_{\mu\nu}^{(\mathbf{k})}(E),
\end{split}
\end{align}
\begin{align}
\begin{split}
    H_{\mu\nu} &= \langle \phi_{\mu,\mathbf{x}}|(1-\hat{\Pi}_{surf}) \hat{H} (1-\hat{\Pi}_{surf})|\phi_{\nu,\mathbf{x}}\rangle \\
    &=\frac{1}{N}\sum_\mathbf{k} \big(\mathbf{H}_{aa}^{(\mathbf{k})}\big)_{\mu\nu},
\end{split}
\end{align}
and
\begin{align}
\begin{split}
    S_{\mu\nu} &= \langle \phi_{\mu,\mathbf{x}}|(1-\hat{\Pi}_{surf}) (1-\hat{\Pi}_{surf})|\phi_{\nu,\mathbf{x}}\rangle \\
    &=\frac{1}{N}\sum_\mathbf{k} \big(\mathbf{S}_{aa}^{(\mathbf{k})}\big)_{\mu\nu}.
\end{split}
\end{align}
Therefore, the local $\hat{H}_{el,\mathbf{x}}$ is equal to the $\mathbf{k}$-point average of $\hat{H}_{el,\mathbf{k}}$. The adsorbate basis in the local $\hat{H}_{el,\mathbf{x}}$ is
\begin{equation}
    (1-\hat{\Pi}_{surf})|\phi_{\mu,\mathbf{x}}\rangle,
\end{equation}
where the surface orbitals in the supercell are projected onto the orthogonal subspace of surface orbitals. $\mathbf{\Delta}(E)$ captures the coupling between the GS-fixS orthogonalized adsorbate orbitals and the surface orbitals, not the coupling between lateral adsorbate orbitals (see \cref{fig:PDOS_orthogonalization}).
\par
Next, we consider GS-fixA orthogonalization. Under GS-fixA orthogonalization, $(\mathbf{\Delta}_\mathbf{k}(E),\mathbf{H}_{aa,\mathbf{k}}, \mathbf{S}_{aa,\mathbf{k}})$ are expressed as
\begin{align}
\begin{split}
    \mathbf{\Delta}_{\mathbf{k},\mu\nu}(E) = \pi \langle \phi_{\mu,\mathbf{k}}|&\hat{H}\hat{\rho}_{s2}(E) \hat{H} |\phi_{\nu,\mathbf{k}}\rangle,
\label{Eq:app_Delta_GS2_k}
\end{split}
\end{align}
\begin{equation}
    \big(\mathbf{H}_{aa,\mathbf{k}}\big)_{\mu\nu} = \langle \phi_{\mu,\mathbf{k}}| \hat{H} |\phi_{\nu,\mathbf{k}}\rangle,
\label{Eq:app_Haa_GS2_k}
\end{equation}
and
\begin{equation}
    \big(\mathbf{S}_{aa,\mathbf{k}}\big)_{\mu\nu} = \langle \phi_{\mu,\mathbf{k}}| \phi_{\nu,\mathbf{k}}\rangle.
\label{Eq:app_Saa_GS2_k}
\end{equation}
Similarly, since the operators in \crefrange{Eq:app_Delta_GS2_k}{Eq:app_Saa_GS2_k} are all block-diagonal in $\mathbf{k}$, the local $(\mathbf{\Delta}(E),\mathbf{H}_{aa}, \mathbf{S}_{aa})$ are the $\mathbf{k}$-point averages of $( \mathbf{\Delta}_\mathbf{k}(E),\mathbf{H}_{aa,\mathbf{k}}, \mathbf{S}_{aa,\mathbf{k}})$. The adsorbate orbital basis in the local $\hat{H}_{el}$ is unchanged, while the surface orbitals are projected to the orthogonal subspace of all adsorbate orbitals, corresponding to the GS-fixA orthogonalization (see \cref{fig:PDOS_orthogonalization}). $\mathbf{\Delta}(E)$ only captures adsorbate-surface interactions, not lateral adsorbate-adsorbate interactions.

\subsubsection{Average over $\text{PDOS}_\mathbf{k}$: with lateral interaction}
If $\text{PDOS}_\mathbf{k}$ is first averaged to obtain the local $\text{PDOS}_\mathbf{x}$, then the $\hat{H}_{el}$ constructed from the local $\text{PDOS}_\mathbf{x}$ will contain lateral adsorbate interactions. 
To show this fact rigorously, we need to show that $\text{PDOS}_\mathbf{x}$ is the local adsorbate PDOS of the supercell, including the surface and all lateral adsorbates.
Therefore, the local adsorbate in a unit cell is treated as the ``adsorbate". All remaining parts of the supercell, including the surface and all other adsorbates, are treated as the ``surface" in the AN model. 
This means that the $\hat{H}_{el}$ constructed from the local $\text{PDOS}_\mathbf{x}$ includes both the interaction with the surface and the lateral interaction with other adsorbates. 
\par
First, we consider the case of GS-fixS orthogonalization, which corresponds to $\text{PDOS}^{\mu\nu}$ (i.e., in the contravariant adsorbate basis with respect to the entire AO basis).
Now, we show that $\text{PDOS}^{\mu\nu}_\mathbf{x}$, the PDOS of a localized adsorbate in the supercell, is equal to
\begin{equation}
    \text{PDOS}^{\mu\nu}_\mathbf{x} = \frac{1}{N}\sum_\mathbf{k}\text{PDOS}^{\mu\nu}_\mathbf{k}.
\label{Eq:app_PDOS_avg}
\end{equation}
We can express $\text{PDOS}^{\mu\nu}_\mathbf{x}$ and $\text{PDOS}^{\mu\nu}_\mathbf{k}$ as
\begin{equation}
    \text{PDOS}^{\mu\nu}_{\mathbf{x}}(E) = \langle \phi^{\mu\mathbf{x}}|\delta(E-\hat{H})|\phi^{\nu\mathbf{x}}\rangle
\end{equation}
and 
\begin{equation}
    \text{PDOS}^{\mu\nu}_{\mathbf{k}}(E) = \langle\phi^{\mu\mathbf{k}}|\delta(E-\hat{H})|\phi^{\nu\mathbf{k}}\rangle,
\end{equation}
where $\hat{H}$ is the supercell Hamiltonian.
Since $\delta(E-\hat{H})$ is block-diagonal in $\mathbf{k}$, it suffices to show that 
\begin{equation}
    |\phi^{\lambda\mathbf{x}}\rangle = \frac{1}{\sqrt{N}}\sum_\mathbf{k} e^{-i\mathbf{k}\cdot\mathbf{x}}|\phi^{\lambda\mathbf{k}}\rangle,
\label{Eq:app_contra_Fourier}
\end{equation}
i.e., that the contravariant orbitals follow the Fourier relation as the covariant orbitals do.
To prove this, we write the Fourier relation between covariant orbitals (\cref{Eq:phi_R_as_sum_phi_k}) as 
\begin{equation}
    |\phi_{\lambda,\mathbf{x}}\rangle = \sum_{\tau,\mathbf{k}} |\phi_{\tau,\mathbf{k}}\rangle T^{\tau,\mathbf{k}}_{\,\,\,\cdot \,\, \lambda,\mathbf{x} },
\label{Eq:app_cov_Fourier_tensor_notation}
\end{equation}
where the change of basis matrix $T$ is given by 
\begin{equation}
    T^{\tau,\mathbf{k}}_{\,\,\,\cdot \,\, \lambda,\mathbf{x} } = \frac{1}{\sqrt{N}}\delta_{\tau,\lambda}e^{-i\mathbf{k}\cdot\mathbf{x}}.
\label{Eq:app_Fouier_COB}
\end{equation}
In \cref{Eq:app_cov_Fourier_tensor_notation}, $\tau$ indexes all AO in a $\mathbf{k}$ point.
From \cref{App:tensor_notation}, the change of basis rule for contravariant orbitals is
\begin{equation}
    |\phi^{\lambda\mathbf{x}}\rangle = \sum_{\tau,\mathbf{k}} |\phi^{\tau\mathbf{k}}\rangle (T^{-1\dagger})_{\tau,\mathbf{k}}^{\,\,\,\cdot \,\, \lambda,\mathbf{x} }.
\label{Eq:app_contravariant_COB_orbital}
\end{equation}
From \cref{Eq:app_Fouier_COB}, we can directly compute $T^{-1\dagger}$ as
\begin{equation}
    (T^{-1\dagger})_{\tau,\mathbf{k}}^{\,\,\,\cdot \,\, \lambda,\mathbf{x} } = \frac{1}{\sqrt{N}}\delta_{\tau,\lambda}e^{-i\mathbf{k}\cdot\mathbf{x}}.
\end{equation}
Substituting this into \cref{Eq:app_contravariant_COB_orbital}, we obtain \cref{Eq:app_contra_Fourier}.
\par
For the case of GS-fixA orthogonalization, we need to show that
\begin{equation}
    \widetilde{\text{PDOS}}^{\mu\nu}_\mathbf{x} = \frac{1}{N}\sum_\mathbf{k}\widetilde{\text{PDOS}}^{\mu\nu}_\mathbf{k},
\end{equation}
which is true if
\begin{equation}
    |\widetilde{\phi}^{\lambda\mathbf{x}}\rangle = \frac{1}{\sqrt{N}}\sum_\mathbf{k} e^{-i\mathbf{k}\cdot\mathbf{x}}|\widetilde{\phi}^{\lambda\mathbf{k}}\rangle.
\label{Eq:app_restricted_contra_Fourier}
\end{equation}
The proof of \cref{Eq:app_restricted_contra_Fourier} follows in a similar manner to the proof of \cref{Eq:app_contra_Fourier}, but restricted to only adsorbate orbitals.

\section{Verifying the PDOS-to-$\hat{H}_{AN}$ procedure using model single-orbital AN Hamiltonians}
\label{App:model_Hamiltonians}

In this section, we verify the procedure of constructing $\hat{H}_{el}$ in two model single-orbital AN Hamiltonians. In the first example, the surface consists of two discrete states. We verify our method of constructing $\Delta(E)$ from PDOS using Lorentzian smearing, and we compare Lorentzian smearing with Gaussian smearing. In the second example, the surface states form a continuum, and $\Delta(E)$ takes a semi-elliptical form. Analytical forms of $\Delta(E)$ and $\text{PDOS}(E)$ in this example can be derived.

\subsection{Three-level Hamiltonian}
We consider an adsorbate coupled to a two-level system (the ``surface"). The total Hamiltonian can be expressed as a $3\times 3$ matrix, i.e., 
\begin{equation}
    \raisebox{0\height}{\includegraphics[width=3.5cm]{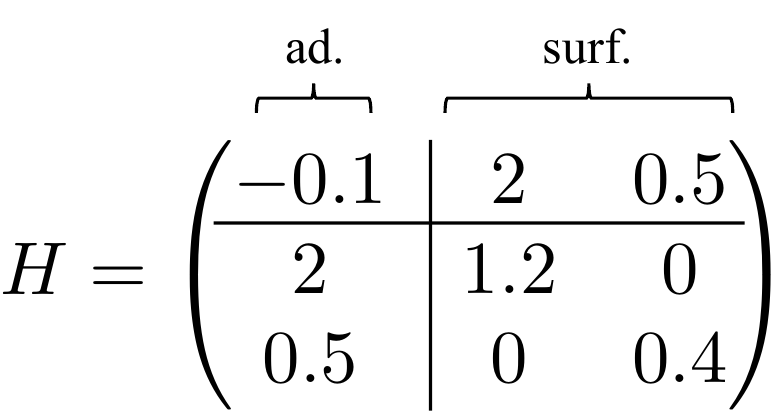}}.
\end{equation}
The $2\times 2$ surface block $\mathbf{H}_{ss}$ is chosen to be diagonal. The independent matrix elements are selected randomly from the interval $(-2, 2)$. The scalar $\text{PDOS}(E)$ and $\Delta(E)$ can be computed directly by diagonalizing $\mathbf{H}$. 
The $\delta$-functions in $\text{PDOS}(E)$ and $\Delta(E)$ are smeared by a Gaussian or a Lorentzian. The Gaussian smearing function is defined as
\begin{equation}
    \delta^{G}_\epsilon(E) = \frac{1}{\sqrt{2\pi\epsilon^2}} e^{-\frac{E^2}{2\epsilon^2}},
\end{equation}
and the Lorentzian smearing function is defined in \cref{Eq:Lorentzian}.
The PDOS contains three peaks at the eigenvalues of $\mathbf{H}$: $-1.63$, $0.44$, and $2.69$. $\Delta(E)$ contains two peaks at the eigenvalues of $\mathbf{H}_{ss}$: 1.2 and 1.4. The smearing width $\epsilon$ for both Gaussian and Lorentzian smearing is taken to be $0.01$. The numerical grid for $E$ ranges from $-200$ to $200$, with a spacing of $10^{-3}$.

\begin{figure}
    \centering
    \includegraphics[width=1\linewidth]{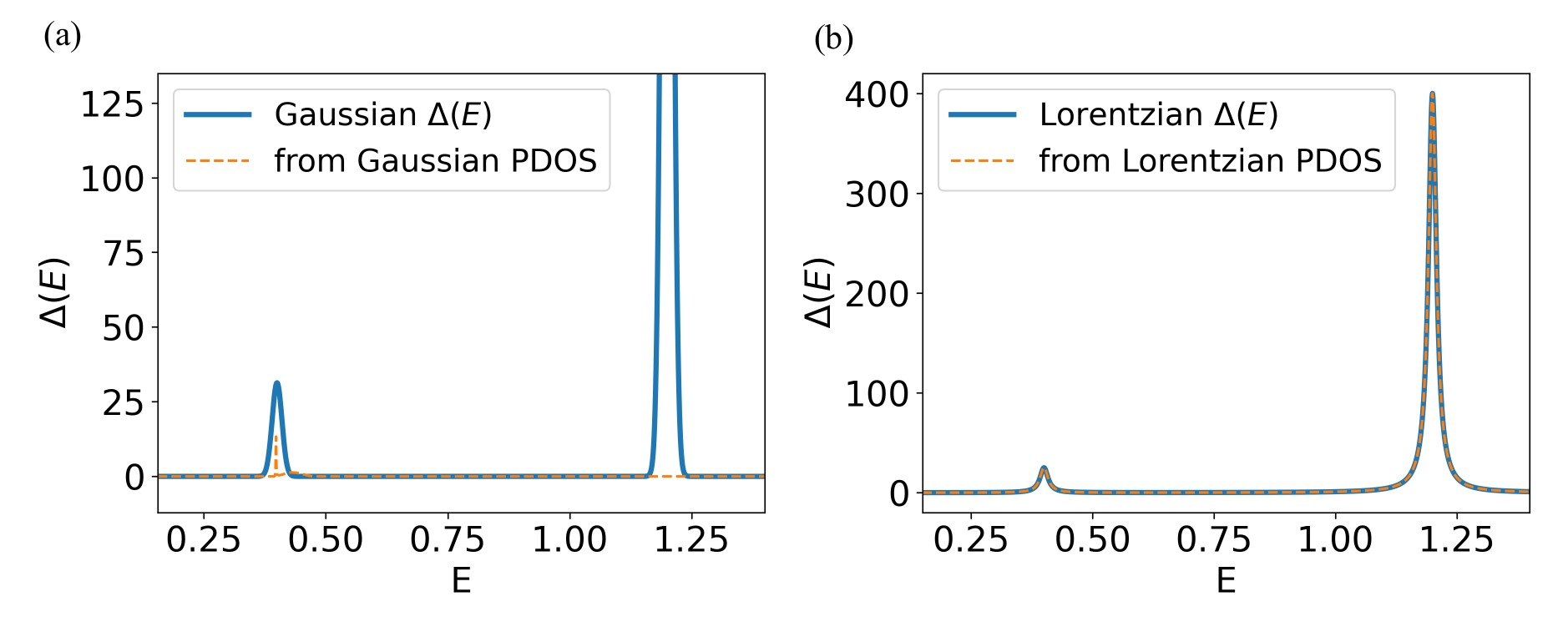}
    \caption{Gaussian vs. Lorentzian smearing in computing $\Delta(E)$ from PDOS. (a) $\Delta(E)$ computed from Gaussian-smeared PDOS fails to match the exact Gaussian-smeared $\Delta(E)$. (b) $\Delta(E)$ computed from Lorentzian-smeared PDOS agrees with the exact Lorentzian-smeared $\Delta(E)$.}
    \label{fig:PDOS_to_Delta_3_level}
\end{figure}

\par
\cref{fig:PDOS_to_Delta_3_level} compares the use of Gaussian vs. Lorentzian smearing in computing the smeared $\Delta(E)$ from the smeared PDOS. We see that the $\Delta(E)$ computed from the Gaussian-smeared PDOS fails to reproduce the Gaussian-smeared $\Delta(E)$, even qualitatively. In contrast, Lorentzian smearing yields quantitative agreement.

\subsection{Single-orbital AN Hamiltonian with semi-elliptical $\Delta(E)$}
\label{App:semi_elliptical_Delta_example}
\begin{figure}
    \centering
    \includegraphics[width=\linewidth]{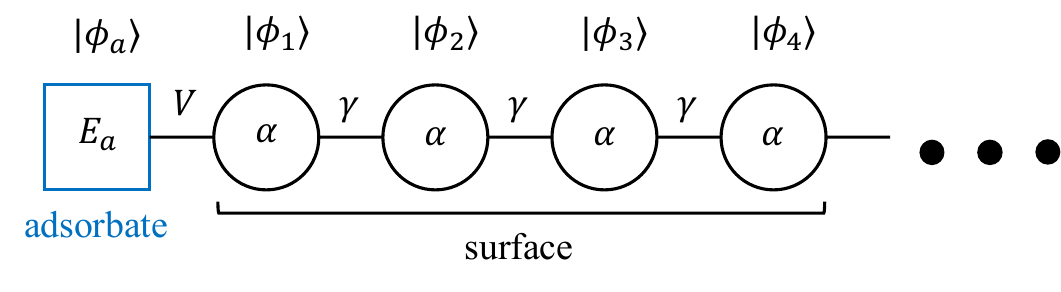}
    \caption{A one-dimensional tight-binding chain used for verifying binding energy expressions. In the infinite chain limit, $\Delta(E)$ takes the semi-elliptical form.\cite{norskov2014fundamental, newns1969}}
    \label{fig:one_plus_zero_d_TB}
\end{figure}

A single-orbital AN Hamiltonian with a semi-elliptical $\Delta(E)$ can be represented by an adsorbate coupled to a one-dimensional tight-binding chain\cite{newns1969} (see \cref{fig:one_plus_zero_d_TB}). We will solve this model analytically and show that $\Delta(E)$ takes the semi-elliptical form.

The full Hamiltonian is 
\begin{equation}
    H = E_a|\phi_a\rangle\langle\phi_a| + V(|\phi_a\rangle\langle \phi_1| + |\phi_1\rangle\langle \phi_a|) + H_s,
\end{equation}
where $|\phi_a\rangle$ is the adsorbate orbital, and $|\phi_i\rangle\,, i=1,2,\cdots$ denotes the orbital of the $i$-th site in the tight-binding chain.
\begin{equation}
    H_s = \sum_{i=1}^N \alpha |\phi_i\rangle\langle \phi_i| + \sum_{i=1}^{N-1} \gamma (|\phi_i\rangle\langle \phi_{i+1}|+|\phi_{i+1}\rangle\langle \phi_i|)
\end{equation}
is the tight-binding Hamiltonian. The orbitals are orthonormal.

The eigenstates of $H_s$ are
\begin{equation}
    |k\rangle = \sqrt{\frac{2}{N+1}}\sum_{m=1}^{N} \sin\big(mk\big) |m\rangle,
\label{Eq:0+1_eigenstate}
\end{equation}
where the wavevector $k$ take the values
\begin{equation}
    k = \frac{n\pi}{N+1}\quad , n=1, 2, \cdots,N.
\end{equation}
In the continuum limit (i.e., $N\rightarrow \infty$), $k$ takes on any value between $0$ and $\pi$.
The eigenenergy of $|k\rangle$ is
\begin{equation}
    E(k) = \alpha + 2\gamma\cos(k).
\label{Eq:0+1_dispersion_relation}
\end{equation}
Inverting the dispersion relation, an energy level $E$ between $-2$ and $2$ corresponds to the wavevectors
\begin{equation}
    k(E) = \pm \cos^{-1}(a),
\end{equation}
where 
\begin{equation}
    a = \frac{E-\alpha}{2\gamma}.
\end{equation}
The hybridization function $\Delta(E)$ is
\begin{align}
\begin{split}
    \Delta(E) &= \pi \sum_k |\langle \phi_a|H|\phi_k\rangle|^2 \, \delta(E-E_k) \\
    &=\pi \sum_k |\langle \phi_a|H|\phi_1\rangle\langle\phi_1|\phi_k\rangle|^2 \, \delta(E-E_k) \\
    &= \pi V^2 \frac{N+1}{\pi} \\
    & \qquad\qquad\int dk\,\Big(\sqrt{\frac{2}{N+1}}\sin(k)\Big)^2 \Big|\frac{dk}{dE}\Big| \delta(k-k_E) \\
    &=\frac{V^2}{\gamma}\sqrt{1-a^2}.
\label{Eq:app_1d_TB_Delta}
\end{split}
\end{align}
This is a semi-ellipse centered at $E=\alpha$, with a width of $2\gamma$ and a height of $V^2/\gamma$.
\par
\begin{figure}
    \centering
    \includegraphics[width=0.7\linewidth]{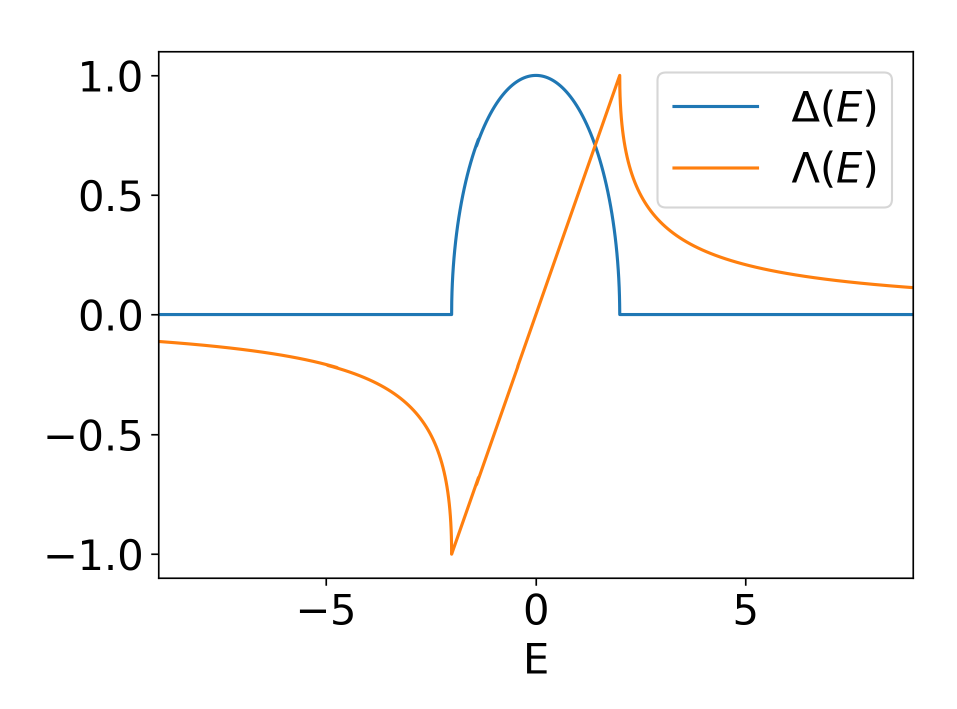}
    \caption{$\Delta(E)$ and $\Lambda(E)$ of a one-dimensional tight binding model without periodic boundary. The parameters are $(\alpha,\gamma,V)=(0,1,1)$. Units are arbitrary.}
    \label{fig:1dTB_Delta_Lambda}
\end{figure}

The Hilbert transform of $\Delta(E)$ (see \cref{Eq:multi_orb_Lambda}) is
\begin{equation}
    \Lambda(E) = 
    \begin{cases}
        \frac{V^2}{\gamma} (a+\sqrt{a^2 - 1}) \qquad, a < -1 \\
        \frac{V^2}{\gamma} a \qquad\qquad\qquad \quad\,\,\,, -1 \leq a \leq 1 \\
        \frac{V^2}{\gamma} (a-\sqrt{a^2 - 1}) \qquad, a > 1 .
    \end{cases}
\label{Eq:app_1d_TB_Lambda}
\end{equation}
Using \cref{Eq:multi_orb_AN_PDOS_from_G}, PDOS is equal to 
\begin{equation}
    \text{PDOS}(E) =-\frac{1}{\pi}\text{Im } \frac{1}{E-E_a-\Lambda(E) + i\Delta(E)}.
\end{equation}
The functions $\Delta(E)$ and $\Lambda(E)$ are plotted in \cref{fig:1dTB_Delta_Lambda} for $(\alpha,\gamma,V)=(0,1,1)$, in arbitrary units.

\begin{figure}
    \centering
    \includegraphics[width=1\linewidth]{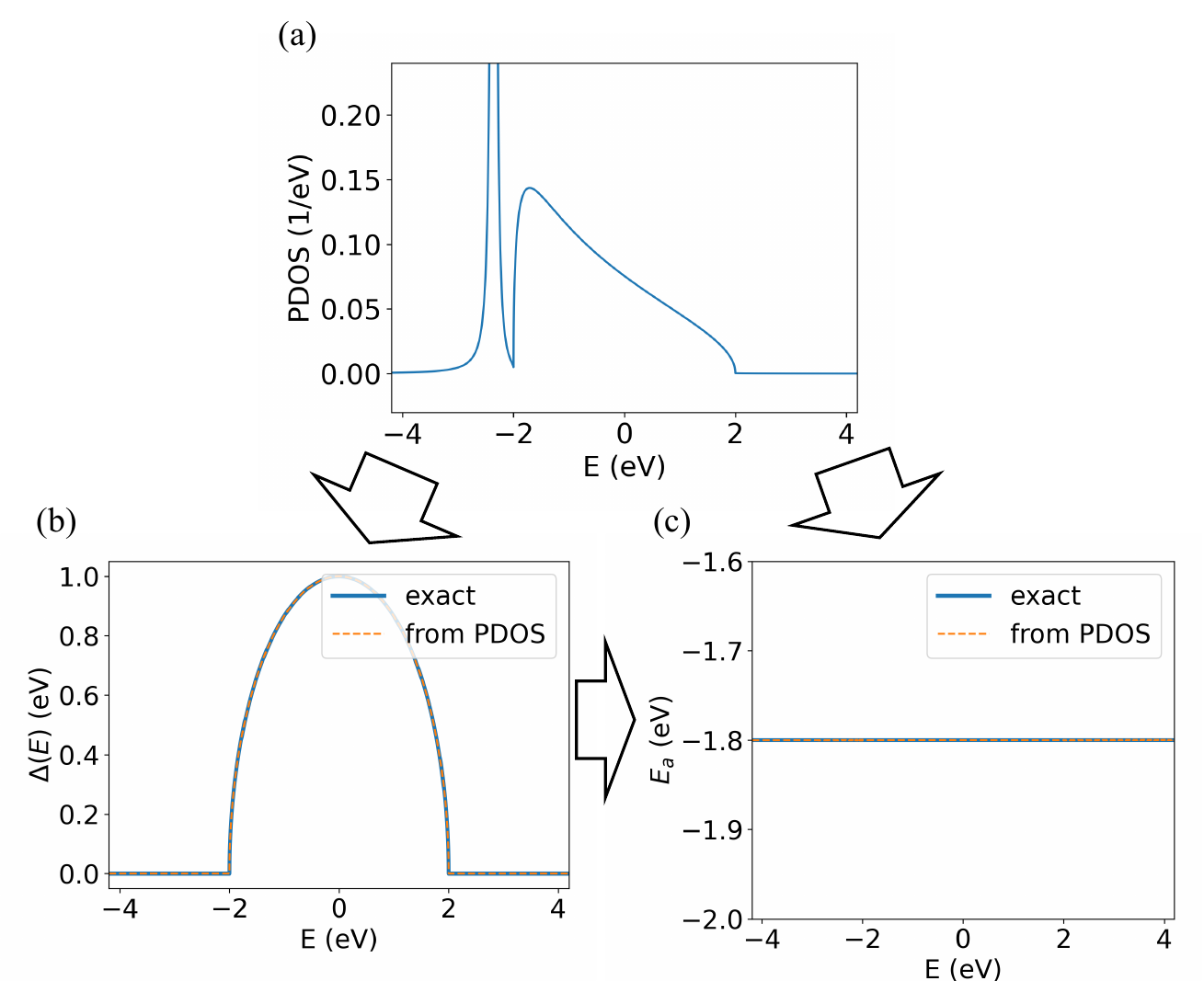}
    \caption{Demonstrating our Kramers-Kronig procedure for obtaining the AN Hamiltonian from the PDOS using the canonical single-orbital AN model with semi-elliptical $\Delta(E)$. (a) PDOS obtained from an exact analytical expression. (b-c) $\Delta(E)$ and $E_a$ (equal to $\mathbf{H}_\text{eff}$ in the single-orbital picture) computed from the PDOS. We obtain quantitative agreement with exact expressions.}
    \label{fig:PDOS_to_Delta_TB}
\end{figure}

\par
In \cref{fig:PDOS_to_Delta_TB}, we choose the parameters to be $(\alpha,\gamma,V,E_a) = (0,1,1,-1.8)$. The energy unit in this model analysis is arbitrary, but for concreteness, we will take the unit to be eV. These eV-range values are typical for modeling atoms adsorbed on transition-metal surfaces.\cite{vojvodic2014electronic} In this parameter regime, there is a discrete bonding state at energy $\approx -2.36$, which lies below the lower band edge at $-2$. Lorentzian smearing with a width of $\epsilon =0.01$ is applied to the PDOS, avoiding singularity due to the $\delta$-function. In \cref{fig:PDOS_to_Delta_TB} (b-c), the differences between the computed quantities and the exact results are much smaller than the smearing width $\epsilon=0.01$. 
When computing the Hilbert transform, we extend the numerical grid points for $E$ to be from $-2000$ to $2000$, with a spacing of $10^{-3}$.

\section{Decomposing $\mathbf{\Delta}(E)$ of CO/Cu using irreducible representations of the $C_{3v}$ point group}
\label{App:representation_group}
\subsection{Symmetry analysis of $\hat{H}_{el}$}
The atomic environment of the CO adsorbate on the atop site of Cu (111) possesses a $C_{3v}$ point group symmetry, whose irreducible representations (irrep) are labeled $A_1$, $A_2$, and $E$.
The $5\sigma$ HOMO (orbital 0) transforms as the one-dimensional $A_1$ irrep, and the $2\pi^*$ LUMOs (orbitals 1 and 2) transform as the two-dimensional irrep $E$.
Every symmetry operation of $C_{3v}$ can be represented by a $3\times3$ matrix in our CO adsorbate basis consisting of the $5\sigma$ HOMO and two $2\pi^*$ LUMOs. This forms a reducible representation of $C_{3v}$.
Due to symmetry, the $3\times3$ matrices $\mathbf{H}_\text{eff}$, $\mathbf{S}_\text{eff}$, and $\mathbf{\Delta}(E)$ commute with the representation of all symmetry operation. We will show that, as a consequence, $\mathbf{H}_\text{eff}$, $\mathbf{S}_\text{eff}$, and $\mathbf{\Delta}(E)$ are diagonal $3\times3$ matrices. Furthermore, the two LUMO diagonal elements are equal.
\par
For generality, consider a matrix $A$, expressed in a basis $B$. The vector space $V$ spanned by the basis $B$ forms a representation of a group $G$, so that for each element $g\in G$, the matrix $T(g)$ represents the group operation $g$ in the vector space $V$. Since $G$ is a symmetry group of the system, 
\begin{equation}
    [A, T(g)]=0 , \forall g\in G.
\label{Eq:app_rep_commutator_A}
\end{equation}
We denote the inequivalent irreducible representations (irrep) of $G$ as $T_1, T_2, \cdots$, whose dimensions are $N_1, N_2, \cdots$. 
In the example of CO/Cu, $A$ can be $\mathbf{H}_\text{eff}$, $\mathbf{S}_\text{eff}$, or $\mathbf{\Delta}(E)$. $B$ is the CO adsorbate orbital basis, i.e., $5\sigma$ and $2\pi^*$. $G$ is the $C_{3v}$ point group. The irreps are $A_1$, $A_2$, and $E$. Their dimensions are 1, 1, and 2, respectively. 
By a change of basis (if necessary), we can decompose the reducible matrix representation $T$ into a direct sum of irreps, i.e., 
\begin{equation}
    T(g) = T_1^{\oplus M_1}(g) \oplus T_2^{\oplus M_2}(g) \oplus \cdots ,\forall g\in G.
\label{Eq:app_direct_sum_irrep}
\end{equation}
In matrix form, $T(g)$ is now block-diagonal with $M_1$ $N_1\times N_1$ blocks, $M_2$ $N_2 \times N_2$ blocks, and so on. We index the basis vectors in this new basis using the triple $(r,m,n)$, where $r$ indexes the inequivalent irrep, $m$ indexes the copy of the irrep $T_r$, and $n$ indexes the basis vector inside the $m$-th copy of the irrep $T_r$. In the CO/Cu example, our adsorbate orbital basis already decomposes $T(g)$ into $A_1(g)\oplus E(g)$. No further basis change is needed.
\par
Expressing $A$ in this basis, we will show two facts:
\begin{itemize}
    \item The matrix elements of $A$ between inequivalent irreps are zero. In other words, if $r_1 \neq r_2$, then
    \begin{equation}
        A_{r_1 m_1 n_1, r_2 m_2 n_2} = 0.
    \end{equation}
    If we take $A$ to be $\mathbf{\Delta}(E)$ in the CO/Cu example, this means that $\Delta_{01}$, $\Delta_{02}$, and their transpose elements are zero. 

    \item The submatrix of $A$ within the irrep $T_r$ is block-diagonal, consisting of $N_r$ degenerate copies of $M_r \times M_r$ blocks. More specifically, 
    \begin{equation}
        A_{r m_1 n, r m_2 n}
    \end{equation}
    is independent of $n$, and if $n_1\neq n_2$, then
    \begin{equation}
        A_{r m_1 n_1, r m_2 n_2} = 0.
    \end{equation}
    In the CO/Cu example, the irrep $E$ has a dimension of 2, so that $\Delta_{12}=\Delta_{21}=0$ and that $\Delta_{11} = \Delta_{22}$.
\end{itemize}
\par
To show the first fact, we show that, for any $m_1$ and $m_2$, the submatrix $A_{r_1 m_1 \bullet, r_2 m_2 \bullet} = 0$. Using the block-diagonal structure of $T(g)$, \cref{Eq:app_rep_commutator_A} implies that, for all $g \in G$,
\begin{equation}
        A_{r_1 m_1 \bullet , r_2 m_2 \bullet} T_{r_2}(g) = T_{r_1}(g)A_{r_1 m_1 \bullet , r_2 m_2 \bullet} .
\label{Eq:app_irrep_proof_first_fact}
\end{equation}
Since $T_{r_1}$ and $T_{r_2}$ are inequivalent irreps, Schur's lemma implies that $A_{r_1 m_1 \bullet , r_2 m_2 \bullet} = 0$.
\par
To show the second fact, we follow the same idea as \cref{Eq:app_irrep_proof_first_fact}, but taking $r_1=r_2=r$. We obtain
\begin{equation}
    A_{r m_1 \bullet, r m_2 \bullet} T_r(g) = T_r(g)A_{r m_1 \bullet, r m_2 \bullet}
\end{equation}
for all $g \in G$. Schur's Lemma then implies that 
\begin{equation}
    A_{r m_1 \bullet, r m_2 \bullet} = \lambda_{r m_1 m_2}I
\end{equation}
is equal to a constant $\lambda_{r m_1 m_2}$ times the $N_r\times N_r$ identity matrix $I$. Therefore, the second fact is proven.

\subsection{Projecting spherical harmonics $Y_{lm}$ onto irreducible representations of point groups}
\label{App:Ylm_to_irreps}
Given a fixed $l$, $\{Y_{lm}|m=-l, \cdots, l\}$ forms an irreducible representation of the group $O(3)$. Since a point group $G$ is a subgroup of $O(3)$, $\{Y_{lm}|m=-l, \cdots, l\}$ forms the basis for a representation (not necessarily irreducible) of a point group. We denote this representation as $T(g)$, where $g \in G$ and $T(g)$ is a $(2l+1)\times(2l+1)$ matrix.
We want to find the basis vectors corresponding to different inequivalent irreps $T_1, T_2, \cdots$ of a point group $G$. The complex spherical harmonics $Y_{lm}$ can be transformed into the real spherical harmonics $\mathcal{Y}_{lm}$ (commonly used in chemistry) via a simple basis transformation.
\par
First, we construct the representation of $G$ in the basis $\{Y_{lm}|m=-l, \cdots, l\}$. All symmetry operations of a point group are composed of rotations and possibly an inversion. The representation of inversion $i$ is
\begin{equation}
    T(i) = (-1)^l.
\end{equation}
We denote a rotation of angle $\theta$ around the axis $\hat{\mathbf{n}}$ as $(\theta,\hat{\mathbf{n}})$.
The representation of $(\theta,\hat{\mathbf{n}})$ is
\begin{equation}
    T(\theta,\hat{\mathbf{n}}) = \exp(-\frac{i}{\hbar}\theta \hat{\mathbf{n}}\cdot \mathbf{L}),
\end{equation}
where $\mathbf{L}=(L_x, L_y, L_z)$ is the angular momentum vector operator, expressed as three $(2l+1)\times (2l+1)$ matrices. Once we have the representation of inversion and a general rotation, we can obtain the representation of all point group elements by composing symmetry operations. 
\par
Next, to find $M_r$, the number of copies of the irrep $T_r$ in $T$, we make use of the orthogonality of characters. We denote the character of a representation $T(g)$ as
\begin{equation}
    \chi_T(g) = \text{Tr}(T(g)).
\end{equation}
$M_r$ is equal to 
\begin{equation}
    M_r = \frac{1}{|G|}\sum_{g\in G} \chi_{T_r}^*(g) \chi_T(g),
\end{equation}
where $|G|$ is the order of the group $G$.
\par
Finally, to find the subspace spanned by the $M_r$ copies of $T_r$, we will show that the projection matrix $\Pi_r$ onto this subspace is given by
\begin{equation}
    \Pi_r = \frac{N_r}{|G|}\sum_{g\in G} \chi^*_{T_r}(g) T(g).
\label{Eq:app_irrep_projector}
\end{equation}
Before proving \cref{Eq:app_irrep_projector}, we first show that, given two irreps, $T_r$ and $T_{r'}$, 
\begin{equation}
    \sum_{g\in G}\chi_{T_r}^*(g) T_{r'}(g) =\frac{|G|}{N_r} \delta_{r r'}.
\label{Eq:app_irrep_projector_lemma}
\end{equation}
$\delta_{rr'}$ is equal to 1 if $T_r$ and $T_{r'}$ are the same irrep, and $\delta_{rr'}$ is equal to 0 if they are inequivalent. 
\par
To prove \cref{Eq:app_irrep_projector_lemma}, we make use of the great orthogonality theorem,\cite{zee2016group} which states that
\begin{equation}
    \sum_{g\in G} T^\dagger_{r}(g)_{ij} T_{r'}(g)_{kl} = \frac{|G|}{N_r}\delta_{r r'}\delta_{il}\delta_{jk}.
\end{equation}
Therefore, the $(i,j)$-th element of the left hand side of \cref{Eq:app_irrep_projector_lemma} becomes
\begin{align}
\begin{split}
    &\sum_{g\in G}\chi_{T_r}^*(g) T_{r'}(g)_{ij} = \sum_{g\in G} \sum_k T_r^\dagger(g)_{kk} T_{r'}(g)_{ij} \\
    &=\sum_k \frac{|G|}{N_r} \delta_{rr'}\delta_{kj}\delta_{ki} = \frac{|G|}{N_r} \delta_{rr'}\delta_{ij}.
\end{split}
\end{align}
Therefore, \cref{Eq:app_irrep_projector_lemma} is proven.
\par
Now, we write $T(g)$ in \cref{Eq:app_irrep_projector} as a direct sum of irreps (see \cref{Eq:app_direct_sum_irrep}). We see that if a vector $v$ is in the subspace corresponding to the irrep $T_r$, then $\Pi_r v = v$. Otherwise, $\Pi_r v = 0$. Therefore, $\Pi_r$ is the projection matrix onto the subspace spanned by all $M_r$ copies of the irrep $T_r$.

\section{Hybridization energy}
\label{App:hybridization_energy}

We treat the isolated Fermi bath (i.e., the isolated surface Hamiltonian) as the reference state and measure the total grand potential (i.e., $\Omega=E-\mu N$)
of the AN Hamiltonian relative to the reference. Given the Hamiltonian $\mathbf{H}$ and overlap $\mathbf{S}$ matrices, we define the hybridization energy at zero temperature as
\begin{equation}
    E_\text{hyb} = \sum_{\substack{E\in \text{eig}(\mathbf{H},\mathbf{S})\\E<\mu}} (E-\mu) - \sum_{\substack{E\in\text{eig}(\mathbf{H}_{ss},\mathbf{S}_{ss})\\ E<\mu}} (E-\mu).
\label{Eq:binding_E_def}
\end{equation}
The notation $\text{eig}(\mathbf{H},\mathbf{S})$ means the energy eigenvalues corresponding to $\mathbf{H}$ and $\mathbf{S}$. $\mathbf{H}_{ss}$ and $\mathbf{S}_{ss}$ are the surface submatrices of $\mathbf{H}$ and $\mathbf{S}$.
The first term is the total energy of the coupled system. The second term represents the energy of the isolated surface.
The binding energy is typically defined as $E_{\text{total}}-E(\text{isolated surface})-E(\text{isolated adsorbate})$. Therefore, the binding energy can be thought of as $E_\text{hyb}$ minus a constant energy of the isolated adsorbate.

\par
$E_\text{hyb}$ can be expressed analytically by noticing that $\text{eig}(\mathbf{H},\mathbf{S})$ are the poles of $\mathbf{G}^{aa}(E)$ (see \cref{Eq:app_Greens_matrix_with_overlap}), which are the zeros of $\det (\mathbf{Q}(E))$. Furthermore, $\text{eig}(\mathbf{H}_{ss}, \mathbf{S}_{ss})$ are the poles of $\det (\mathbf{Q}(E))$ (see \crefrange{Eq:app_multi_orb_phi0}{Eq:app_multi_orb_AN_K}).
\par
For generality, we will compute the quantity
\begin{equation}
    \Delta f = \sum_{\substack{E\,\in\, \text{zeros of}\\ \det\mathbf{Q}(E)}} f(E)\, - \sum_{\substack{E\,\in\, \text{poles of} \\ \det\mathbf{Q}(E)}} f(E).
\label{Eq:binding_f_general}
\end{equation}
In the special case $f(E) = (E-\mu)\theta(\mu-E)$, we obtain $E_\text{hyb}$ in \cref{Eq:binding_E_def}.  
In the following, we will let $f(E)$ be a generic function and express $\Delta f$ in terms of $f(E)$ and $\det\mathbf{Q}(E)$. Afterward, we then substitute the appropriate functions for $f(E)$ to obtain various physical quantities.

\par
\begin{figure}[H]
    \centering
    \includegraphics[width=\linewidth]{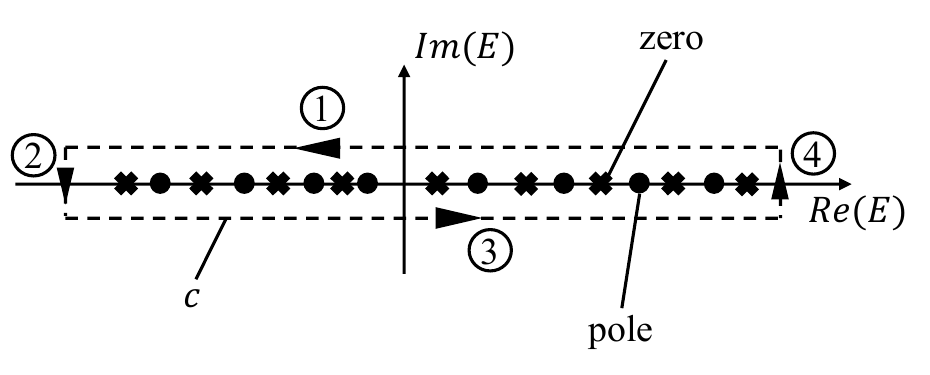}
    \caption{Contour integration of \cref{Eq:single_orbital_Delta_f_contour_int}. The contour encloses the zeros and poles of $\det\mathbf{Q}(E)$ that we consider. }
    \label{fig:zeros_and_poles}
\end{figure}

To evaluate $\Delta f$, we use the argument principle in complex analysis,\cite{brown2009complex} which implies that
\begin{align}
\begin{split}
&\Big(\sum_{\substack{E\,\in\,\text{zeros of}\\ \det\mathbf{Q} \text{ inside } c}} f(E) \Big)\,- \Big(\sum_{\substack{E\,\in\,\text{poles of}\\ \det\mathbf{Q} \text{ inside } c}} f(E)\Big) \\
&= \frac{1}{2\pi i}\int_c \, f(E)\frac{(\det\mathbf{Q}(E))'}{\det\mathbf{Q}(E)} dE.
\label{Eq:argument_principle}
\end{split}
\end{align}
Here, $c$ represents a closed contour, oriented counter-clockwise, in the complex plane of the variable $E$. The notation $(\det\mathbf{Q}(E))'$ denotes the complex derivative $d (\det\mathbf{Q}(E))/dE$. 
Choosing the contour $c$ to enclose the zeros and poles we want to sum over, we have
\begin{equation}
    \Delta f = \frac{1}{2\pi i}\int_c \, f(E)\frac{(\det\mathbf{Q}(E))'}{\det\mathbf{Q}(E)} dE .
\label{Eq:single_orbital_Delta_f_contour_int}
\end{equation}
The contour $c$ is depicted in \cref{fig:zeros_and_poles}.

\par
The rectangular contour is divided into four segments.
We can make the width (i.e., line segments 2 and 4) of the contour infinitesimally small, so that these two line segments do not contribute to the contour integral. The integral along segment $1$ is
\begin{equation}
    I_1 = -\frac{1}{2\pi i}\lim_{\epsilon\rightarrow 0^+}\int_{E_1}^{E_2} f(E+i\epsilon) \frac{(\det\mathbf{Q}(E+i\epsilon))'}{\det\mathbf{Q}(E+i\epsilon)} dE,
\label{Eq:int_seg_1}
\end{equation}
where $E_1$ and $E_2$ are the minimum and the maximum of $\text{Re}(E)$ along the contour.
The integral along segment $3$ is
\begin{equation}
    I_3 = \frac{1}{2\pi i}\lim_{\epsilon\rightarrow 0^+}\int_{E_1}^{E_2} f(E-i\epsilon) \frac{(\det\mathbf{Q}(E-i\epsilon))'}{\det\mathbf{Q}(E-i\epsilon)} dE.
\label{Eq:int_seg_3}
\end{equation}
Following the derivation of \cref{App:Deriving_Gaa_multi_orb_AN} and changing $+i\epsilon$ to $-i\epsilon$, we see that 
\begin{align}
\begin{split}
    &\lim_{\epsilon\rightarrow0^+} \mathbf{Q}(E-i\epsilon)\\ &= \lim_{\epsilon\rightarrow0^+} (E-i\epsilon)\mathbf{S}_{aa}-\mathbf{H}_{aa}-\mathbf{K}(E)-i\mathbf{\Delta}(E) \\
    &= \Big(\lim_{\epsilon\rightarrow0^+} \mathbf{Q}(E+i\epsilon)\Big)^\dagger.
\end{split}
\end{align}
Taking the determinants, we see that in the limit $\epsilon\rightarrow0^+$,
\begin{equation}
    \det\mathbf{Q}(E-i\epsilon) = \big(\det\mathbf{Q}(E+i\epsilon)\big)^*.
\end{equation}
If the observable $f(E)$ is continuous inside the contour and if $f(E)$ is real-valued for real-valued $E$, then $I_1 = I_3^*$. \cref{Eq:single_orbital_Delta_f_contour_int} now becomes
\begin{align}
\begin{split}
    \Delta f &= I_1+I_3 = 2\text{ Re}(I_1) \\
     &= -\frac{1}{\pi}\lim_{\epsilon\rightarrow 0^+}\text{Im}\int^{E_2}_{E_1} f(E) \frac{(\det\mathbf{Q}(E+i\epsilon))'}{\det\mathbf{Q}(E+i\epsilon)} dE\\
     &= -\frac{1}{\pi}\text{Im}\int^{E_2}_{E_1} f(E) \frac{(\det\mathbf{Q}(E))'}{\det\mathbf{Q}(E)} dE.
\end{split}
\end{align}
In the last line, we have used the convention that $\mathbf{Q}(E)$ means the limit $\lim_{\epsilon\rightarrow0^+}\mathbf{Q}(E+i\epsilon)$ (see \cref{App:Deriving_Gaa_multi_orb_AN}).
Integrating by parts, we have
\begin{align}
\begin{split}
    \Delta f &= \frac{1}{\pi}\int^{E_2}_{E_1} f'(E)\text{Im}\big(\ln\det\mathbf{Q}(E)\big)dE \\
    &\quad+\frac{1}{\pi}f(E_1)\text{Im}\big(\ln\det\mathbf{Q}(E_1)\big) \\
    &\quad -\frac{1}{\pi} f(E_2)\text{Im}\big(\ln\det\mathbf{Q}(E_2)\big) \\
    &= \frac{1}{\pi} \Big(\int^{E_2}_{E_1} f'(E)\arg\big(\det\mathbf{Q}(E)\big)dE \\
    & \qquad +f(E_1)\arg\big(\det\mathbf{Q}(E_1)\big) \\
    &\qquad -f(E_2)\arg\big(\det\mathbf{Q}(E_2)\big)\Big),
\label{Eq:I_explicit}
\end{split}
\end{align}
where $\text{arg}(z)$ is the argument of the complex number $z$.
Expressing $\Delta f$ in terms of the hybridization energy density function
\begin{equation}
    \eta(E) = \frac{1}{\pi}\Big(\arg(\det \mathbf{Q}(E)) - \arg(\det \mathbf{Q}(-\infty))\Big)
\label{Eq:eta_def_app}
\end{equation}
(see \cref{Eq:eta_def}), we have
\begin{align}
\begin{split}
    \Delta f = &\int^{E_2}_{E_1} f'(E)\eta(E) dE \\
    &+ f(E_1)\eta(E_1)-f(E_2)\eta(E_2).
\label{Eq:app_Delta_f_in_eta}
\end{split}
\end{align}
\par
Now, we substitute different forms of $f(E)$ into \cref{Eq:app_Delta_f_in_eta} to obtain physical quantities. At zero temperature, only energy levels below $\mu$ contribute to the sum in \cref{Eq:single_orbital_Delta_f_contour_int}. Therefore, we take $E_1=-\infty$ and $E_2 = \mu$, so that the contour encloses only the energy levels below $\mu$. Taking $f(E) = E$, the energy change (i.e., not relative to $\mu$) is
\begin{align}
\begin{split}
    \Delta E &=  \int_{-\infty}^\mu \eta(E) dE -\mu\eta(\mu) .
\label{Eq:app_DeltaE}
\end{split}
\end{align}
We have used the fact that $\eta(-\infty) = 0$ (see \cref{Eq:eta_def_app}).
\par
The change in occupation number $N$ is obtained by taking $f(E) = 1$, so that 
\begin{align}
\begin{split}
    \Delta N 
    &=-\eta(\mu).
\label{Eq:app_DeltaN}
\end{split}
\end{align}
The hybridization energy $E_{\text{hyb}}$ is the change in grand potential, which is equal to 
\begin{equation}
    E_\text{hyb} = \Delta E - \mu\Delta N.
\end{equation}
Combining \cref{Eq:app_DeltaE,Eq:app_DeltaN}, we see that 
\begin{equation}
    E_\text{hyb} = \int^\mu_{-\infty} \eta(E) dE .
\label{Eq:app_DeltaF}
\end{equation}

\par
At finite temperature $\beta = 1/(k_B T)$ and chemical potential $\mu$, the grand potential of an energy level $E$ is
\begin{equation}
    \Omega(E) = -k_B T \ln\big( 1+e^{-\beta(E-\mu)} \big).
\label{Eq:app_finite_temperature_grand_free_energy}
\end{equation}
Therefore, the change in grand-canonical free energy is obtained by taking $f(E)=\Omega(E)$ and setting the integration bounds to be $(E_1,E_2) = (-\infty, \infty)$, i.e., 
\begin{align}
\begin{split}
    E_{\text{hyb}, \text{ finite T}} &= \int^\infty_{-\infty} \frac{d\Omega}{dE}\eta(E)dE.
\label{Eq:app_Delta_F_finite_temp}
\end{split}
\end{align}
We have used the fact that $\Omega(\infty) = 0$. 
As the temperature approaches zero, $\Omega(E)\rightarrow \theta(\mu-E)(E-\mu)$, and \cref{Eq:app_Delta_F_finite_temp} reduces to \cref{Eq:app_DeltaF}.

\subsection{Comparing zero-temperature grand potential $\Omega(R)$ at fixed $\mu$ to energy $E(R)$ at fixed $N$ }

We note that the total energy in the AN model is taken to be the grand potential $\Omega(R)=E(R)-\mu N(R)$ at fixed $\mu$, while the DFT energy $E_\text{DFT}$ is computed as the energy $E(R)$ at fixed $N$. In this section, we show that when the surface slab is infinitely thick, the grand potential difference $\Delta \Omega(R)$ and the energy difference $\Delta E(R)$ have exactly the same functional dependence on $R$. 
\par
\begin{figure}
    \centering
    \includegraphics[width=0.6\linewidth]{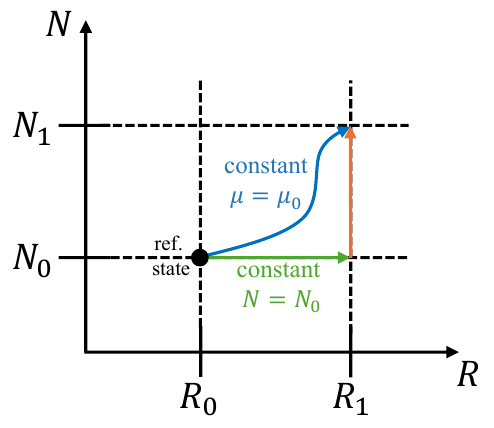}
    \caption{Constant $\mu$ vs. constant $N$ paths in the thermodynamic state space.}
    \label{fig:thermodynamic_cycle}
\end{figure}

Consider the energy and grand potential differences for moving the adsorbate coordinate from a reference $R_0$ to $R_1$. Let the reference state at $R_0$ have an electron number of $N_0$ and a chemical potential of $\mu=\partial E(N_0,R_0)/\partial N=\mu_0$. The state of the system is specified by the pair $(N,R)$ (see \cref{fig:thermodynamic_cycle}). The change in energy $E$ at fixed $N$ is
\begin{equation}
    \Delta E(R_1)|_{N=N_0} = E(N_0, R_1)-E(N_0,R_0).
\label{Eq:DeltaE_R1}
\end{equation}
The path of constant $N$ is shown in green in \cref{fig:thermodynamic_cycle}.
Denoting the grand potential as $\Omega=E-\mu N$, the change in grand potential $F$ at fixed $\mu$ is
\begin{equation}
    \Delta \Omega(R_1)|_{\mu=\mu_0} = \Omega(N_1,R_1) - \Omega(N_0,R_0).
\label{Eq:DeltaF_R1}
\end{equation}
The path of constant $\mu$ is shown in blue in \cref{fig:thermodynamic_cycle}. $N_1$ is the electron number corresponding to $R=R_1$ and $\mu=\mu_0$. 

Using the Legendre relation
\begin{equation}
    \Omega = E-\mu N,
\end{equation}
\cref{Eq:DeltaF_R1} becomes
\begin{align}
\begin{split}
    \Delta \Omega(R_1)|_{\mu=\mu_0} =& E(N_1,R_1)-E(N_0,R_1) \\
    &+E(N_0,R_1)-E(N_0,R_0) \\
    &-\mu_0(N_1-N_0).
\label{Eq:app_Delta_F_R1}
\end{split}
\end{align}
We have decomposed the path of constant $\mu$ into a path of constant $N$ (second line) and a path of constant $R$ (first line). The paths are illustrated in \cref{fig:thermodynamic_cycle}. Since $\mu(N,R) = \partial E(N,R)/\partial N$, we can rewrite the first line as an integral, i.e., 
\begin{align}
\begin{split}
    E(N_1,R_1)-E(N_0,R_1) = \int^{N_1}_{N_0}\mu(N,R_1)\, dN.
\end{split}
\end{align}
The second line of \cref{Eq:app_Delta_F_R1} is equal to $\Delta E|_{N=N_0}$ (see \cref{Eq:DeltaE_R1}). Therefore, \cref{Eq:app_Delta_F_R1} becomes
\begin{align}
\begin{split}
    \Delta \Omega(R_1)|_{\mu=\mu_0} =& \Delta E(R_1)|_{N=N_0} + \\
    &\int^{N_1}_{N_0} \mu(N,R_1)-\mu_0 \,dN.
\label{Eq:DeltaF_vs_DeltaE}
\end{split}
\end{align}
Note that this is the finite-difference generalization of the differential identity 
\begin{equation}
    \left.\frac{\partial \Omega}{\partial R}\right|_{\mu} = \left.\frac{\partial E}{\partial R}\right|_N,
\end{equation}
which is due to the Legendre transform between $E$ and $\Omega$.
In \cref{Eq:DeltaF_vs_DeltaE}, since $\mu_0$ is equal to $\mu(N_1,R_1)$, and $N_1-N_0$ is of order 1, the integral term in \cref{Eq:DeltaF_vs_DeltaE} scales as $\sim |\partial \mu/\partial N|\sim 1/N$. Therefore, in the limit of an infinitely thick slab (i.e., $N\rightarrow \infty$), the integral term in \cref{Eq:DeltaF_vs_DeltaE} vanishes, and $\Delta \Omega|_{\mu=\mu_0} = \Delta E|_{N=N_0}$. 

\subsection{Special case: single-orbital AN Hamiltonian}
In the single-orbital AN model, $Q(E)$ is a $1\times1$ matrix, so $\det Q(E)=Q(E)$. The imaginary part of $Q(E)$ is $\Delta(E)+\epsilon$, which is always positive. Therefore, $\arg(Q(E))$ is in the range $[0, \pi]$. When $E=-\infty$, $Q(E)$ has a negative real part and an infinitesimal positive imaginary part $\epsilon$. Therefore, $\arg(Q(-\infty)) = \pi$. Combining these facts, we have
\begin{align}
\begin{split}
    \eta(E) &= \arg(Q(E)) - \pi \\
    &=\lim_{\epsilon\rightarrow 0^+}\tan^{-1}_{[-\pi, 0]}\Big(\frac{\Delta(E) + \epsilon}{E-E_a-\Lambda(E)}\Big).
\end{split}
\end{align}
The range of $\tan^{-1}_{[-\pi,0]}$ is taken to be $[-\pi, 0]$. 
Setting the chemical potential $\mu$ to be 0, we obtain the binding energy expression for the conventional single-orbital AN Hamiltonian, i.e.,
\begin{equation}
    E_{\text{binding}} = \frac{1}{\pi} \int_{-\infty}^0  \tan^{-1}_{[-\pi,0]}\big( \frac{\Delta(E)}{E-E_a-\Lambda(E)} \big) dE \, - \, E_a.
\label{Eq:common_bindingE_1}
\end{equation}
This expression has been widely used in catalysis to rationalize binding energy trends.\cite{newns1969, norskov2014fundamental,vojvodic2014electronic,vijay2022limits, wang2020bayesian}
In \cref{Eq:common_bindingE_1}, we have subtracted the isolated adsorbate energy $E_a$ and omitted the spin multiplicity factor of 2. 
The range of $\tan^{-1}$ is defined to be $[-\pi, 0]$, instead of the conventional choice of $(-\pi/2, \pi/2)$.
This leads to an ambiguity when $\Delta(E)=0$ because $\tan^{-1}(0)$ can be chosen as $-\pi$ or $0$. By comparison, our expression of $E_\text{AN}$ based on the $\arg$ function is free of ambiguity.
\par
Sometimes, the single-orbital binding energy is taken to be\cite{santos2006model, santos2009model}
\begin{equation}
    E_{\text{binding}}' = \int^\mu_{-\infty} E \cdot  \text{PDOS}(E) \, dE - E_a. 
\label{Eq:single_orbital_local_E}
\end{equation}
This only accounts for the energy component projected onto the adsorbate. It omits the contribution from the surface energy change. 

\par
\begin{figure}[H]
    \centering
    \includegraphics[width=0.8\linewidth]{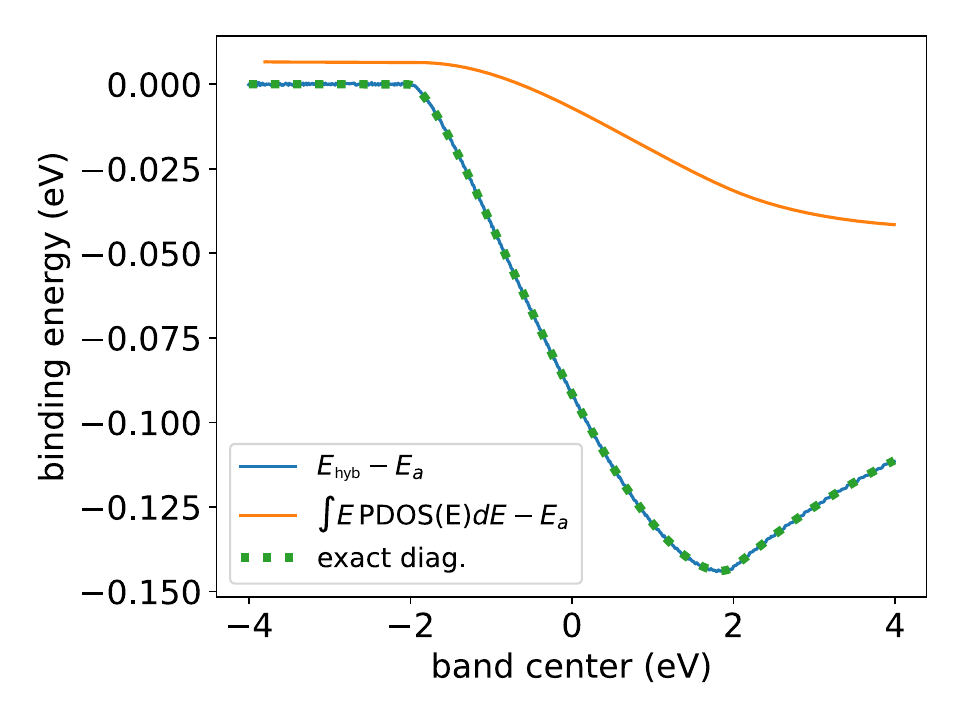}
    \caption{Comparing different binding energy expressions using an AN model with semi-elliptical $\Delta(E)$, where exact diagonalization is possible. \cref{Eq:single_orbital_bindingE,Eq:common_bindingE_1} reproduce the exact result, while \cref{Eq:single_orbital_local_E} misses the contribution due to surface energy change.}
    \label{fig:compare_bindingE_scan_alpha}
\end{figure}

\cref{fig:compare_bindingE_scan_alpha} compares the binding energy expressions (\cref{Eq:common_bindingE_1,Eq:single_orbital_local_E}) with exact diagonalization for a tight-binding AN model with a semi-elliptical $\Delta(E)$, described in \cref{App:semi_elliptical_Delta_example}.
The parameter $\gamma$ is taken to be 1 eV, resulting in a bandwidth of $4\gamma =4$ eV, which is in the range of the bandwidth for typical transition metals.\cite{vojvodic2014electronic} Following Ref.~\onlinecite{vojvodic2014electronic}, we set $\mu=0$ eV, $V = 1$ eV, and $E_a=-5$ eV.  
We vary $\alpha$, the band center, while keeping all other parameters fixed. 
\par
We perform exact diagonalization on a tight-binding chain with $N=1000$. The eigenvalues are summed directly using \cref{Eq:binding_E_def}. To avoid numerical discontinuities, we apply a small temperature $k_B T = 0.02$ eV to smear the energy levels. Specifically, we take $f(E)$ in \cref{Eq:binding_f_general} to be the grand free energy with a small finite temperature (see \cref{Eq:app_finite_temperature_grand_free_energy}).
We see that the binding energy expression of \cref{Eq:common_bindingE_1} produces the exact binding energy, while \cref{Eq:single_orbital_local_E} results in much weaker binding. Notice that when $\mu$ lies inside the surface band (i.e., band center $\in(-2, 2)$ eV), the binding energy decreases as the band center decreases, consistent with the d-band theory.\cite{norskov2014fundamental}

\subsection{Bypassing an orthogonalization energy}
\label{App:bypassing_ortho_E}
\begin{figure*}
    \centering
    \includegraphics[scale=0.35]{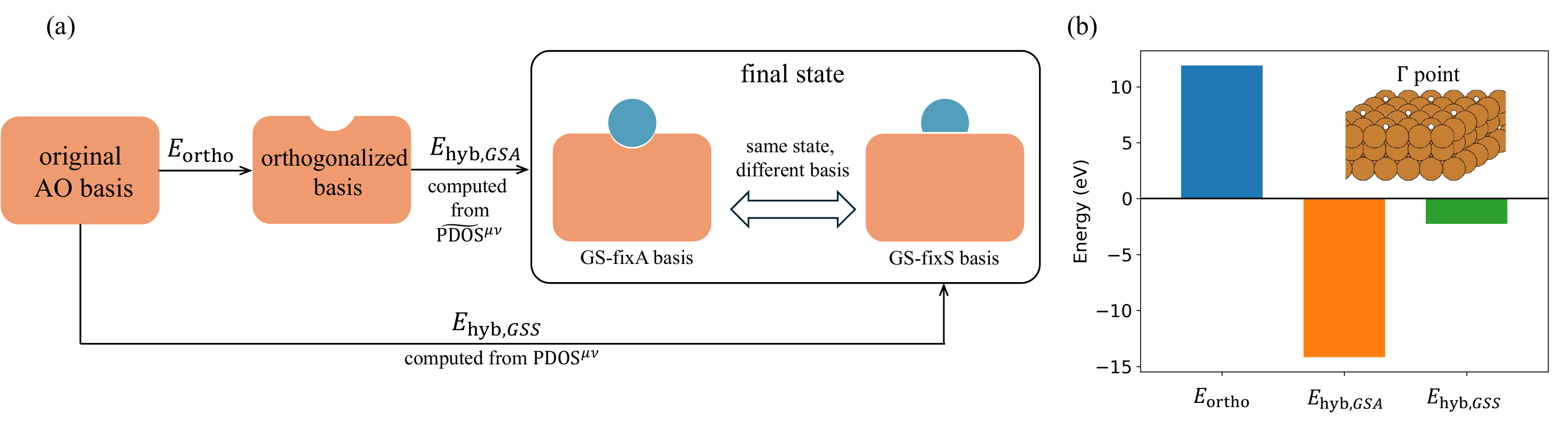}
    \caption{(a) Using $\text{PDOS}^{\mu\nu}$ to compute the hybridization energy $E_\text{hyb}$ allows one to bypass an energy penalty $E_{\text{ortho}}$ due to orthogonalizing surface orbitals. (b) For example, at $\Gamma$ point of H on Cu, $E_{\text{ortho}}$ and $E_{\text{hyb},GSA}$ have large magnitudes and opposite signs, leading to large cancellation of energy. $E_{\text{hyb},GSS}=E_{\text{ortho}}+E_{\text{hyb},GSA}$, obtained directly from $\text{PDOS}^{\mu\nu}$, only has a small magnitude.}
    \label{fig:remove_E_ortho}
\end{figure*}

The hybridization energy $E_\text{hyb}$ in \cref{Eq:single_orbital_bindingE} is computed from $\mathbf{Q}(E)$, which can be obtained from $\text{PDOS}^{\mu\nu}$ or $\widetilde{\text{PDOS}}^{\mu\nu}$.
As described in \cref{App:PDOS_orthogonalization_correspondence}, the two different PDOS correspond to GS-fixS and GS-fixA orthogonalization schemes.
Since $E_\text{hyb}$ is the difference between the coupled adsorbate-surface system and the isolated surface subsystem, we see that $E_{\text{hyb},GSS}$ (i.e., the $E_\text{hyb}$ computed from $\text{PDOS}^{\mu\nu}$, which corresponds to GS-fixS) is the difference between the final energy and the isolated surface in the original AO basis. On the other hand, $E_{\text{hyb},GSA}$ (i.e., the $E_\text{hyb}$ computed from $\widetilde{\text{PDOS}}^{\mu\nu}$, which corresponds to GS-fixA) is the difference between the final energy and the isolated surface in the GS-fixA orthogonalized basis. There is an energy difference between the isolated surface in the original basis and in the GS-fixA orthogonalized basis.
This energy difference (usually positive) is a type of orthogonalization energy $E_{\text{ortho}}$ (see \cref{fig:remove_E_ortho} (a)). Hence, we write
\begin{equation}
    E_{\text{hyb},GSS} = E_{\text{ortho}} + E_{\text{hyb},GSA}.
\end{equation}
 
\par
Previous studies of adsorption energy trends\cite{vijay2022limits, wang2020bayesian, hammer1996co} estimate the orthogonalization energy using a two-state model. They model the d-band as a single state that overlaps with the single adsorbate orbital. The accuracy of this approach is difficult to assess, and this approach does not generalize easily to the multi-orbital AN model. Using the contravariant $\text{PDOS}^{\mu\nu}$ in the non-orthogonal basis to compute $E_\text{hyb}$, one can bypass the orthogonalization energy term, so that the total energy does not contain $E_{\text{ortho}}$ as a contribution.
\par
As an example, we compute $E_{\text{ortho}}$, $E_{\text{hyb},GSS}$, and $E_{\text{hyb},GSA}$ at the $\Gamma$ point of H on Cu(111) (see \cref{fig:remove_E_ortho} (b)). We see that $E_{\text{ortho}}$ and $E_{\text{hyb},GSA}$ have large magnitudes and opposite signs, resulting in large cancellation of energy. Their sum, $E_{\text{hyb},GSS}$, only has a small magnitude. Therefore, computing $E_{\text{hyb},GSS}$ directly from $\text{PDOS}^{\mu\nu}$ can reveal the binding energy contribution more clearly, without the additional factor of $E_{\text{ortho}}$.

\section{Isolated CO molecule}
\label{Sec:isolated_CO}
Unlike the H/Cu example, the CO adsorbate on Cu has an internal nuclear degree of freedom (i.e., a vibrational degree of freedom).
We first consider the limiting case in which the adsorbate is infinitely far away from the surface. This provides insight into how the AN Hamiltonian depends on the vibrational degree of freedom in the absence of adsorbate-surface coupling. The molecular DFT calculation is performed on Q-Chem\cite{epifanovsky2021software} without periodic boundary conditions. The calculation parameters are the same as in the CO/Cu calculation.
\par
\begin{figure}
    \centering
    \includegraphics[width=0.7\linewidth]{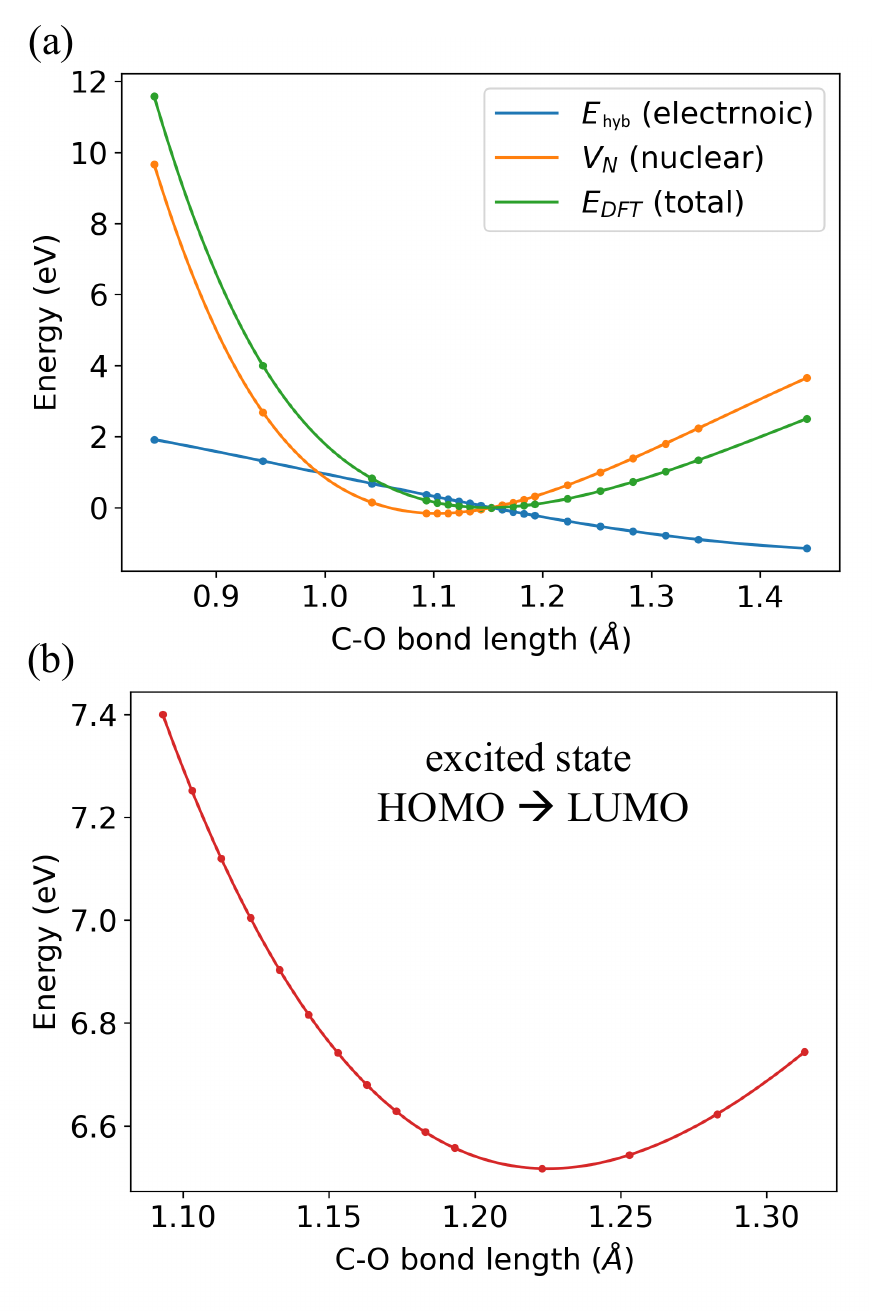}
    \caption{Mapping the DFT electronic structure of an isolated CO molecule to the AN Hamiltonian without adsorbate-surface coupling. (a) Within the AN Hamiltonian, the ground state DFT energy is decomposed into an electronic part $E_\text{hyb}$ and a nuclear part $V_N$. (b) The excited-state energy is derived from the AN Hamiltonian. The equilibrium bond length is slightly longer than the ground-state equilibrium bond length and matches reasonably well with the experimental value.}
    \label{fig:CO_isolated}
\end{figure}

\cref{fig:CO_isolated} (a) shows the energy contribution at various bond lengths. We have set the reference state to be the optimized geometry $R_0$, with a bond length of $\approx 1.15\, \AA$ (compared to the experimental value of $1.128 \,\AA$\cite{nist_CO}). $E_\text{DFT}$ and $E_\text{hyb}$ are set to 0 in this reference state.
We focus on the frontier orbitals, i.e., the $5\sigma$ HOMO and the two degenerate $2\pi^*$ LUMOs. Since there is no adsorbate-surface coupling, $E_\text{hyb}$ is equal to the HOMO orbital energy times 2 (spin multiplicity), shifted by a constant so that $E_\text{hyb}(R_0) = 0$. Near the equilibrium $R_0$, $E_\text{hyb}$ decreases and $V_N$ increases as the bond length increases. 
\par
In the AN Hamiltonian with independent electrons, the lowest excited state is obtained by simply promoting an electron from the HOMO to the LUMO. Therefore, in this simplified picture, the excited state energy is equal to the ground state energy $E_\text{DFT}$ plus the HOMO-LUMO gap. The excited state energy is also equal to the sum of $V_N$, the energy of an electron in the HOMO, and the energy of an electron in the LUMO.
\cref{fig:CO_isolated} (b) shows the excited state energy as a function of bond length. The equilibrium bond length of $\approx 1.22\,\AA$ is slightly longer than the ground state equilibrium bond length. The equilibrium energy of the excited state is found to be $\approx 6.5$ eV. These values match reasonably well with the experimental values of $1.24\,\AA$ and $8.1$ eV.\cite{nist_CO}

\section{Electronic dynamics}
\label{App:electronic_dynamics}

We let the system be in the multi-electron grand-canonical equilibrium state before $t=0$. At temperature $\beta=1/k_BT$ and chemical potential $\mu$, the equilibrium state is described by 
\begin{equation}
    \rho_{eq} = \frac{e^{-\beta(\hat{H}_{el}-\mu \hat{N})}}{\text{Tr}(e^{-\beta(\hat{H}_{el}-\mu \hat{N})})}.
\end{equation}
$\hat{H}_{el}$ is the electronic part of the second-quantized AN Hamiltonian (see \cref{Eq:H_tot}). $\hat{N}$ is the electron number operator.
\par
Let an electron in adsorbate orbital $\mu$ be excited to adsorbate orbital $\nu$ at time $t=0$. The population of adsorbate orbital $\sigma$ is then measured as a function of time as the system returns to equilibrium.
The initial state at $t=0$ is equal to 
\begin{equation}
    \rho(0) = \frac{a^\dagger_\nu a_\mu \rho_{eq}a^\dagger_\mu a_\nu}{\text{Tr}(a^\dagger_\nu a_\mu \rho_{eq}a^\dagger_\mu a_\nu)}.
\label{Eq:rho_0_photo_excited}
\end{equation}
This state can be prepared from $\rho_{eq}$, for example, via pump-probe spectroscopy, where only the signal contribution due to the $\mu\rightarrow \nu$ photo-excited state is post-selected.

The population of orbital $\sigma$ at time $t$ is given by
\begin{align}
\begin{split}
    p_{\sigma}(t) &= \text{Tr}(a^\dagger_\sigma a_\sigma U(t)\rho(0)U^\dagger(t)) \\
    &= \text{Tr}(a^\dagger_\sigma(t)a_\sigma(t)\rho(0)),
\label{Eq:app_p_sigma_0}
\end{split}
\end{align}
where $U(t) = \exp(-i\hat{H}_{el}t/\hbar)$ is the time-evolution operator. In the second line, $a_\nu(t) = U^\dagger(t) a_\nu U(t)$ is the Heisenberg picture operator. Substituting \cref{Eq:rho_0_photo_excited} into \cref{Eq:app_p_sigma_0}, we have
\begin{align}
\begin{split}
    p_\sigma(t) = \frac{\langle a^\dagger_\mu a_\nu a^\dagger_\sigma(t) a_\sigma(t)a^\dagger_\nu a_\mu\rangle}{\langle a^\dagger_\mu a_\nu a^\dagger_\nu a_\mu\rangle},
\label{Eq:app_p_sigma_1}
\end{split}
\end{align}
where $\langle A\rangle = \text{Tr}(A\rho_{eq})$ means the equilibrium expectation value.
\par
Using Wick's theorem,\cite{rammer2011quantum} the denominator of \cref{Eq:app_p_sigma_1} is
\begin{align}
\begin{split}
    \langle a^\dagger_\mu a_\nu a^\dagger_\nu a_\mu\rangle = |\langle a^\dagger_\mu a_\nu \rangle |^2 + n_\mu (1-n_\nu),
\label{Eq:app_p_sigma_denominator}
\end{split}
\end{align}
where $n_\tau = \langle a^\dagger_\tau a_\tau\rangle$.
To evaluate the expectation value $\langle a^\dagger_\tau a_\lambda\rangle$, we express $\{a_\mu\}$ (corresponding to the orthonormal AO $\{|\overline{\phi}_\mu\rangle\}$) as linear combinations of $\{a_i\}$ (corresponding to the orthonormal MO $\{|\psi_i\rangle\}$). Since both the AO basis and the MO basis are orthonormal, the coefficient matrix $C_{\mu i}$ is unitary, and
\begin{equation}
    \langle \overline{\phi}_\mu | = \sum_i C_{\mu i} \langle \psi_i |.
\label{Eq:app_AO_as_linear_combination_MO}
\end{equation}
Therefore, 
\begin{equation}
    a_\mu = \sum_i C_{\mu i} a_i.
\label{Eq:app_time_evolved_MO_fermion_operator}
\end{equation}
Since the AO basis is orthonormal here, there is no need to distinguish between covariant and contravariant indices.
The expectation value $\langle a^\dagger_\tau a_\lambda\rangle$ now becomes
\begin{align}
\begin{split}
    \langle a^\dagger_\tau a_\lambda\rangle &= \sum_{ij} C_{\lambda j} C^*_{\tau i} \langle a_i^\dagger a_j\rangle \\
    &= \sum_i C_{\lambda i} C^*_{\tau i} n_F(E_i) \\
    &=\int dE\, \text{PDOS}_{\lambda\tau}(E) n_F(E).
\label{Eq:two_point_correlation_no_time}
\end{split}
\end{align}
We have defined $n_F(E)$ as the Fermi-Dirac occupation number, i.e.,
\begin{equation}
    n_F(E) = \frac{1}{e^{\beta(E-\mu)}+1}.
\end{equation}
The numerator in \cref{Eq:app_p_sigma_1} is
\begin{align}
\begin{split}
    &\langle a^\dagger_\mu a_\nu a^\dagger_\sigma(t) a_\sigma(t) a^\dagger_\nu a_\mu\rangle \\
    =&
    n_\sigma \langle a^\dagger_\mu a_\nu a^\dagger_\nu a_\mu\rangle - (1-n_\nu)|\langle a^\dagger_\sigma(t) a_\mu\rangle|^2 \\
    &+n_\mu |\langle a_\sigma(t)a^\dagger_\nu\rangle|^2 + 2\text{Re}\big(\langle a^\dagger_\mu a_\nu\rangle\langle a^\dagger_\sigma(t)a_\mu\rangle \langle a_\sigma(t) a^\dagger_\nu\rangle\big).
\label{Eq:app_p_sigma_numerator}
\end{split}
\end{align}
Similar to \cref{Eq:two_point_correlation_no_time}, one can show that 
\begin{equation}
    \langle a^\dagger_\tau(t)a_\lambda\rangle = \int dE\,\text{PDOS}_{\lambda\tau}(E) n_F(E) e^{iEt/\hbar}
\label{Eq:two_point_correlation_1}
\end{equation}
and 
\begin{equation}
    \langle a_\tau(t)a^\dagger_\lambda\rangle = \int dE\,\text{PDOS}_{\tau\lambda}(E) (1-n_F(E)) e^{-iEt/\hbar}.
\label{Eq:two_point_correlation_2}
\end{equation}
Combining \cref{Eq:app_p_sigma_1,Eq:app_p_sigma_denominator,Eq:app_p_sigma_numerator}, we obtain
\begin{widetext}
\begin{align}
\begin{split}
    p_\sigma(t) = n_\sigma + \frac{ - (1-n_\nu)|\langle a^\dagger_\sigma(t) a_\mu\rangle|^2 +n_\mu |\langle a_\sigma(t)a^\dagger_\nu\rangle|^2 + 2\text{Re}\big(\langle a^\dagger_\mu a_\nu\rangle\langle a^\dagger_\sigma(t)a_\mu\rangle \langle a_\sigma(t) a^\dagger_\nu\rangle\big)}{|\langle a^\dagger_\mu a_\nu \rangle |^2 + n_\mu (1-n_\nu)}.
\label{Eq:electronic_dynamics_p_sigma}
\end{split}
\end{align}
\end{widetext}
\par
Using the fact that the two-point correlation functions (\cref{Eq:two_point_correlation_1,Eq:two_point_correlation_2}) vanish as $t\rightarrow \infty$, one can show that $p_\sigma(\infty) = n_\sigma$. Therefore, the correct equilibrium population is obtained as $t \rightarrow \infty$. If $\mu\neq\nu$, then at $t=0$, $p_\mu(0)=0$ and $p_\nu(0)=1$, as expected. 
\par
\begin{figure}
    \centering
    \includegraphics[width=0.9\linewidth]{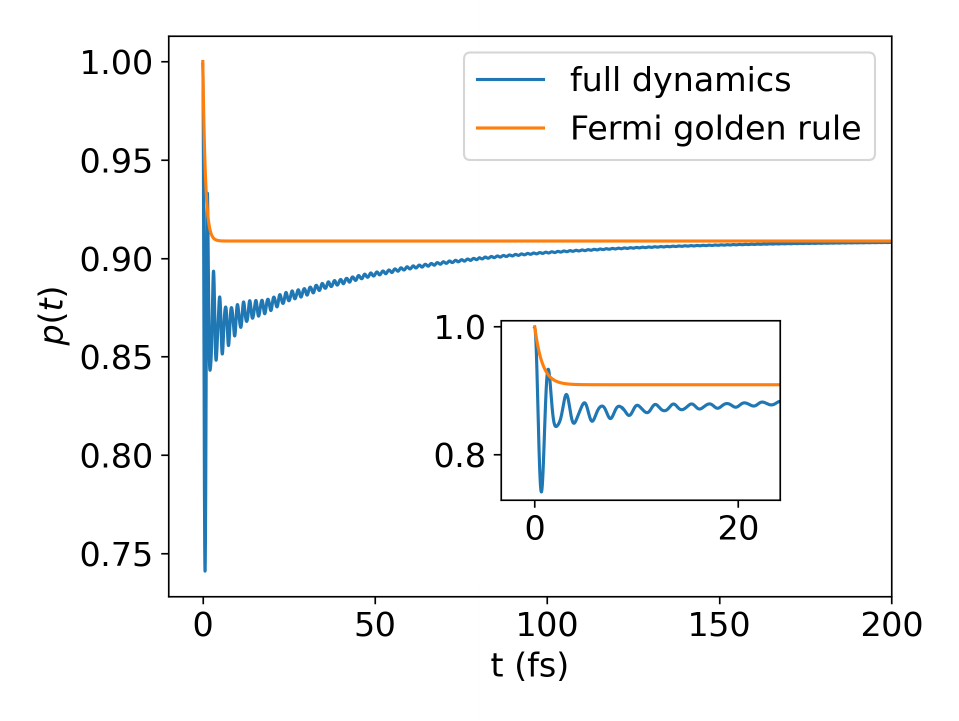}
    \caption{Population of the adsorbate electronic state $p(t)$ for a single-orbital AN model with a semi-elliptical $\Delta(E)$. We compare the full dynamics with the Fermi golden rule. The Fermi golden rule does not capture population oscillations or the slow-decaying component.}
    \label{fig:exact_electronic_dynamics_vs_FGR}
\end{figure}

\par
Taking the example of a single-orbital AN model with a semi-elliptical $\Delta(E)$, described in \cref{App:semi_elliptical_Delta_example}, we plot the full dynamics and the exponential decay due to the Fermi golden rule in \cref{fig:exact_electronic_dynamics_vs_FGR} (c). All three orbitals ($\mu$, $\nu$, and $\sigma$) in \cref{Eq:electronic_dynamics_p_sigma} are taken to be the single adsorbate orbital. Physically, we fill the adsorbate orbital from $\rho_{eq}$ at $t=0$ and track the adsorbate population as a function of time. The parameters are taken to be $(\alpha,\gamma,V,E_a,\mu)=(0, 1, 1, -1.8, 0)\text{ eV}$. The tunneling time from the Fermi golden rule is $1/\Gamma_{el}\approx 0.8$ fs. We plot the Fermi golden rule dynamics as an exponential decay from 1 to the equilibrium value of $\approx 0.91$.
The full dynamics shows a fast-decaying component on the order of 1 fs, followed by a slow-decaying component that reaches the equilibrium value. An oscillatory background exists throughout the full dynamics.
The slow decay indicates that the electron may be trapped in a localized state weakly coupled to the rest of the system.

\section{Vibrational relaxation rate via Fermi's golden rule}
\label{App:single_orbital_vib_relaxation_rate}

We will use Fermi's golden rule to show how the vibrational relaxation rate and electronic friction can be obtained from the AN model. We consider only one nuclear degree of freedom here. Let the nuclear potential $V_N(R)$ be a harmonic potential centered at $R_0$, then we can write the total Hamiltonian as 
\begin{equation}
    \hat{H}_{tot} = \hbar\Omega b^\dagger b + \hat{H}_{el}(R_0) + \sqrt{\frac{\hbar}{2m\Omega}}(b+b^\dagger) \hat{H}_{el}'(R_0),
\label{Eq:H_electron_vibration_coupling}
\end{equation}
where
$b$ is the bosonic annihilation operator for the vibrational mode, $m$ is the effective mass of the vibrational mode, and $\Omega$ is the vibrational mode frequency. $\hat{H}_{el}'$ is the derivative $d \hat{H}_{el}/dR$ at $R=R_0$. From \cref{Eq:H_tot}, $\hat{H}_{el}'$ is equal to
\begin{align}
\begin{split}
    \hat{H}_{el}'(R_0) = & \sum_{\mu,\nu\in\text{ad.}}a_\mu^\dagger (\overline{\mathbf{H}}'+[\mathbf{X},\overline{\mathbf{H}}])_{\mu\nu} a_\nu \\
    &+ \sum_{\substack{\mu\in\text{ad.}\\ \sigma\in\text{surf.}}} a_\mu^\dagger \big( \overline{\mathbf{V}}'+\mathbf{X}\overline{\mathbf{V}}\big)_{\mu \sigma} a_\sigma+\text{h.c.}
\label{Eq:H_el_derivative}
\end{split}
\end{align}
The overbar on $\overline{\mathbf{H}}$ and $\overline{\mathbf{V}}$ is a reminder that the underlying orbitals $\{|\overline{\phi}_\mu\rangle\}$ are orthonormal. Orthonormality is required in order for the fermionic operators to satisfy the canonical anti-commutation relations.
In \cref{Eq:H_el_derivative}, $\overline{\mathbf{H}}'$ and $\overline{\mathbf{V}}'$ are the matrix derivatives $d\overline{\mathbf{H}}/dR$ and $d\overline{\mathbf{V}}/dR$. $X$ is the derivative coupling matrix
\begin{equation}
    X_{\mu\nu} = \langle\overline{\phi}_\mu|\frac{d}{dR}\overline{\phi}_\nu\rangle.
\end{equation}
If we assume that the adsorbate orbital basis is large enough to capture $d\overline{\phi}_\nu/dR$, then
\begin{equation}
    \frac{d}{dR} a^\dagger_\nu = \sum_\mu a^\dagger_{\mu} X_{\mu\nu}.
\end{equation}
Orthonormality of the orbitals implies that $\mathbf{X}$ is anti-Hermitian, i.e., $\mathbf{X}^\dagger = -\mathbf{X}$.
Since $R$ is the nuclear coordinate of the adsorbate, we assume the surface orbitals $|\overline{\phi}_\sigma\rangle$ are independent of $R$. 

Treating the last term of \cref{Eq:H_electron_vibration_coupling} as the perturbation that couples electronic excitations to nuclear degrees of freedom, we derive the Fermi golden rule rates for transitions between different vibrational states $|n\rangle$. The initial electronic state is the thermal state of $\hat{H}_{el}(R_0)$. The excitation of a vibrational level is coupled to the de-excitation of an MO, and vice versa.

We write the Hamiltonian in \cref{Eq:H_electron_vibration_coupling} as the sum of 
\begin{equation}
    \hat{H}_0 = \hbar\Omega b^\dagger b + \hat{H}_{el}(R_0) 
\end{equation}
and
\begin{equation}
    \hat{H}_1 = \sqrt{\frac{\hbar}{2m\Omega}}(b+b^\dagger) \hat{H}_{el}'(R_0).
\end{equation}
Treating $\hat{H}_1$ as the perturbation, the Fermi golden rule rate for the vibrational transition $|1\rangle \rightarrow |0\rangle$ can be expressed in terms of the correlation function.~\cite{nitzan2024chemical}
\begin{equation}
    \gamma_0 = \frac{1}{\hbar^2}\int^\infty_{-\infty} d\tau \,\text{Tr}\Big(\langle 0|\hat{H}_1(\tau)|1\rangle\rho_{el}\langle 1 | \hat{H}_1|0\rangle\Big),
\label{Eq:app_gamma_0_correlation_0}
\end{equation}
where $\hat{H}_1(\tau)=e^{i\hat{H}_0\tau/\hbar}\hat{H}_1 e^{-i\hat{H}_0\tau/\hbar}$. $\rho_{el}$ is the initial electronic thermal state with respect to $\hat{H}_{el}(R_0)$ and chemical potential $\mu$. 
Simplifying \cref{Eq:app_gamma_0_correlation_0}, we have
\begin{equation}
    \gamma_0 = \frac{1}{2m\Omega\hbar}\int^\infty_{-\infty}d\tau \, e^{i\Omega\tau/\hbar} \Big\langle \hat{H}_{el}'(\tau)\hat{H}_{el}'\Big\rangle,
\end{equation}
where $\langle A\rangle$ denotes the expectation value with respect to $\rho_{el}$, and $\hat{H}_{el}(\tau) = e^{i\hat{H}_{el}\tau/\hbar}\hat{H}_{el}'e^{-i\hat{H}_{el}\tau/\hbar}$.
To further simply, we express $\hat{H}_{el}'$ in terms of the MO basis $\{|\psi_i\rangle\}$ of $\hat{H}_{el}$. $\hat{H}_{el}'(\tau)$ is written as
\begin{equation}
    \hat{H}_{el}'(\tau) = \sum_{i,l} \langle \psi_i|\hat{H}_{el}'|\psi_l\rangle e^{i(E_i-E_l)\tau/\hbar} a^\dagger_i a_l .
\end{equation}
Note that on the left-hand side, $\hat{H}_{el}'$ is a second-quantized operator, while on the right-hand side, $\hat{H}_{el}'$ is the corresponding single-particle operator. 
Therefore, $\gamma_0$ becomes
\begin{align}
\begin{split}
    \gamma_0 &= \frac{\pi}{m\Omega}\sum_{i,l} \delta(E_i+\hbar\Omega-E_l) n_F(E_i)(1-n_F(E_l))\\
    &\quad \langle \psi_i|\hat{H}_{el}'|\psi_l\rangle\langle\psi_l|\hat{H}_{el}'|\psi_i\rangle.
\label{Eq:single_orbital_vib_relax_rate_0}
\end{split}
\end{align}
We see that the vibrational transition $|1\rangle\rightarrow|0\rangle$ is accompanied by an electronic excitation $|\psi_l\rangle \rightarrow |\psi_i\rangle$ between the MO of $\hat{H}_{el}(R_0)$. 
\par
We define an energy-resolved density operator $\rho(E)$ as
\begin{align}
\begin{split}
    \hat{\rho}(E) &= \sum_i |\psi_i\rangle\langle \psi_i| \delta(E-E_i) \\
    &= \frac{1}{-2\pi i}(\hat{G}(E)-\hat{G}^\dagger(E)).
\label{Eq:P_in_terms_of_G}
\end{split}
\end{align}
Note that $\hat{\rho}(E)$ is different from the $\text{PDOS}(E)$ defined in \cref{Eq:multi_orb_PDOS_def} because $\hat{\rho}(E)$ acts on the entire space, while $\text{PDOS}^{\mu\nu}(E) = \langle \phi^\mu|\hat{\rho}(E)|\phi^\nu\rangle$ is equal to $\hat{\rho}(E)$ projected onto adsorbate orbitals. 
Now, $\gamma_0$ becomes
\begin{align}
\begin{split}
    \gamma_0 &= \frac{\pi}{m\Omega}\sum_i n_F(E_i)(1-n_F(E_i+\hbar\Omega)) \\
    &\qquad\text{Tr}(\hat{H}_{el}' \hat{\rho}(E_i+\hbar\Omega)\hat{H}_{el}' |\psi_i\rangle\langle\psi_i|)\\
    &=\frac{\pi}{m\Omega} \int dE\, n_F(E)(1-n_F(E+\hbar\Omega)) \\
    &\qquad \text{Tr}(\hat{H}_{el}' \hat{\rho}(E+\hbar\Omega)\hat{H}_{el}' \hat{\rho}(E)).
\end{split}
\end{align}
\par
We now make two approximations. First, we assume that $\hat{\rho}(E)$ is close to $\hat{\rho}(E+\Omega)$, so that
\begin{equation}
    \hat{\rho}(E)\approx \hat{\rho}(E+\hbar\Omega)\approx \hat{\rho}(E+\frac{\hbar\Omega}{2}).
\end{equation}
Second, we assume that $\hat{\rho}(E+\hbar\Omega/2)$ is proportional to a pure state $|\psi_0\rangle\langle\psi_0|$. This assumption is reasonable because in the absence of an adsorbate, the surface part $\hat{H}_\text{bath}$ has $\hat{\rho}(E_\sigma)$ proportional to the pure state $|\phi_\sigma\rangle\langle\phi_\sigma|$. However, this approximation can break down when different degenerate symmetry sectors of $\hat{H}_{el}$ mix with each other under the action of $\hat{H}_{el}'$. This can occur, for example, if moving $R$ away from $R_0$ breaks some crystal symmetry. Assuming the validity of these approximations, we have
\begin{align}
\begin{split}
    \gamma_0 &\approx \frac{\pi}{m\Omega}\int dE\, n_F(E)(1-n_F(E+\hbar\Omega)) \\
    &\qquad\qquad\qquad\qquad\text{Tr}\Big(\hat{\rho}(E+\frac{\hbar\Omega}{2})\hat{H}_{el}'\Big)^2 .
\label{Eq:single_orbital_vib_relax_rate_1}
\end{split}
\end{align}
\par
Expressing the operators inside the trace as matrices and using \cref{Eq:H_el_derivative}, we have
\begin{align}
\begin{split}
    \text{Tr}(\hat{\rho}(E)\hat{H}_{el}') = & \text{Tr}\Big(\bm{\rho}_{aa}(E) (\mathbf{H}'+[\mathbf{X},\mathbf{H}])\Big) \\
    & + \text{Tr}\Big(\bm{\rho}_{sa}(E)(\mathbf{V}_{as}'+\mathbf{X}\mathbf{V}_{as}) \Big) \\
    &+\text{Tr}\Big(\bm{\rho}_{as}(E)(\mathbf{V}_{sa}'-\mathbf{V}_{sa}\mathbf{X}) \Big),
\label{Eq:single_orbital_vib_relax_Tr_P_Hp}
\end{split}
\end{align}
where $\bm{\rho}_{aa}(E)$, $\bm{\rho}_{sa}(E)$, and $\bm{\rho}_{as}(E)$ are submatrices of the matrix representation of $\hat{\rho}(E)$. $a$ represents the adsorbate states, and $s$ represents the surface states. We have dropped the overbar in $\overline{\mathbf{H}}$ and $\overline{\mathbf{V}}$ for notational simplicity.
Expressing $\bm{\rho}(E)$ in terms of $\mathbf{G}(E)$ using \cref{Eq:P_in_terms_of_G},
\begin{align}
\begin{split}
    \text{Tr}(\hat{\rho}(E)\hat{H}_{el}') &= \\
    \qquad -\frac{1}{\pi} \text{Im }\Bigg(&\text{Tr}\Big(\mathbf{G}_{aa}(E) (\mathbf{H}'+[\mathbf{X},\mathbf{H}])\Big) \\
    &+\text{Tr}\Big(\mathbf{G}_{sa}(E)(\mathbf{V}_{as}'+\mathbf{X}\mathbf{V}_{as}) \Big) \\
    &+\text{Tr}\Big(\mathbf{G}_{as}(E)(\mathbf{V}_{sa}'-\mathbf{V}_{sa}\mathbf{X}) \Big)\Bigg).
\label{Eq:app_Tr_P_Hp_2}
\end{split}
\end{align}
\par
From \cref{App:Deriving_Gaa_multi_orb_AN}, the submatrices of $\mathbf{G}(E)$ are given by
\begin{equation}
    \mathbf{G}_{aa}(E) = \mathbf{Q}(E)^{-1},
\label{Eq:app_Gaa_phi_vib_relax}
\end{equation}
\begin{equation}
    \mathbf{G}_{sa}(E) = \lim_{\epsilon\rightarrow 0^+} (E-\mathbf{H}_{ss}+i\epsilon)^{-1} \mathbf{V}_{sa} \mathbf{Q}(E)^{-1},
\end{equation}
and
\begin{equation}
    \mathbf{G}_{as}(E) = \lim_{\epsilon\rightarrow 0^+} \mathbf{Q}(E)^{-1}\mathbf{V}_{as}(E-\mathbf{H}_{ss}+i\epsilon)^{-1}.
\label{Eq:app_G_ak_mu}
\end{equation}
\cref{Eq:app_G_ak_mu} is obtained from a different form of the Dyson equation
\begin{equation}
    \mathbf{G} = \mathbf{G}_0 + \mathbf{G}\mathbf{V}\mathbf{G}_0
\end{equation}
(to be compared with \cref{Eq:App_modified_Dyson}). Since the orbitals are orthonormal, there is no overlap and
\begin{equation}
    \mathbf{Q}(E) = E-\mathbf{H}-\mathbf{\Lambda}(E)+i\mathbf{\Delta}(E).
\end{equation}
\par
Substituting \crefrange{Eq:app_Gaa_phi_vib_relax}{Eq:app_G_ak_mu} into \cref{Eq:app_Tr_P_Hp_2}, we see that 
\begin{align}
\begin{split}
    &\text{Tr}(\hat{\rho}(E)\hat{H}_{el}')  \\
    &= -\frac{1}{\pi} \text{Im Tr}\Big(\mathbf{Q}(E)^{-1}\Big(\mathbf{H}'+\mathbf{\Lambda}'(E)-i\mathbf{\Delta}'(E)\\
    &\qquad\qquad\qquad\qquad\qquad+[\mathbf{X},\mathbf{H}+\mathbf{\Lambda}(E)-i\mathbf{\Delta}(E)]\Big)\Big) \\
    &= \frac{1}{\pi} \text{Im Tr}\Big(\mathbf{Q}(E)^{-1}\Big(\mathbf{Q}'(E)+[\mathbf{X},E-\mathbf{Q}(E)]\Big)\Big).
\label{Eq:app_single_orbital_vib_relax_Im_G_Hp}
\end{split}
\end{align}
$\mathbf{Q}'(E)$ denotes the derivative $\partial\mathbf{Q}(E)/\partial R$.
Using the invariance of trace under cyclic permutation, one can show that
\begin{equation}
    \text{Tr}(\mathbf{Q}(E)^{-1}[\mathbf{X}, E-\mathbf{Q}(E)]) = 0.
\end{equation}
Therefore, 
\begin{align}
\begin{split}
    \text{Tr}(\hat{\rho}(E)\hat{H}_{el}')&=\frac{1}{\pi}\text{Im Tr}\big(\mathbf{Q}(E)^{-1}\mathbf{Q}'(E)\big) \\
    &= \frac{1}{\pi}\text{Im}\frac{\partial}{\partial R} \ln\big(\det \mathbf{Q}(E)\big)\\
    &= \frac{\partial}{\partial R} \frac{1}{\pi} \arg \big( \det \mathbf{Q}(E)\big).
\end{split}
\end{align}
The equality between the first and the second lines follows from Jacobi's formula.\cite{bellman1997introduction}
Using the definition of the hybridization energy density function (see \cref{Eq:eta_def})
\begin{equation}
    \eta(E) = \frac{1}{\pi} \Big(\arg\big(\det \mathbf{Q}(E)\big) - \arg\big(\det \mathbf{Q}(-\infty)\big)\Big),
\end{equation}
we have
\begin{equation}
    \text{Tr}(\hat{\rho}(E)\hat{H}_{el}') = \frac{\partial \eta(E)}{\partial R}.
\label{Eq:app_Tr_P_Hp_d_eta_dR}
\end{equation}
\par
Substituting \cref{Eq:app_Tr_P_Hp_d_eta_dR} into \cref{Eq:single_orbital_vib_relax_rate_1}, we obtain the vibrational relaxation rate from $|1\rangle$ to $|0\rangle$ as
\begin{align}
\begin{split}
    \gamma_0 \approx \frac{\pi}{m\Omega}\int dE\, n_F(E)&(1-n_F(E+\hbar\Omega)) \\
    &\Big(\frac{\partial \eta(E+\hbar\Omega/2)}{\partial R}\Big)^2 .
\label{Eq:single_orbital_vib_relax_rate_2}
\end{split}
\end{align}
Similarly, the vibrational relaxation rate from $|0\rangle$ to $|1\rangle$ is 
\begin{align}
\begin{split}
    \gamma_{0\rightarrow 1} &\approx \frac{\pi}{m\Omega}\int dE\, (1-n_F(E))n_F(E+\hbar\Omega) \\
    &\qquad\qquad\qquad\qquad\Big(\frac{\partial \eta(E+\hbar\Omega/2)}{\partial R}\Big)^2 \\
    &=\gamma_0 e^{-\beta\hbar\Omega}.
\label{Eq:single_orbital_vib_relax_rate_heating_01}
\end{split}
\end{align}
Since $n_F(E)(1-n_F(E+\hbar\Omega))$ is nonzero in a small range around $\mu$, if we further approximate $\partial \eta(E+\hbar\Omega/2)/\partial R$ as $\partial \eta(\mu)/\partial R$, then \cref{Eq:single_orbital_vib_relax_rate_2} becomes
\begin{equation}
    \gamma_0\approx \frac{\pi\hbar}{m}(n_B(\hbar\Omega)+1)\Big(\frac{\partial \eta(\mu)}{\partial R}\Big)^2,
\label{Eq:app_gamma0}
\end{equation}
where
\begin{equation}
    n_B(E) = \frac{1}{e^{\beta E}-1}
\label{Eq:App_Bose_Einstein_num}
\end{equation}
is the Bose-Einstein occupation number with no chemical potential. We have used the identity
\begin{equation}
    \int dE\, n_F(E)(1-n_F(E+\hbar\Omega)) = \hbar\Omega(n_B(\hbar\Omega) + 1).
\end{equation}
\par
From the matrix elements of the ladder operators $b$ and $b^\dagger$, one can show that a general cooling rate $\gamma_{n\rightarrow n-1}$ from $|n\rangle$ to $|n-1\rangle$ is
\begin{equation}
    \gamma_{n\rightarrow n-1} = n \gamma_0.
\end{equation}
A general heating rate $\gamma_{n-1\rightarrow n}$ from $|n-1\rangle$ to $|n\rangle$ is
\begin{equation}
    \gamma_{n-1\rightarrow n} = n e^{-\beta\hbar\Omega}\gamma_0,
\end{equation}
satisfying detailed balance.

\begin{figure}[H]
    \centering
    \includegraphics[width=\linewidth]{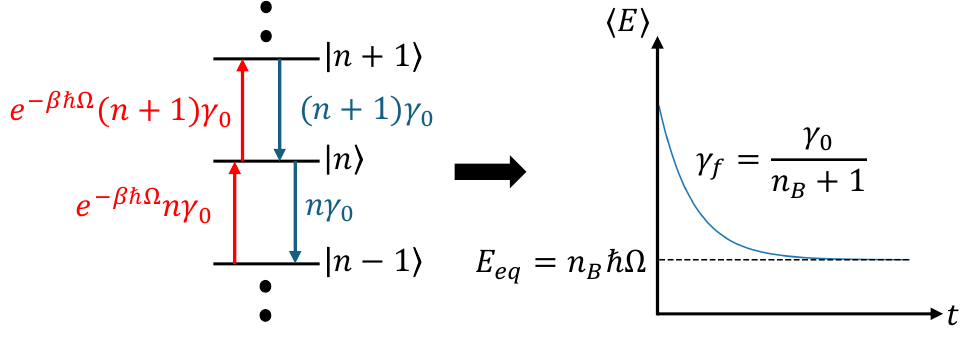}
    \caption{The transition rates between different vibrational states lead to energy dissipation with a rate of $\gamma_f = \gamma_0/(n_B + 1)$. The decay rate $\gamma_f$ is identified as the electronic friction, which is independent of the vibrational frequency $\Omega$.}
    \label{fig:vib_relax_to_friction}
\end{figure}

To obtain the electronic friction from the vibrational relaxation rates, we compare the energy dissipation dynamics from the rate equation model and that from a classical Langevin model.
The transition rates between different vibrational levels are depicted in \cref{fig:vib_relax_to_friction}. Under these transition rates, it can be shown that the average energy of the harmonic oscillator is given by,~\cite{nitzan2024chemical}
\begin{equation}
    \frac{d\langle E\rangle}{dt} = -\gamma_f\langle E\rangle + e^{-\beta\hbar\Omega}\gamma_0\hbar\Omega,
\label{Eq:average_E_rate_eqn}
\end{equation}
where the decay rate is
\begin{equation}
    \gamma_f = \frac{\gamma_0}{n_B(\hbar\Omega)+1}.
\label{Eq:vib_relax_gamma0}
\end{equation}
The qualitative behavior of $\langle E(t)\rangle$ in \cref{Eq:average_E_rate_eqn} is shown in \cref{fig:vib_relax_to_friction}. The average energy decays exponentially to the equilibrium energy $n_B(\hbar\Omega)\hbar\Omega$, with a decay rate of $\gamma_f$.
\par
To connect this energy dissipation rate of a quantum harmonic oscillator to classical friction, we consider the energy dissipation in the classical Langevin model. 
The Fokker-Planck equation for a Langevin particle with a mass of $m$ and a friction coefficient of $\gamma_c$ in a one-dimensional potential $V(x)$ is~\cite{nitzan2024chemical}
\begin{align}
\begin{split}
    \frac{\partial P(x,v,t)}{\partial t} = &\Big[ -v\frac{\partial}{\partial x}+\frac{1}{m}\frac{\partial V}{\partial x}\frac{\partial}{\partial v} \\
    &+\gamma_c \frac{\partial}{\partial v}(v+\frac{k_B T}{m}\frac{\partial}{\partial v}) \Big]P(x,v,t).
\end{split}
\end{align}
$P(x,v,t)$ is the probability distribution in position $x$ and velocity $v$.
The time evolution of the expectation value of an observable 
\begin{equation}
    \langle A\rangle = \int dxdv\, A(x,v)P(x,v,t)
\end{equation}
can be obtained using integration by parts, resulting in 
\begin{align}
\begin{split}
    \frac{\partial \langle A\rangle}{\partial t} = &\langle v\frac{\partial A}{\partial x}\rangle - \frac{1}{m}\langle \frac{\partial V}{\partial x} \frac{\partial A}{\partial v}\rangle  \\
    &- \gamma \langle v \frac{\partial A}{\partial v}\rangle +\frac{\gamma k_B T}{m}\langle\frac{\partial^2 A}{\partial v^2}\rangle.
\end{split}
\end{align}
Substituting in the total energy $A(x,v)=\frac{mv^2}{2}+V(x)$, we see that
\begin{equation}
    \frac{d\langle E\rangle}{dt} = -\gamma_c\langle mv^2\rangle +\gamma_c k_B T.
\label{Eq:Langevin_energy_dissipation_KE}
\end{equation}
In a harmonic potential, we expect the average kinetic energy to be the same as the average potential energy. Therefore, we replace $\langle mv^2\rangle$ in \cref{Eq:Langevin_energy_dissipation_KE} with $\langle E\rangle$.
Comparing \cref{Eq:average_E_rate_eqn,Eq:Langevin_energy_dissipation_KE}, we see that we can identify $\gamma_f$ in the rate equation model as the Langevin friction coefficient $\gamma_c$, so that \cref{Eq:Langevin_energy_dissipation_KE} is equal to the high-temperature limit of \cref{Eq:average_E_rate_eqn}. Combining \cref{Eq:app_gamma0,Eq:vib_relax_gamma0}, we obtain the electronic friction as
\begin{equation}
    \gamma_f = \frac{\pi\hbar}{m}\Big( \frac{\partial \eta(\mu,R)}{\partial R}\Big)^2.
\label{Eq:app_electronic_friction_no_smear}
\end{equation}
This equals the vibrational energy dissipation rate.
Notice that the electronic friction is independent of the vibrational frequency $\Omega$.
We note that Ref.~\onlinecite{brandbyge1995electronically} has derived the electronic friction in the case of a single-orbital AN Hamiltonian using path integral. The result is equal to \cref{Eq:electronic_friction_no_smear} smeared by a Fermi function, i.e., 
\begin{equation}
    \gamma_f = \frac{\pi\hbar}{m} \int dE\, \frac{\partial n_F(E)}{\partial E} \Big( \frac{\partial \eta(E,R)}{\partial R}\Big)^2.
\label{Eq:Brandbyge_electronic_friction}
\end{equation}

\end{document}